\input mtexsis
\def\M{{\cal M}}
\def\cont{{\,{\underline{\;\;\,}}{\kern -0.23em}|\;}}
\def\4{{\scriptscriptstyle 0}}
\def\x{{\vec{\rm x}}}
\def\boh{{\rm I{\kern -0.20em}H{\kern -0.20em}I}}
\def\bom{{\rm I{\kern -0.20em}M{\kern -0.20em}I}}
\def\real{{\rm I{\kern -0.20em}R}}
\def\comps{{\rm C{\mkern -8.5mu}
           {{\vrule height 7.8pt width 0.85pt depth -0.83pt}}}{\mkern 8.5mu}}

\def\proj{{\rm I{\kern -0.20em}P}}
\def\ident{{1{\kern -0.36em}1}}

\def\subd{{}_{{}_{\scriptscriptstyle {\cal D}}}}
\def\subk{{}_{{}_{\scriptscriptstyle {\cal K}}}}
\def\subA{{}_{{}_{\scriptscriptstyle {\cal A}}}}
\def\subo{{}_{{}_{\scriptscriptstyle {\rm O}}}}

\def\subn{{}_{{}_{\scriptscriptstyle {\rm N}}}}
\def\suba{{}_{{}_{\scriptscriptstyle 1}}}
\def\subb{{}_{{}_{\scriptscriptstyle 2}}}
\def\semidirect{{\ooalign{\hfil\raise.07ex\hbox{s}\hfil\crcr\mathhexbox20D}}}
\def\longlongrightarrow{\relbar\joinrel\longrightarrow}
\def\longlonglongrightarrow{\relbar\joinrel\longlongrightarrow}
\def\Tbf{\sixteenpoint\bf}
\def\boxit#1{\vbox{\hrule\hbox{\vrule\kern3pt\vbox{\kern-13pt#1\kern-13pt}
              \kern3pt\vrule}\hrule}}
\def\mapright#1{\smash{\mathop{\longlonglongrightarrow}\limits^{#1}}}
\def\mapdown#1{\Bigg\downarrow\rlap{$\vcenter{\hbox{$\scriptstyle#1$}}$}}

\referencelist

\reference{Kuchar}
K.~Kucha\v r, \journal Phys. Rev.;D22,1285 (1980)
\endreference

\reference{Large}
R.~Penrose, {\sl The Large, the Small and the Human Mind} (Cambridge
University Press, Cambridge, 1997), pp 90-92
\endreference

\reference{Time}
K.~Kucha\v r, in {\sl Proceedings of the 4th Canadian Conference on General
Relativity and Relativistic Astrophysics}, edited by G.~Kunstatter,
D.~Vincent, and J.~Williams (World Scientific, Singapore, 1992), pp 211-314;
and C.J.~Isham, in {\sl Integrable Systems, Quantum Groups, and Quantum Field
Theories}, edited by L.A.~Ibort and M.A.~Rodriguez (Kluwer Academic Publishers,
London, 1993), pp 157-288
\endreference

\reference{Cartan}
\'E.~Cartan, \journal Ann. \'Ecole Norm. Sup.;40,325 (1923)
\endreference

\reference{Friedrichs}
K.~Friedrichs, \journal Math. Annalen;98,566 (1927)
\endreference

\reference{Trautman}
A.~Trautman, \journal Comptes Rendus Acad. Sci.;257,617 (1963);
and A.~Trautman in {\sl Lectures in General Relativity}, edited by S.~Deser
and K.W.~Ford (Prentice-Hall, Englewood Cliffs, N.J., 1965), pp 1-248
\endreference

\reference{Havas}
P.~Havas, \journal Rev. Mod. Phys.;36,938 (1964)
\endreference

\reference{Dombrowski}
H.D.~Dombrowski and K.~Horneffer, \journal Math. Zeitschr.;86,291 (1964)
\endreference

\reference{Kunzle, 72}
H.P.~K\"unzle, \journal Ann. Inst. H. Poincar\'e;17,337 (1972)
\endreference

\reference{Dixon}
W.G.~Dixon, \journal Commun. Math. Phys.;45,167 (1975)
\endreference

\reference{Kunzle, 76}
H.P.~K\"unzle, \journal Gen. Relativ. Gravit.;7,445 (1976)
\endreference

\reference{Ehlers, 81}
J.~Ehlers,  in {\sl Grundlagenprobleme der modernen Physik}, edited by
J.~Nitsch et al. (Bibliographisches Institut, Mannheim, 1981), pp 65-84;
J.~Ehlers, in {\sl Logic, Methodology and Philosophy of Science
VII}, edited by B.~Marcus et al. (Elsevier Science Publishers B.V., 1986),
pp 387-403; and J.~Ehlers, in {\sl Classical Mechanics and Relativity:
Relationship and Consistency}, edited by G.~Ferrarese (Naples, Bibliopolis,
1991), pp 95-106
\endreference

\reference{Malament}
D.B.~Malament, in {\sl From Quarks to Quasars}, edited by R.G.~Colodny 
(University of Pittsburgh Press, Pittsburgh, 1986), pp 181-201
\endreference

\reference{Kretschmann}
E.~Kretschmann, \journal Annalen der Physik (Lpz.);53,575 (1917)
\endreference

\reference{Norton}
J.D.~Norton, \journal Rep. Prog. Phys.;56,791 (1993)
\endreference

\reference{Stachel}
J.~Stachel, in {\sl Philosophical Problems of the Internal and External
Worlds: Essays on the Philosophy of Adolf Gr\"unbaum}, edited by
J.~Earman, A.~Janis, G.~Massey and N.~Rescher (University of Pittsburgh Press,
Pittsburgh, PA, 1993), pp 129-160
\endreference

\reference{Einstein}
A.~Einstein, K\"oniglich Preussische Akademie der Wissenschaften
(Berlin). Sitzungsberichte: 1030-85 (1914)
\endreference

\reference{Peebles}
P.J.E.~Peebles, {\sl The Large-Scale Structure of the Universe} (Princeton
University Press, Princeton, New Jersey, 1980)
\endreference

\reference{Brauer}
U.~Brauer, A.~Rendall, and O.~Reula, \journal
Class. Quantum Grav.;11,2283 (1994)
\endreference

\reference{Ruede}
C.~R\"uede and N.~Straumann, \journal Helv. Phys. Acta;70,318 (1997)
\endreference

\reference{Buchert}
J.~Ehlers and T.~Buchert, \journal Gen. Relativ. Gravit.;29,733 (1997)
\endreference

\reference{Witten, 88}
E.~Witten, \journal Nuclear Physics;B311,46 (1988)
\endreference

\reference{Ashtekar, 96}
A.~Ashtekar and M.~Pierri, \journal J. Math. Phys.;37,6250 (1996)
\endreference

\reference{Christian}
J.~Christian, {\sl On Definite Events in a Generally Covariant Quantum
World}, Oxford University Preprint (1994)
\endreference

\reference{Shadows}
R.~Penrose, {\sl Shadows of the Mind}
(Oxford University Press, Oxford, 1994), p 339, and references therein
\endreference

\reference{Penrose, GR}
R.~Penrose, \journal Gen. Relativ. Gravit.;28,581 (1996), p 592
\endreference

\reference{Haag, 87}
K.~Fredenhagen and R.~Haag, \journal Comm. Math. Phys.;108, 91 (1987)
\endreference

\reference{Wald}
R.M.~Wald, {\sl General Relativity}
(University of Chicago Press, Chicago, 1984)
\endreference

\reference{Carter}
B.~Carter and I.M.~Khalatnikov, \journal Rev. Math. Phys.;6,277 (1994)
\endreference

\reference{Duval and Kunzle, 78}
C.~Duval and H.P.~K\"unzle, \journal Rep. Math. Phys.;13,351 (1978)
\endreference

\reference{Mala-New}
D.B.~Malament, \journal Philosophy of Science;62,489 (1995)
\endreference

\reference{Trumper}
M.~Trumper, \journal Ann Phys. (N.Y.);149,203 (1983)
\endreference

\reference{Duval-93}
C.~Duval, \journal Class. Quantum Grav.;10,2217 (1993)
\endreference

\reference{Duval and Kunzle}
C.~Duval and H.P.~K\"unzle, \journal Gen. Relativ. Gravit.;16,333 (1984)
\endreference

\reference{Earman}
J.~Earman, {\sl World enough and Spacetime} (MIT Press, Cambridge,
Massachusetts, 1989), Chapter 2
\endreference

\reference{Milne}
E.A.~Milne, \journal Quart. J. Math. (Oxford Series);5,64 (1934)
\endreference

\reference{Hawking}
S.W.~Hawking and G.F.R.~Ellis, {\sl The Large Scale Structure of Space-time}
(Cambridge University Press, Cambridge, 1973)
\endreference

\reference{Dirac}
R.~Arnowitt, S.~Deser, and C.~Misner, \journal Phys. Rev.;116,1322 (1959);
P.A.M.~Dirac, {\sl Lectures on Quantum Mechanics} (Belfer Graduate School of
Science, Yeshiva University, New York, 1964), Chapters 3 and 4;
K.~Kucha\v r, \journal J. Math. Phys.;17,801 (1976); C.J.~Isham and
K.~Kucha\v r, \journal Annals of Physics;164,288 (1985)
\endreference

\reference{Kuchar, Boston}
K.~Kucha\v r, in {\sl Conceptual Problems of Quantum Gravity}, edited by
A.~Ashtekar and J.~Stachel (Birkh\"auser, Boston, 1991), pp 141-168
\endreference

\reference{De Bievre}
S.~De Bi\'evre, \journal Class. Quantum Grav.;6,731 (1989)
\endreference

\reference{Schweber}
S.S.~Schweber, {\sl An Introduction to Relativistic Quantum Field Theory}
(Row, Peterson and Company, Evanston, Illinois, 1961), Chapter 6
\endreference

\reference{Brown}
L.S.~Brown, {\sl
Quantum Field Theory} (Cambridge University Press, Cambridge, 1992), Chapter 2
\endreference

\reference{Gross}
F.~Gross, {\sl Relativistic Quantum Mechanics and Field Theory} (John Wiley and
\& Sons, New York, 1993), Chapter 7
\endreference

\reference{Lee}
J.~Lee and R.M.~Wald, \journal J. Math. Phys.;31,725 (1990)
\endreference

\reference{Barnich}
G.~Barnich, M.~Henneaux, and C.~Schomblond, \journal Phys. Rev.;D44,R939 (1991)
\endreference

\reference{Torre}
C.G.~Torre, \journal J. Math. Phys.;33,3802 (1992)
\endreference

\reference{Goenner}
H.F.M.~Goenner, \journal Gen. Relativ. Gravit.;16,513 (1984)
\endreference

\reference{remark}
C.~Duval and H.P.~K\"unzle, in {\sl Semantical Aspects of
Spacetime Theories}, edited by U.~Majer and H.-J.~Schmidt 
(B.I. Wissenschaftsverlag, Mannheim, 1994), pp 113-129
\endreference

\reference{Woodhouse}
N.M.J.~Woodhouse, {\sl Geometric Quantization} (Clarendon Press, Oxford, 1991)
\endreference

\reference{Witten}
C.~Crnkovic and E. Witten, in {\sl Three Hundred Years of Gravitation},
edited by S.W.~Hawking and W.~Israel (Cambridge University Press, Cambridge,
1987), pp 676-684
\endreference

\reference{Ashtekar}
A.~Ashtekar, L.~Bombelli, and O.~Reula, in {\sl Mechanics, Analysis and
Geometry: 200 Years after Lagrange}, edited by M.~Francaviglia (Elsevier
Science Publishers B.V., 1991), pp 417-450
\endreference

\reference{Abraham}
R.~Abraham and J.E.~Marsden, {\sl Foundations of Mechanics}, Second edition
(Benjamin, Reading, Massachusetts, 1978)
\endreference

\reference{Arnold}
V.I.~Arnold, {\sl Mathematical Methods of Classical Mechanics}
(Springer-Verlag, New York, 1989)
\endreference

\reference{Wald-94}
R.M.~Wald, {\sl Quantum Field Theory in Curved Spacetime and Black Hole
Thermodynamics} (University of Chicago Press, Chicago, 1994)
\endreference

\reference{Chernoff}
P.R.~Chernoff, \journal Hadronic Journal;4,879 (1981)
\endreference

\reference{Haag}
R.~Haag, {\sl Local Quantum Physics} (Springer-Verlag, Berlin, 1992)
\endreference

\reference{Bohr}
N.~Bohr and L.~Rosenfeld,
\journal Mat. Fys. Medd. K. Dan. Vidensk. Selsk.;12,No.8 (1933)
\endreference

\reference{Fell}
J.M.G.~Fell, \journal Trans. Am. Math. Soc.;94,365 (1960)
\endreference

\reference{Okun-1991}
L.B. Okun, \journal Sov. Phys. Usp.;34,818 (1991)
\endreference

\reference{Christian-2001}
J. Christian, in {\sl Physics Meets Philosophy at the Planck Scale},
edited by C. Callender and N. Huggett (Cambridge University Press,
Cambridge, England, 2001)
\endreference

\reference{Gorelik-1994}
G. E. Gorelik and V. Ya. Frenkel, {\sl Matvei Petrovich Bronstein
and Soviet Theoretical Physics in the Thirties} (Birkh\"auser Verlag,
Basel, 1994)
\endreference

\endreferencelist

\titlepage
\hoffset=-0.35cm
\voffset=1.0cm
\superrefsfalse
\headline={\hfil}
\footline={\hfil}
\title
{\Tbf{Exactly ~Soluble ~Sector ~of ~Quantum ~Gravity}}
\endtitle
\vskip 0.25in
\center
Joy ~Christian
\smallskip\tenpoint
{\it Wolfson College, University of Oxford, Oxford OX2 6UD, United Kingdom}
\smallskip\tenpoint
{\it E-mail} : {\it joy.christian@wolfson.oxford.ac.uk}
\endcenter
\vskip 0.30in
\baselineskip 0.5cm

\bigskip

\abstract
{\tenpoint Cartan's spacetime reformulation of the Newtonian theory of gravity
is a generally-covariant Galilean-relativistic limit-form of Einstein's theory
of gravity known as the Newton-Cartan theory. According to this theory, space
is flat, time is absolute with instantaneous causal influences, and the
degenerate `metric' structure of spacetime remains fixed with two mutually
orthogonal non-dynamical metrics, one spatial and the other temporal. The
{\it spacetime} according to this theory is, nevertheless, {\it curved}, duly
respecting the principle of equivalence, and the non-metric gravitational
connection-field is {\it dynamical} in the sense that it is determined
by matter distributions.
Here, this generally-covariant but Galilean-relativistic theory of gravity
with a possible non-zero cosmological constant, viewed as a parameterized
gauge theory of a gravitational vector-potential minimally
coupled to a complex Schr\"odinger-field (bosonic or fermionic), is
successfully
cast --- for the first time --- into a manifestly covariant Lagrangian form.
Then, exploiting the fact that Newton-Cartan
spacetime is intrinsically globally-hyperbolic with a fixed causal structure,
the theory is recast both into a {\it constraint-free} Hamiltonian form in
3+1-dimensions and into a manifestly covariant reduced phase-space form with
{\it non-degenerate} symplectic structure in 4-dimensions. Next, this
Newton-Cartan-Schr\"odinger system is non-perturbatively quantized using the
standard C${^*}$-algebraic technique combined with the geometric procedure of
manifestly covariant phase-space quantization. The ensuing unitary quantum
field theory of Newtonian gravity coupled to Galilean-relativistic matter is
not only generally-covariant, but also {\it exactly soluble} and --- thanks to
the immutable causal structure of the Newton-Cartan spacetime --- free of all
conceptual and mathematical difficulties usually encountered in quantizing
Einstein's theory of gravity. Consequently, the resulting theory of quantized
Newton-Cartan-Schr\"odinger system constitutes a perfectly consistent
Galilean-relativistic sector of
the elusive full quantum theory of gravity coupled to relativistic matter,
regardless of what ultimate form the latter theory eventually takes.}
\endabstract

\bigskip
\bigskip
\bigskip
\bigskip
\bigskip
\bigskip

\center
Journal-ref: {\sl Physical Review} D56, No. 8, 15 Oct. 1997, pp 4844-4877
\endcenter
\endtitlepage

\hoffset=-0.35cm
\voffset=1.0cm
\superrefsfalse
\footline={\hfil}
\headline={\tenrm\hfil\folio}
\pageno=2
\parindent 0.75cm
\baselineskip 0.75cm
\input epsf

\section{Introduction}

The primary aim of this paper is to demonstrate that the principle of
equivalence by itself is not responsible for the conceptual and mathematical
difficulties encountered in constructing a viable quantum theory of gravity;
rather, it is the conjunction of this principle with the conformal structure
(i.e., the field of light-cones) of the general-relativistic spacetime which
resists subjugation to the otherwise well-corroborated canonical rules of
quantization. This elementary fact, of course, has been
duly appreciated by the workers in the field at least implicitly since the
earliest days of attempts to quantize Einstein's theory of gravity. Curiously
enough, however, except for a partial illustration of this state of affairs by
Kucha\v r\cite{Kuchar}, so far the problem of explicitly constructing a
generally-covariant but Galilean-relativistic quantum theory of gravity ---
i.e., a generally-covariant quantum field theory of spacetime with degenerate
structure of `flattened' light-cones but unique
{\it dynamical}{\parindent 0.40cm
\baselineskip 0.53cm\footnote{$^{\scriptscriptstyle 1}$}{\ninepoint{\hang
The adjective `dynamical' here and below simply refers to the mutability
of spacetime structure dictated by evolving distributions of matter. It only
refers to the fact that, even in a Galilean-relativistic theory, spacetime
is not fixed {\it a priori}. In particular, unlike the case in general
relativity, it does not refer to any transverse propagation degrees of
freedom associated with the spacetime structure since nonrelativistic gravity
does not possess such a freedom. See subsection 5.4, however, for a discussion
on the {\it longitudinal} degrees of freedom of the gravitational field.\par}}}
connection --- has been completely neglected\cite{Large}.
Here we set out to construct such
a `nonrelativistic' theory and explicitly demonstrate that, unlike the case of
relativistic quantum gravity, there are no insurmountable
conceptual or mathematical
difficulties in achieving this goal. In particular, we show that the
Galilean-relativistic limit-form (see Figure 1) of the as-yet-untamed full
quantum theory of gravity interacting with matter
is {\it exactly soluble}, and that in this very classical
(`c$\;=\infty$') domain `the problem of time'\cite{Time} --- the well-known
central stumbling block encountered in quantizing Einstein's gravity --- and
related problems of causality, along with other conceptual and mathematical
difficulties, are nonexistent.

\midfigure{monolith}
\hrule
\vskip 0.85cm
\centerline{\epsfysize=10cm \epsfbox{q-gra-fig.eps}}
\vskip 0.65cm
\hrule
\bigskip
\smallskip
\parindent 0.77cm 
\baselineskip 0.5cm
{\narrower\smallskip
{\vbox to 9.8cm{\elevenpoint\noindent {\bf Figure 1:}
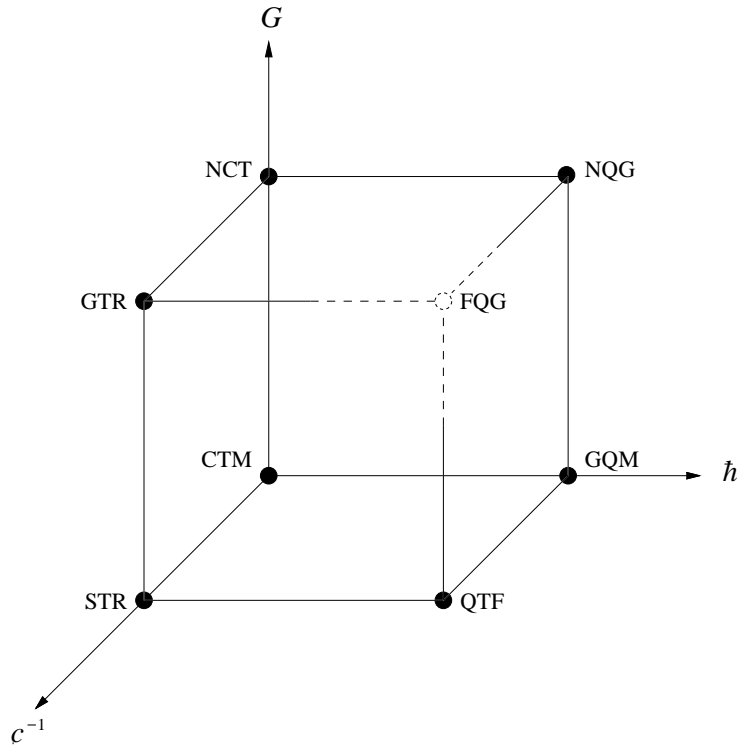
The great dimensional monolith of physics indicating the fundamental role
played by the three universal
constants $G$ (the Newton's gravitational constant),
$\hbar$ (the Planck's constant of quanta divided by $2\pi$), and $c$ (the
absolute upper bound on the speed of causal influences) in various basic
physical theories. These theories, appearing at the eight vertices of the cube,
are: CTM = Classical Theory of Mechanics, STR = Special Theory of Relativity,
GTR = General Theory of Relativity, NCT = (classical) Newton-Cartan Theory, NQG
= (generally-covariant) Newtonian Quantum Gravity (constructed in
this paper), GQM = Galilean-relativistic Quantum Mechanics, QTF = Quantum
Theory of (relativistic) Fields, and FQG = the elusive Full-blown Quantum
Gravity. If FQG turns out to require some additional fundamental constants ---
like the constant ${\alpha'\equiv(2\pi T)^{-1}}$ of the string theory
controlling the string tension ${T}$, for example ---
then, of course, the above
representation of the foundational theories would be inadequate, and
additional axes corresponding to such constants ${\alpha_i\,}$,
${i = {\scriptstyle 1, 2,\dots,} n\,}$, would have to be added
to the diagram, making it a ${3+n}$ dimensional hypercube. 
In that case, NQG, in particular, would be a limit-form of FQG
with respect to total ${n+1}$ limits, ${\alpha_i\rightarrow 0}$
and ${c\rightarrow\infty}$, in conjunction. (A diagrammatic illustration of
existing and future physical theories in terms
of the fundamental constants was first given by Kucha\v r in the form of a
pyramid in \Ref{Kuchar}. This was adapted in \Ref{Christian}, from which
Penrose was inspired to (characteristically) unfold the pyramid into a cube.
The figure here, in turn, has been inspired by the cubic version
presented by Penrose in his 1994-95 Tanner lectures delivered at the University
of Cambridge, UK.
These lectures are now published in \Ref{Large}.)}}\smallskip}
\endfigure

\parindent 0.75cm
\baselineskip 0.75cm

The nonrelativistic limit-form of Einstein's theory of gravity is a
spacetime reformulation of the Newtonian theory of gravity --- the so-called
Newton-Cartan theory. With the hindsight of Einstein's theory it is
clear that gravitation should be treated as a consequence of the curving of
spacetime rather than as a force-field even at the Newtonian level because the
principle of equivalence is equally compatible with the Newtonian spacetime.
A geometrical description of the Newtonian spacetime explicitly incorporating
the principle of equivalence at the classical (non-quantal) level was 
given by Cartan\cite{Cartan} and Friedrichs\cite{Friedrichs} soon after the
completion of Einstein's theory, and later further developed by many 
authors\cite{Trautman}\cite{Havas}\cite{Dombrowski}\cite{Kunzle,
72}\cite{Dixon}. The outcome of such a reformulation of the Newtonian
theory is a theory whose qualitative
features lie in-between those of special relativity with its completely fixed
spacetime background and general relativity with no background structure
whatsoever. Unlike the latter two well-known theories, Newton-Cartan theory
has two fixed and degenerate metrics --- a temporal metric and a spatial
metric, vaguely resembling the fixed Minkowski metric as far as their
non-dynamical character is concerned --- and a non-metric but
{\it dynamical}${{\,}^{\scriptscriptstyle 1}}$
connection-field mimicking Einstein's metric connection-field to some
extent. Thus, unlike the static and flat Galilean spacetime, and analogous to
the mutable general-relativistic spacetime, the
generally-covariant{\parindent 0.39cm
\baselineskip 0.53cm\footnote{$^{\scriptscriptstyle 2}$}{\ninepoint{\hang The
philosophical dispute over
the meaning of the principle of general covariance begun almost immediately
after the completion of Einstein's theory of gravity\cite{Kretschmann} and
persists today\cite{Norton}. Einstein, for instance, read through Stachel's
eyeglasses\cite{Stachel}, would not view Newton-Cartan theory as a
genuinely generally-covariant theory because, unlike general relativity,
it does not avert the `hole
argument'\cite{Einstein}. As far as this paper is concerned, however, we can
afford to refrain from the controversy and join the bandwagon calling
Newton-Cartan spacetime generally-covariant simply because it respects the
principle of equivalence and precludes
existence of global inertial frames of reference. Moreover, at least in
the framework followed in this paper, the non-metric Newton-Cartan theory is as
diffeomorphism-invariant as Einstein's metric theory of gravity (cf. equation
\Ep{gene-cove}). (See also footnote ${\scriptstyle 5}$.)\par}}}
Newton-Cartan spacetime is dynamical, curved by the Newtonian gravity, and
requires no {\it a priori} assumption of a global inertial frame.
Consequently, the transition
from Galilean spacetime to this general-{\it non}relativistic Newton-Cartan
spacetime drastically changes the {\it qualitative} features of the
Galilean-relativistic physics by elevating the status of the affine connection
from that of an absolute element --- given once and for all --- to a 
dynamical quantity determined by the distribution of matter. Furthermore,
it is this Newton-Cartan theory of gravity with its mutable spacetime which is
in general the true Galilean-relativistic limit-form of
Einstein's general theory of
relativity\cite{Dixon}\cite{Kunzle, 76}\cite{Ehlers, 81}\cite{Malament},
and not the Newtonian theory on the immutable background of flat Galilean
spacetime (cf. subsection 2.3 below). In summary, Cartan's geometric
reformulation of Newton's theory of gravity makes it a {\it local} field
theory analogous to general relativity, and, as a result, the instantaneous
gravitational interactions between gravitating bodies can now be understood
as propagating continuously through the curvature of the region of
spacetime among them.

To understand these stipulations in detail, in section 2 we accumulate various
scattered results to provide a coherent --- but by no means exhaustive ---
review of the classical (non-quantal) Newton-Cartan theory of gravity. Then,
in section 3, we review the previous work on one-particle Schr\"odinger quantum
mechanics on the curved Newton-Cartan spacetime initiated by
Kucha\v r\cite{Kuchar}, and extend it to a Galilean-relativistic quantum field
theory on such a spacetime. Next, in section 4, we derive,
{\it for the first time},
the complete classical Newton-Cartan-Schr\"odinger theory (i.e., the theory of
classical Newton-Cartan field interacting with a Galilean-relativistic matter)
from extremizations of a single
diffeomorphism-invariant action functional. This functional --- defined on an
arbitrary measurable region of spacetime --- is carefully selected to allow
recasting of the
theory into a constraint-free Hamiltonian form in 3+1 dimensions, as well as
into a manifestly covariant reduced phase-space form in 4-dimensions (where the
phase-space is viewed as the space of solutions of the equations of motion
modulo gauge-transformations). Thus obtained covariant and constraint-free
phase-space then paves the way in section 5 for a straightforward quantization
of the Newton-Cartan-Schr\"odinger
system using the standard C${^*}$-algebraic techniques and the
associated representation theory. The resulting local quantum field
theory describes an {\it exactly} soluble interacting matter-gravity system.

In addition to the primary aim discussed in the beginning of this Introduction,
there are various other motivations for the present exercise which we enumerate
here (counting the primary aim as (1)):

\item{(2)} As we shall see in the next section, one of the set of gravitational
           field equations of Newton-Cartan theory closely resembles
           Einstein's field equation ${G_{\mu\nu}\,=\,8\pi\,T_{\mu\nu}}$
           relating geometry to matter. In particular, it dictates that the
           connection-field of Newton-Cartan gravity must be determined by the
           distribution of matter. Therefore, consistency requires that
           this connection-field must be quantized along with matter since it
           participates in the dynamical unfolding of the combined system.

\item{(3)} It is well-known that Newtonian models play quite a significant role
           in cosmology. In particular, they are useful in studying structure
           formation in the early universe\cite{Peebles} and provide useful
           insights for the relativistic case. Recently, the need to generalize
           Newtonian cosmology to Newton-Cartan cosmology has been recognized,
           and a considerable progress has been made in this
           direction\cite{Brauer}\cite{Ruede}\cite{Buchert}. In this context,
           then, the relevance of an exactly soluble quantum theory of
           Newton-Cartan gravity interacting with matter cannot be
           overestimated.

\item{(4)} It is fair to say that we know very little about the quantum gravity
           proper (i.e., FQG in the Figure 1). Therefore, insights coming form
           any quarters which enable us to better understand the difficulties
           of constructing the final theory should be welcomed. It is with this
           attitude that the recent exercises on exactly soluble gravitating
           systems in reduced spacetime dimensions\cite{Witten, 88} and/or
           reduced symmetries\cite{Ashtekar, 96} have been carried out, and
           used as probes to investigate the conceptual problems of the full
           quantum gravity. Here we do not reduce either symmetries or the
           spacetime dimensions, but instead provide an exactly soluble system
           in the full 4-dimensional setting --- albeit only in the
           Galilean-relativistic limit of the full theory.

\item{(5)} In \Ref{Christian} we have argued, on both group theoretical and
           physical grounds, that, since Newton-Cartan symmetries --- duly
           respecting the principle of equivalence --- and not
           Galilean symmetries are the true spacetime symmetries of the
           nonrelativistic quantum domain, any discussion on the conceptual
           issues like the `measurement problem' in this domain must be carried
           out within the Newton-Cartan framework. In fact, it was argued,
           Penrose-type speculations of gravitationally induced
           state-reduction\cite{Shadows} might greatly
           benefit from analyzing the relevant physical systems within this
           framework. It is gratifying to note that this suggestion has already
           attracted at least
           Penrose's attention\cite{Shadows}\cite{Penrose, GR}. However, the
           complete framework for such deep conceptual issues must be the fully
           quantized Newton-Cartan gravity interacting with matter --- and
           hence the present work.

\item{(6)} The theory constructed in this paper provides a selection criterion
           for any exotic, top-down approach (e.g., the superstring approach)
           to the final `theory of everything.' Clearly, any
           general-relativistic exotic theory would
           lose its physical relevance if it does not reduce to the Newtonian
           quantum gravity interacting with Schr\"odinger-fields in the
           Galilean-relativistic limit. Therefore, any future theory must
           reduce to NQG of Figure 1 in the `${c\rightarrow\infty}$' limit,
           as much as it should reduce to GTR in the `${\hbar\rightarrow 0}$'
           limit and QTF in the `${G\rightarrow 0}$' limit.

\item{(7)} Finally, the existence of NQG opens up a completely novel direction
           of research in the full quantum gravity. In majority of 
           orthodox approaches to quantum gravity the direction of research has
           been to go from GTR to FQG (cf. Figure 1) --- i.e., the program has
           been to quantize the general theory of relativity. Somewhat less
           popular and less explored program is to go from QTF to FQG --- i.e.,
           to general-relativize the quantum theory of fields\cite{Haag, 87}.
           The existence of NQG opens up a third possibility, that of starting
           from NQG and arriving at FQG by undoing the
           `${c\rightarrow\infty}$' limit --- i.e., by special-relativizing
           the Newton-Cartan quantum gravity.

\section{Spacetime approach to Newtonian gravity}

In this section we review the covariant, spacetime reformulation of the
Newtonian theory of gravity, and, thereby, set the notations and conventions
to be used in the following sections. Most of the ideas presented in this and
the next section are not new, but it is the manner in which they are organized
here that makes them conducive to the fruitful results of the later sections.

\subsection{General Galilean spacetime}

We begin with the familiar spacetime structure ${(\M;\;t_{\alpha},\;
h^{\alpha\beta},\;\nabla_{\!\alpha})}$ presupposed by the usual
Galilean-relativistic dynamics delineated in Penrose's abstract index
notation\cite{Wald} using the greek
alphabet\cite{Cartan}\cite{Kunzle, 72}\cite{Malament}.
{\it Spacetime} --- the arena in which physical events and processes take
place --- is represented by a real, contractible, and differentiable Hausdorff
4-manifold $\M$ without boundaries. 
Unlike the case in general-relativistic spacetimes, here
spatial and temporal measures on $\M$ are not taken to be soldered into a
single semi-Riemannian metric, but appear as two distinct geometric entities.
A smooth, never vanishing covariant vector field ${t_{\alpha}}$ on $\M$ is
defined which induces a degenerate
{\it temporal metric} ${t_{\mu\nu}=t_{(\mu\nu)}:=t_{\mu}t_{\nu}\,}$ 
of signature (+ 0 0 0) specifying durations of processes occurring between
events, and hence induces a degenerate `cone structure' on the tangent space
${T_x\M}$ at each point ${x\in\M\,}$ (here the parentheses indicate
symmetrization with respect to the enclosed indices).
A vector ${\xi^{\mu}}$ at a point on $\M$ is said to be {\it timelike} if
${\sqrt{t_{\mu\nu}\xi^{\mu}\xi^{\nu}}>0\,}$, {\it spacelike} if
${\sqrt{t_{\mu\nu}\xi^{\mu}\xi^{\nu}}=0\,}$, and {\it future-directed} if
${t_{\mu}\xi^{\mu}>0\,}$. Between spacelike vectors ${\xi^{\alpha}}$ on $\M$
with vanishing `temporal lengths' --- i.e.,
${\sqrt{t_{\mu\nu}\xi^{\mu}\xi^{\nu}}=0\,}$ --- there is defined an inner
product with the help of a smooth, symmetric, contravariant
vector field ${h^{\alpha\beta}=h^{(\alpha\beta)}}$ on $\M$ which
serves as a degenerate {\it spatial metric} of signature (0 + + +), and
indirectly assigns lengths to these vectors: ${|\xi|:=\sqrt{h^{\mu\nu}
\lambda_{\mu}\lambda_{\nu}}\,}$, where ${h^{\mu\nu}\lambda_{\nu}=\xi^{\mu}}$
with an arbitrary choice of ${\lambda_{\mu}}$. As we have done here,
the metric tensor ${h^{\mu\nu}}$ can be used to raise indices;
however, since it is not invertible, it cannot be
used to lower indices. Thus, the distinction between covectors and
contravectors has much greater significance in this general
Galilean-relativistic spacetime than in the semi-Riemannian
general-relativistic spacetime. The {\it affine structure} of the spacetime
$\M$ is represented by a smooth derivative operator ${\nabla_{\!\alpha}}$ 
introducing a (not necessarily `flat') torsion-free linear connection
${\Gamma}$ on $\M$. The two tensor fields ---
${\tau := t_{\alpha}dx^{\alpha}}$
measuring the proper time of world lines and ${h :=
h^{\alpha\beta}\partial_{\alpha}\otimes\partial_{\beta}}$ inducing a 3-metric
on the null space of ${\tau\,}$ --- are mutually orthogonal, 
$$
\tau\cont h \;=\; 0 \;=\; h^{\alpha\beta}t_{\beta}\EQN orthogonal
$$
(the kernel of ${h}$ generating the span of the 1-form ${\tau\,}$), and taken
to be compatible with the derivative operator:
$$
\nabla_{\!\alpha}h^{\beta\gamma} = 0 = \nabla_{\!\alpha}t_{\beta}\;.
\EQN integrab
$$

A Galilean spacetime is orientable as there exists a 4-volume element for
the structure ${(\M;\;t_{\alpha},\;h^{\alpha\beta},\;\nabla_{\!\alpha})}$.
Given the structure
${(\M;\;t_{\alpha},\;h^{\alpha\beta},\;\nabla_{\!\alpha})}$
satisfying the conditions \Ep{integrab}, a continuous, nowhere vanishing
spacetime measure form on $\M$ with tensor components
${{\scriptstyle{\cal E}}_{\alpha\beta\gamma\delta} = 
{\scriptstyle{\cal E}}_{[\alpha\beta\gamma\delta]}}$
can be derived\cite{Carter} such that
${\nabla_{\!\mu}\,{\scriptstyle{\cal E}}_{\alpha\beta\gamma\delta} = 0}$
(where the square brackets indicate
anti-symmetrization with respect to the bracketed indices).
If one defines an oriented {\it Galilean frame}
at a point ${x\in\M}$ as a basis ${\{e_i\}}$ of the tangent space
${T_x\M}$ such that ${e^{\alpha}_it_{\alpha} = \delta^0_i}$ and 
${h^{\mu\nu}\theta^i_{\mu}\theta^j_{\nu} = \delta^i_a\delta^{ab}\delta^j_b}$
(where ${\scriptstyle{\alpha,\,i,\,j\,=\,0,\dots,3}}$ and
${\scriptstyle{a,\,b\,=\,1,\,2,\,3}}$), with ${\{\theta^i\}}$ being
the dual basis of ${\{e_i\}}$, then for any such Galilean frame ${\{e_i\}}$
the canonical 4-volume element can be defined by\cite{Kunzle, 76}
$$
\wp\,d^4x
\;:=\;{1\over{4!}}\;{\scriptstyle{\cal E}}_{\alpha\beta\gamma\delta}\;
dx^{\alpha}\wedge dx^{\beta}\wedge dx^{\gamma}\wedge dx^{\delta}
\;:=\;\theta^0\wedge\theta^1\wedge\theta^2\wedge\theta^3\,.\EQN vol
$$

The compatibility of the temporal metric ${\nabla_{\!\alpha}t_{\beta} = 0}$,
giving the condition ${\nabla_{\lbrack\gamma}\,t_{\delta\rbrack} = 0\,}$, 
at least locally allows the relation ${t_{\beta} = \nabla_{\!\beta}t}$ for some
time function $t\,$. Since $\M$ is contractible by
definition, the Poincar\'e lemma allows one to
define the absolute time function also globally by a map ${t: \M\rightarrow
{\real}\,}$, foliating the spacetime {\it uniquely} into one-parameter family
of (not necessarily flat) spacelike hypersurfaces{\parindent 0.40cm
\baselineskip 0.53cm\footnote{$^{\scriptscriptstyle 3}$}{\ninepoint{\hang
As is well-known, hypersurfaces are best represented by embeddings. A
one-parameter family of embeddings in the present context is a map
${{{}^{(t)}\!e}\,:\,\Sigma\rightarrow\M}$ which takes a point ${x^a}$ from the
spatial submanifold ${\Sigma}$ of ${\M}$ to a point ${x^{\alpha}(x^a,\,t)}$
in the spacetime, where ${t\in\real}$ labels the leaves of this foliation.
A hypersurface, then, is an equivalence class of such embeddings modulo
diffeomorphisms of the submanifold ${\Sigma\,}$ (cf. subsection 4.2).\par}}}
of simultaneity. One 
can use this time-function $t$ as an affine parameter for arbitrary timelike
curves representing the worldlines of test particles. 

Since the compatibility condition ${\nabla_{\!\mu}t_{\nu} = 0}$ leads to the
relation ${t_{\gamma}\,\Gamma_{\!\lbrack\mu\;\>\nu\rbrack}^{\,\;\>\gamma}\,=\,
\partial_{\lbrack\mu}\,t_{\nu\rbrack}\,}$, it is clear that a torsion-free
connection,
$$
\Gamma_{\!\lbrack\mu\;\>\nu\rbrack}^{\,\;\>\gamma}\;=\;0\;,\EQN
$$
is admissible only if
$$
\partial_{\lbrack\mu}\,t_{\nu\rbrack}\;=\;0\,;\;\;\;\;\;\;{\rm i.e.,}\;\;
d\tau\;=\;0\;.\EQN
$$
In fact, it can be shown\cite{Kunzle, 72} that the closed-ness of ${\tau}$ is
both necessary and sufficient condition for the existence of a torsion-free
connection as a part of the Galilean structure.
Moreover, without the condition  that the 1-form ${\tau}$ is closed, the
structure ${(\M;\;{\tau},\;{h}\,)}$ is {\it not integrable}. A
linear symmetric Galilean connection satisfying the compatibility conditions
\Ep{integrab} exists if and only if the structure ${(\M;\;{\tau},\;{h}\,)}$
is integrable. However, given integrability, such a
symmetric connection is unique {\it only} up to an arbitrary 2-form
${F = {1\over 2}\;F_{\mu\nu}\;dx^{\mu}\wedge\,dx^{\nu}\,}$, and can be
decomposed as\cite{Duval and Kunzle, 78}:
$$
\Gamma_{\mu\;\>\nu}^{\;\>\gamma} \;=\;
\Gamma_{\!(\mu\;\>\nu)}^{\,\;\>\gamma} \;= \!\!\!\!\!
\buildchar{\>\;\;\;\;\Gamma_{\mu\;\>\nu}^{\;\>\gamma}}{u}{}\;
+\; t_{\lparen\mu}\,F_{\nu\rparen\sigma} h^{\sigma\gamma}\;,
\EQN decomposed
$$
where\cite{Trumper}
$$
\buildchar{\>\;\;\;\;\Gamma_{\mu\;\>\nu}^{\;\>\gamma}}{u}{} \;:=\;\, 
h^{\gamma\sigma}\Leftcases{30pt}\partial_{\lparen\mu}\!\!\!\!\!\!\!
\buildchar{\>\;\;\;\;h_{\nu\rparen\sigma}}{u}{}
-{\scriptstyle{1\over 2}}\partial_{\sigma}\!\!\!\!\!\!
\buildchar{\;\;\;\;h_{\mu\nu}}{u}{}\Rightcases{30pt} +
u^{\gamma}\partial_{\lparen\mu}t_{\nu\rparen}\;\;\;\,\EQN u-part
$$
with ${u = u^{\alpha}\partial_{\alpha}}$ representing
an arbitrary unit timelike vector-field,
$$
t_{\alpha}u^{\alpha}\;=\;1\;=\;\tau\cont u\;,\EQN u-norm
$$
interpreted as the four-velocity of an adscititious `\ae ther-frame'
and ${{\!\!\!\!\!\!\buildchar{\;\;\;\;h_{\mu\nu}}{u}{}} =
{\!\!\!\!\!\!\!\!\!\!\buildchar{\;\;\;\;\;\;h_{\lparen\mu\nu\rparen}}{u}{}}}$
representing the associated spatial projection
field {\it relative to} $u^{\alpha}$ defined by
$$
{\buildchar{\;\;\;\;h_{\mu\rho}}{u}{}}u^{\rho} := 0 
\;\;\;\;\;\;{\rm and}\;\;\;\;\;\;
{\!\!\!\!\!{\buildchar{\;\;\;\;h_{\mu\rho}}{u}{}}}
h^{\rho\nu} := \delta_{\mu}^{\;\;\nu} - t_{\mu}
u^{\nu}=:{\!\!\!\buildchar{\>\;\;\delta_{\mu}^{\;\;\nu}}{u}{}}\;.\EQN inverse
$$
Throughout the paper, a letter (e.g., $u$) on the
top of a quantity indicates gauge-dependence of the quantity (e.g., dependence
of the quantity ${\!\!\!\!\!\!\buildchar{\;\;\;\;h_{\mu\nu}}{u}{}}$
on the `gauge' $u$). This relative projection field
${\!\!\!\!\!\!\buildchar{\;\;\;\;h_{\mu\nu}}{u}{}}$ introducing a fiducial
`absolute rest'\cite{Malament} may be used to lower indices, but, of course,
only relative to ${u\,}$; for, under an \ae ther-frame transformation of the
form
$$
u^{\alpha}\;\mapsto\;{\tilde u}^{\alpha}\;:=\;u^{\alpha}\;+\;h^{\alpha\sigma}
{\rm w}_{\sigma}\EQN
$$
for a boost covector ${{\rm w}_{\sigma}\,}$, the relative projection field
undergoes a nontrivial boost transformation:
$$
{\!\!\!\!\!\!\buildchar{\;\;\;\;h_{\mu\nu}}{u}{}}\;\mapsto\;
{\!\!\!\!\!\!\buildchar{\;\;\;\;h_{\mu\nu}}{{\tilde u}}{}}\;:=\;
{\!\!\!\!\!\!\buildchar{\;\;\;\;h_{\mu\nu}}{u}{}}\;-\;2\,t_{\lparen\mu}
{\!\!\!\!\!\!\!\buildchar{\>\;\;\;\;h_{\nu\rparen\alpha}}{u}{}}\,
h^{\alpha\sigma}{\rm w}_{\sigma}\;+\;t_{\mu\nu}\,
h^{\alpha\sigma}{\rm w}_{\alpha}{\rm w}_{\sigma}\;.\EQN rel-trans
$$
In other words,
$$
{{\partial\,\;}\over{\partial u^{\sigma}}}
{\!\!\!\!\!\buildchar{\;\;\;\;h_{\mu\nu}}{u}{}}\;=\;-\;2\,t_{\lparen\mu}
{\!\!\!\!\!\!\!\buildchar{\>\;\;\;\;h_{\nu\rparen\alpha}}{u}{}}
\;,\EQN ink-rat
$$
which follows from the definition \Ep{inverse}.

The special connection ${\buildchar{\Gamma}{u}{}}$ is
unique and symmetric, and such that the Galilean observer
$u^{\alpha}$ associated with it is geodetic, ${u^{\sigma}\,\!\!\!\!
{\buildchar{\>\;\nabla_{\!\sigma}}{u}{}}u^{\alpha} = 0}$,
and curl-free, ${h^{\sigma\lbrack\mu}\,\!\!\!\!
{\buildchar{\>\;\nabla_{\!\sigma}}{u}{}}u^{\nu\rbrack} = 0\,}$.
Using these two properties of ${u^{\alpha}}$, and the above definition of
its relative projection field
${\!\!\!\!\!\!\buildchar{\;\;\;\;h_{\mu\nu}}{u}{}\,}$, one can express the
2-form $F$ in terms of the covariant derivative of
this observer ${u^{\alpha}}$\cite{Duval and Kunzle, 78}:
$$
F_{\mu\nu} \;=\; -\,2
{\!\!\!\!\!\!\buildchar{\;\;\;\;\;h_{\sigma\lbrack\mu}}{u}{}}
\nabla_{\!\nu\rbrack}\,u^{\sigma}\,.\EQN useful
$$
For future purposes, we also observe that the spatial projection field
${\!\!\!\!\!\!\buildchar{\;\;\;\;h_{\mu\nu}}{u}{}}$ is {\it not} compatible
with the covariant derivative operator ${\nabla_{\!\sigma}}$ in general, but,
instead, its covariant derivative depends on the covariant rate
of change of the observer-field ${u^{\alpha}}$\cite{Duval and Kunzle, 78}:
$$
\nabla_{\!\sigma}{\!\!\!\!\!\!\buildchar{\;\;\;\;h_{\mu\nu}}{u}{}}\;\;=\;\;-\,
2\,\Leftcases{23pt}\nabla_{\!\sigma}u^{\alpha}\Rightcases{23pt}{\!\!\!\!\!\!\!
\buildchar{\>\;\;\;\;h_{\alpha\lparen\mu}}{u}{}}t_{\nu\rparen}\;\,,\EQN cmpti
$$
where the covariant derivative of ${u^{\alpha}}$ may be decomposed
as\cite{Mala-New}
$$
\nabla_{\!\mu}u^{\nu}\;=\;{\!\!\!\!\!\!\buildchar{\;\;\;\;h_{\mu\alpha}}{u}{}}
h^{\sigma\lparen\alpha}\nabla_{\!\sigma}u^{\nu\rparen}\;+\;
{\!\!\!\!\!\!\buildchar{\;\;\;\;h_{\mu\alpha}}{u}{}}
h^{\sigma\lbrack\alpha}\nabla_{\!\sigma}u^{\nu\rbrack}\;+\;
t_{\mu}u^{\sigma}\nabla_{\!\sigma}u^{\nu}\;.\EQN dcmpti
$$
The first term of this decomposition is a measure of the lack of `rigidity'
of the field ${u^{\alpha}}$, and, hence, in the case of the rigid
Galilean frame it vanishes identically:
${h^{\sigma\lparen\mu}\nabla_{\!\sigma}u^{\nu\rparen} \equiv 0}$.
Further, since the Galilean observer ${u^{\alpha}}$ 
is curl-free and geodetic with respect to the associated
connection ${\buildchar{\Gamma}{u}{}\,}$, the remaining two terms of the
equation \Ep{dcmpti} also vanish in this special case giving the uniformity
property
$$
{\,\!\!\!\!\buildchar{\>\;\nabla_{\!\mu}}{u}{}}u^{\nu}\;=\;0\;;\EQN uniformity
$$
i.e., ${u^{\nu}}$ is covariantly constant with respect to the special
connection ${\buildchar{\Gamma}{u}{}\,}$.
As a result of this property, equation \Ep{cmpti} yields the compatibility
relation for the relative spatial projection field
${\!\!\!\!\!\!\buildchar{\;\;\;\;h_{\mu\nu}}{u}{}\,}$,
$$
{\,\!\!\!\!\buildchar{\>\;\nabla_{\!\sigma}}{u}{}}
{\!\!\!\!\!\!\buildchar{\;\;\;\;h_{\mu\nu}}{u}{}}\;=\;0\,,\EQN con-uniformity
$$
in this special case of unique derivative operator
${\,\!\!\!\!\buildchar{\>\;\nabla_{\!\sigma}}{u}{}}$
associated with the \ae ther-frame ${u^{\alpha}}$.

The curvature tensor
corresponding to the symmetric Galilean connection \Ep{decomposed} --- defined
by ${R^{\,\sigma}_{\;\;\,\beta\,\gamma\,\delta}\,f_{\!\sigma} \,=\, 2\,
\nabla_{\!\lbrack\gamma}\nabla_{\!\delta\rbrack}\,f_{\!\beta}\,}$,
or, equivalently,
by ${R^{\,\alpha}_{\;\;\,\sigma\,\gamma\,\delta}\,g^{\sigma} \,=\,-\, 2\,
\nabla_{\!\lbrack\gamma}\nabla_{\!\delta\rbrack}\,g^{\alpha}\,}$, for arbitrary
vectors ${f_{\!\alpha}}$ and ${g^{\alpha}}$ --- observes
the symmetries\cite{Kunzle, 72}\cite{Duval and Kunzle, 78}
$$
h^{\sigma\lparen\beta}
R^{\,\alpha\rparen}_{\;\;\;\>\sigma\,\gamma\,\delta} = 0\;\;\;\;
{\rm and}\;\;\;\;
t_{\sigma}R^{\,\sigma}_{\;\;\,\beta\,\gamma\,\delta} = 0
\EQN properties-add
$$
by virtue of the metric compatibility conditions, in addition to the usual
constraints
$$
R^{\,\alpha}_{\;\;\;\beta\lparen\gamma\,\delta\rparen} = 0\,,\;\;\;
R^{\,\alpha}_{\;\;\;\lbrack\,\beta\,\gamma\,\delta\,\rbrack} = 0\,,\;\;\;
{\rm and}\;\;\;
R^{\,\alpha}_{\;\;\,\beta\,\lbrack\gamma\,\delta\,;\lambda\rbrack}
= 0\,\EQN properties-use
$$
satisfied by any curvature tensor, where ${;\!\lambda}$ denotes the covariant
derivative ${\nabla_{\!\lambda}\,}$. It
can be easily verified that the contracted Bianchi identities of the form
$$
\nabla_{\!\nu}(R^{\mu\nu} \,-\, {\scriptstyle{1\over 2}}\,R\,h^{\mu\nu})
\;=\;0\EQN bianchi
$$
hold in the general Galilean spacetime as a consequence of
the last of these constraints, where ${R_{\lparen\mu\nu\rparen}=R_{\mu\nu} :=
R^{\,\sigma}_{\;\;\,\mu\,\nu\,\sigma}}$ and ${R:=h^{\mu\nu}R_{\mu\nu}}$ are
the corresponding Ricci tensor and Ricci scalar, respectively.

Finally, we close this subsection by reemphasizing that, although an
integrable Galilean
structure ${(\M;\;t_{\alpha},\;h^{\alpha\beta},\;\nabla_{\!\alpha})}$
is completely specified by the four conditions
$$
h^{\alpha\beta}t_{\beta}\;=\;0\,,\;\;\;\;
\nabla_{\!\alpha}h^{\beta\gamma}\;=\;0\,,\;\;\;\;
\nabla_{\!\alpha}t_{\beta}\;=\;0\,,\;\;\;\;{\rm and}\;\;\;\;
\partial_{\lbrack\alpha}\,t_{\beta\rbrack}\;=\;0\;,
\EQN three-cond
$$
the Galilean connection \Ep{decomposed} remains under-determined, in general,
by an arbitrary 2-form.

\subsection{Specializing to the Newton-Cartan spacetime}

Now, Cartan's spacetime reformulation of the Newtonian theory of gravity
can be motivated in exact analogy with Einstein's theory of gravity. 
The analogy works because the 
universal equality of the inertial and the passive gravitational masses is
independent of the relativization of time, and hence is equally valid
at the Galilean-relativistic level. 
As a result, it is possible to parallel
Einstein's theory and reconstrue the trajectories of (only) gravitationally
affected particles as geodesics of a unique, `{\it non-flat}' connection
$\Gamma$ satisfying 
$$
{a}^i\;:=\;{{d^2x^{i}}\over{dt^2\>}} \;+\, {{\Gamma_{j\;\>k}^{\;\>i}}}\,
{{dx^{j}}\over{dt\>}}{{dx^{k}}\over{dt\>}} \;=\; 0\EQN eqa-mot
$$
in a coordinate basis, or, equivalently,
$$
{a}^{\alpha}\;:=\;v^{\sigma}\nabla_{\!\sigma}v^{\alpha}\;=\;0\EQN
$$
in general, such that
$$
\Gamma_{\nu\;\>\lambda}^{\;\>\mu} \,\;\equiv \!\!\!\!\!
\buildchar{\>\;\;\;\;\Gamma_{\nu\;\>\lambda}^{\;\>\mu}}{v}{}
\,+\!\!\!\!\!\!\!
\buildchar{\>\;\;\;\;G_{\nu\;\>\lambda}^{\;\>\mu}}{v}{} 
\;:=\!\!\!\!
\buildchar{\>\;\;\;\;\Gamma_{\nu\;\>\lambda}^{\;\>\mu}}{v}{} +\,
h^{\mu\alpha}{\!\!\buildchar{\>\;\nabla_{\!\alpha}}{v}{}}\,
{\buildchar{\Phi}{v}{}}\;t_{\nu\lambda}\;\EQN gauge-depend
$$
with ${\buildchar{\Phi}{v}{}}$ representing the Newtonian gravitational
potential relative to the freely falling observer field $v$, ${\!\!\!\!\!
\buildchar{\>\;\;\;\;\Gamma_{\nu\;\>\lambda}^{\;\>\mu}}{v}{}\,}$ representing
the coefficients of the corresponding `flat' connection (i.e., one whose
coefficients can be made to vanish in a suitably chosen linear coordinate
system), and
$$
\buildchar{\>\;\;\;\;G_{\nu\;\>\lambda}^{\;\>\mu}}{v}{} \,:=\,
h^{\mu\alpha}{\!\!\buildchar{\>\;\nabla_{\!\alpha}}{v}{}}\,
{\buildchar{\Phi}{v}{}}\;t_{\nu\lambda}\EQN gravi-poten
$$
representing the traceless gravitational field tensor associated with
the Newtonian potential\cite{Carter}.
The conceptual superiority of this geometrization of Newtonian gravity
is reflected in the trading of the two `gauge-dependent'
quantities ${\buildchar{\Gamma}{v}{}}$ and ${\buildchar{G}{v}{}}$
in favor of their gauge-independent sum ${\Gamma}$.
Physically, it is the `curved' connection ${\Gamma}$ rather than any `flat'
connection ${\buildchar{\Gamma}{v}{}}$ that can be determined by local
experiments. The potential ${\buildchar{\Phi}{v}{}}$ and the `flat'
connection ${\buildchar{\Gamma}{v}{}}$
do not have an independent existence; they exist
only relative to an arbitrary choice of inertial frame. Given
the `curved' connection $\Gamma$, its associated
curvature tensor works out to be
$$
R^{\,\alpha}_{\>\;\;\beta\,\gamma\,\delta} \,=\, 2\>t_{\beta}\>
h^{\alpha\lambda}\!\!\!\!\!\!\!
{\buildchar{\;\;\;\;\;\Phi_{;\lambda;\lbrack\gamma}}{v}{}}\,
t_{\delta\rbrack}\;,\EQN curvature
$$
where ${;\alpha}$ represent covariant derivatives
${\!\!\!\!\buildchar{\>\;\nabla_{\!\alpha}}{v}{}}$
corresponding to the `flat' connection ${\buildchar{\Gamma}{v}{}}\,$.
Although written in terms of gauge-dependent quantities, this
expression for the curvature tensor is, of course, gauge-invariant.

The non-flat connection $\Gamma$ in equation \Ep{eqa-mot} is still compatible
with the temporal and  spatial metrics ${\tau}$ and ${h}$:
${\nabla_{\!\alpha}t_{\beta} = 
\nabla_{\!\alpha}h^{\beta\gamma} = 0}$. This non-uniqueness of the compatible
connections is due to the degenerate nature of the metrics 
${\tau}$ and ${h}$ --- i.e., due 
to the non-semi-Riemannian nature of the Galilean structure
${(\M;\;{\tau},\;{h},\;\nabla)}$.
Unlike in the semi-Riemannian spacetime structures of the special and general
theories of relativity, 
here the covariant derivative operator ${\nabla_{\!\alpha}}$ is not fully
determined; as noted in the previous subsection, the connection \Ep{decomposed}
is determined by the compatibility conditions
\Ep{integrab} {\it only} up to an arbitrary 2-form.
Therefore, in addition to the structure ${(\M;\;{\tau},\;{h}\>)}$,
a connection must be specified to construct a
completely geometrized, spacetime formulation of the Galilean-relativistic
physics. In 
the case of Newtonian theory of gravity, geometrization can be easily achieved
by taking those compatible connections $\Gamma$ for the structure
${(\M;\,{\tau},\,{h},\,\nabla)}$ whose curvature tensor, in addition
to the properties \Ep{properties-add} and \Ep{properties-use},
satisfies\cite{Kunzle, 72}
$$
R^{\,\alpha\;\;\>\gamma}_{\,\;\;\>\beta\;\;\>\delta}
= R^{\,\gamma\;\;\>\alpha}_{\,\;\;\>\delta\;\;\>\beta}\;,\EQN constraint
$$
where ${R^{\,\alpha\;\;\>\gamma}_{\,\;\;\>\beta\;\;\>\delta}\;\equiv\;
h^{\gamma\sigma}R^{\,\alpha}_{\;\;\>\beta\,\sigma\,\delta}\;}$.
This extra condition roughly expresses the curl-freeness of the Newtonian
gravitational field, and, together with the constraints \Ep{three-cond},
{\it uniquely} specifies the `curved' Newton-Cartan connection; it is clearly
satisfied by any Galilean connection that is obtained as a limit of the
torsion-free Lorentzian connection.
Equivalently, one can specify the Newton-Cartan connection by
requiring that the 2-form $F$ in equation \Ep{decomposed} be
closed\cite{Kunzle, 72}\cite{Duval and Kunzle, 78}: ${dF = 0}$.
Then, in the light of the identity ${d^{\,2}=0\,}$, the 2-form $F$ may be
expressed in terms of an arbitrary 1-form $A$ as 
$$
F_{\mu\nu} \;=\; 2\,\nabla_{\lbrack\mu}A_{\nu\rbrack}
\;=\; 2\,\partial_{\lbrack\mu}A_{\nu\rbrack}\,.\EQN 2-form
$$
Clearly, this expressibility of the 2-form $F$ as an exterior derivative of
an arbitrary 1-form is at least a {\it sufficient} condition for the connection
\Ep{decomposed} to be Newton-Cartan. Since $\M$ is
contractible by definition, however, the Poincar\'e
lemma implies that \Ep{2-form} is also a {\it necessary} condition for the
specification of the Newton-Cartan connection. Comparison of equations
\Ep{2-form} and \Ep{useful} shows that, at least locally, the
intrinsic condition \Ep{constraint} can be equivalently expressed in terms of
the fields ${u^{\nu}}$ and ${A_{\mu}}$ as\cite{Duval and Kunzle}
$$
\nabla_{\lbrack\mu}A_{\nu\rbrack}\;+\;
{\!\!\!\!\!\!\buildchar{\;\;\;\;\;h_{\sigma\lbrack\mu}}{u}{}}
\nabla_{\!\nu\rbrack}\,u^{\sigma}\;=\;0\,. \EQN one-Lagrangian
$$
It is this convenient local form of the condition which will be useful in
latter sections.

Physically, the covariant vector ${A_{\mu}}$ may be interpreted as
the `vector-potential' representing the combination of gravitational and 
inertial effects with respect to the Galilean observer ${u = u^{\alpha}
\partial_{\alpha}}$. Given the `\ae ther-frame' based observer
${u^{\alpha}}$, an
{\it arbitrary} observer may be characterized by a normalized four-velocity
vector $v^{\alpha}$, $v^{\alpha}t_{\alpha} = 1$, such that the 
difference vector ${h^{\alpha\sigma}A_{\sigma} := u^{\alpha} - v^{\alpha}}$
describes how the motion of this arbitrary observer deviates from
that of the `\ae ther-frame'. In terms of $u$ and $A$ any Newton-Cartan
connection can be affinely decomposed as
$$
\Gamma_{\alpha\;\>\beta}^{\;\>\gamma} \;= \!\!\!\!\!\!
{\buildchar{\>\;\;\;\;\Gamma_{\alpha\;\>\beta}^{\;\>\gamma}}{u}{}}\;
+ \!\!\!\!\!\!\buildchar{\>\;\;\;\;\Gamma_{\alpha\;\>\beta}^{\;\>\gamma}}{A}{}
\;\;,\;\;\EQN affinely
$$
where
$$
\buildchar{\>\;\;\;\;\Gamma_{\alpha\;\>\beta}^{\;\>\gamma}}{A}{}\;
:= \;t_{\lparen\alpha}
[\partial_{\beta\rparen}A_{\sigma} - \partial_{\sigma}A_{\beta\rparen}]
h^{\sigma\gamma}\,.\EQN A-part
$$
By comparing this decomposition in terms of the `gauge' $(u,\,A)$ with that
in terms of the generic `gauge' ${(v,\,{\buildchar{\Phi}{v}{}})}$ (cf.
equation \Ep{gauge-depend}) we find the relation
$$
\EQNalign{v^{\alpha}        &=\, u^{\alpha} - h^{\alpha\sigma}A_{\sigma}\cr
  {\buildchar{\Phi}{v}{}}\; &=\, {1\over 2}\,h^{\mu\nu}A_{\mu}A_{\nu} -
                                     A_{\sigma}u^{\sigma}\EQN two-gauges\cr}
$$
between the two gauges\cite{Duval-93}. With ${{A^2}/2}$ being the `Coriolis'
rotational potential\cite{Kuchar}, we recognize that the observer $v$ is in
4-rotation with respect to the \ae ther ${u\,}$. Conversely, the
vector-potential $A$ may be expressed in terms of the observer field $v$ and
its relative Newtonian gravitational potential ${\buildchar{\Phi}{v}{}}$
as\cite{Duval-93}
$$
A_{\mu}\,\;=\,\;-\,
{\!\!\!\!\!\!\buildchar{\;\;\;\;h_{\mu\alpha}}{u}{}}\,v^{\alpha}\;\,+\,\;
[{\buildchar{\Phi}{v}{}}\;\,-\,{1\over 2}\;
{\!\!\!\!\!\!\buildchar{\;\;\;\;h_{\nu\sigma}}{u}{}}\;v^{\nu}\,v^{\sigma}
]\;t_{\mu}\;.\EQN
$$

We noted in the previous subsection that given a Galilean structure
${(\M,\,h,\,\tau)}$ and an observer field $u$, there exists an associated
unique torsionless connection ${\buildchar{\Gamma}{u}{}}$ with respect to
which the observer is geodetic and curl-free. It turns out that in the case
of Newton-Cartan connection defined by \Ep{constraint}, the converse is also
true\cite{Dombrowski}: for every Newton-Cartan connection ${\Gamma}$, at least
locally there exists a unit, timelike, nonrotating, and freely-falling
observer $u$ such that ${\Gamma\,=\,{\buildchar{\Gamma}{u}{}}\,}$. For such
an observer, the uniformity property \Ep{uniformity} immediately leads to
$$
u^{\sigma}\,R^{\,\alpha}_{\;\;\,\sigma\,\gamma\,\delta}\;=\;
-\,2\,\nabla_{\!\lbrack\gamma}\nabla_{\!\delta\rbrack}\,u^{\alpha}\;=\;0
\EQN path-in-u
$$
implying that its parallel-transport around a small closed curve is
path-independent. But, of course, for any other timelike observer
${{\widetilde u}^{\sigma}\,=\,u^{\sigma}\,+\,h^{\sigma\mu}\,
{\rm w}_{\mu}\,}$, with ${\rm w}$ being an arbitrary 1-form, we would
instead have
$$
{\widetilde u}^{\sigma}\,R^{\,\alpha}_{\;\;\,\sigma\,\gamma\,\delta}\;=\;
h^{\sigma\mu}\,{\rm w}_{\mu}\,
R^{\,\alpha}_{\;\;\;\sigma\,\gamma\,\delta}\;\equiv\;
{\rm w}_{\mu}\,R^{\,\alpha\mu\;\;\>}_{\;\;\>\;\;\>\gamma\delta}
\;\not=\;0\,,\EQN
$$
in general, unless a further restriction,
${h^{\mu\sigma}\,R^{\,\alpha}_{\;\;\;\sigma\,\gamma\,\delta}\,=\,0\,}$, is
imposed on the curvature tensor. That is, parallel-transport of an arbitrary
observer around a small closed curve is path-dependent in general unless
${R^{\,\alpha\mu\;\;\>}_{\;\;\>\;\;\>\gamma\delta}\,=\,0\,}$ everywhere.
This innocuous looking extra condition turns out to be conceptually quite
significant as far as a clearer understanding of the limit-relation between
Einstein's and Newton's theories of gravity is concerned. We shall devote the
entire next subsection to elaborate on its significance.

As in the general theory of relativity, spacetime becomes
{\it dynamical}${{\,}^{\scriptscriptstyle 1}}$ under
this geometrical reformulation of Newton's theory: the affine structure
of the spacetime --- i.e., the connection $\Gamma$ --- crucially depends
on the distribution of matter $\rho$, and
participates in the unfolding of physics rather than being a
passive backdrop for the unfolding. This mutability of the 
Newton-Cartan spacetime is captured in a generalized Newton-Poisson
equation which dynamically correlates the
curvature of spacetime with the presence of matter:
$$
R_{\alpha\beta} \;=\; 4\pi G\;\rho\;t_{\alpha\beta} \;=:\;
4\pi G\,P_{\alpha\beta}\;,\EQN Poisson
$$
where ${R_{\lparen\alpha\beta\rparen} = R_{\alpha\beta} := 
R^{\gamma}_{\;\>\alpha\beta\gamma}}$ is the Ricci
tensor corresponding to
the connection $\Gamma$, $G$ is the Newtonian gravitational 
constant, and ${P_{\mu\nu}=
\rho\,t_{\mu\nu}}$ are the tensor components of the 
momentum tensor defined on $\M$
(seen to be as such by writing ${P^{\alpha\sigma} := \rho\,
u^{\alpha}u^{\sigma}}$ and 
then lowering the indices by ${t_{\mu\nu}}$:
${P_{\mu\nu}}$ = ${t_{\mu}t_{\alpha}t_{\nu}t_{\sigma}P^{\alpha\sigma}}$).
The constraint \Ep{Poisson} on the Ricci tensor is a Galilean-relativistic
limit form of Einstein's equation, and, unlike the possible foliation of the
general Galilean spacetime in terms of arbitrary (i.e., possibly curved)
spacelike hypersurfaces discussed in the previous subsection, its presence 
in this specialized Newton-Cartan spacetime
necessitates that the hypersurfaces of simultaneity remain 
copies of ordinary Euclidean three-spaces:
${h^{\mu\alpha}h^{\nu\beta}R_{\alpha\beta}=0}$ because
${h^{\mu\nu}t_{\nu}=0\,}$\cite{Malament}. More generally ${R^{\mu\nu}}$ may
be defined as
$$
R^{\mu\nu}t_{\mu\alpha}t_{\nu\beta}\;:=\;R_{\alpha\beta}\;,\EQN upper-R
$$
subject to the consistency condition
${{\!\!\!\!\!\!\buildchar{\;\;\;\;h_{\mu\nu}}{u}{}}R^{\mu\nu}:=R=
h^{\mu\nu}R_{\mu\nu}\,}$.

In contrast to the general theory of relativity, here the mass 
density $\rho$ is the only source of the gravitational field. However,
without loss of consistency, {\it spacelike} tensor fields, say
${s^{\alpha}}$ and ${S^{\alpha\sigma}=S^{(\alpha\sigma)}\,}$, may be added to
the matter side of the above field equation as long as the resulting
mass-momentum-stress tensor (or matter tensor, for short)
$$
M^{(\alpha\sigma)}\;=\;M^{\alpha\sigma} \;:=\; P^{\alpha\sigma}\;-\;
2\,u^{\lparen\alpha}s^{\sigma\rparen}\;-\; S^{\alpha\sigma}\; \EQN mass-tens
$$
describing continuous matter distributions satisfies the conservation condition
$$
\nabla_{\!\alpha}M^{\alpha\sigma}\;=\;0\,.\EQN mass-cons
$$
As a matter of fact, expression \Ep{mass-tens} is the only consistent
possible nonrelativistic limit-form of the relativistic stress-energy
tensor\cite{Kunzle, 76}.
Note that, unlike the case in general relativity, here this conservation
law {\it must be postulated independently} to allow the derivation of
Newtonian equations of motion \Ep{eqa-mot} in a manner analogous to the
derivation of Einsteinian equations of motion from the general relativistic
conservation law. In other words, at least for now, the matter conservation
condition \Ep{mass-cons} must be viewed as a separate field equation of the
theory. As we shall see, however, it can be {\it derived} using only the
principle of general covariance in a variational
approach discussed in the subsection 2.5 below. 

As in Einstein's theory, the field equation \Ep{Poisson} is generalizable
by an additional cosmological term:
$$
R_{\mu\nu}\;+\;\Lambda\,t_{\mu\nu} \;=\; 4\pi G\,M_{\mu\nu}\;,\EQN cosmology-1
$$
where ${\Lambda\,}$, the enigmatic cosmological parameter, is a spacetime
constant. It is important
to note that, analogous to the general relativistic case, this is the
{\it only} admissible generalization of the field equations compatible with the
hypotheses that (1) gravitation is a manifestation of spacetime geometry and
(2) Newtonian mechanics is valid in the absence of gravitation\cite{Dixon}.
Not surprisingly, however, it is possible to relax the
restriction of spacetime constancy on ${\Lambda}$ if the matter conservation
condition \Ep{mass-cons} is concurrently relaxed.

\subsection{An additional constraint on the curvature tensor}

As far as the Newton-Cartan theory is viewed as a limiting form of Einstein's
theory of gravity in which special relativistic effects become negligible,
equations \Ep{three-cond}, \Ep{constraint},
and \Ep{cosmology-1} constitute the complete set of
gravitational field equations of the theory\cite{Kunzle, 72}.
The logical structure of the covariant Newtonian theory, however, is
flexible enough\cite{Dixon} to allow an additional constraint on the
curvature tensor, namely
$$
h^{\lambda\sigma}\,R^{\,\alpha}_{\;\;\;\sigma\,\gamma\,\delta}
\;\equiv\;R^{\,\alpha\lambda\;\;\>}_{\;\;\>\;\;\>\gamma\delta}
\;=\; 0\EQN additional
$$
(or, equivalently, ${t_{\lbrack\sigma}
R^{\,\alpha}_{\;\;\;\beta\rbrack\,\gamma\,\delta}\,=\,0\,}$\cite{Trautman}),
implying that parallel-transport of spacelike vectors around a small closed
curve is path-independent. Since this restriction, in physical terms, implies
that ``the rotation axes of freely falling, neighboring gyroscopes do not
exhibit relative rotations in the course of time'' (no {\it relative} rotations
for timelike geodesics), in essence it just asserts a standard of rotation:
the existence of `absolute rotation' {\it \`a la} Newton (``{\it Gesetz der
Existenz absoluter Rotation}''\cite{Ehlers, 81}). In fact,
Newton-Cartan theory commonly discussed in the literature with its usual
set of field equations \Ep{three-cond}, \Ep{constraint}, and \Ep{cosmology-1}
is, strictly speaking, a slight generalization of Newton's original theory of
gravity (although, {\it it} constitutes the true Galilean-relativistic limit of
Einstein's theory in general, and {\it not} the classical Newtonian theory of
gravity\cite{Dixon}\cite{Kunzle, 76}\cite{Ehlers, 81}\cite{Malament}).
Put differently, unlike the usual Newton-Cartan field equations the condition
\Ep{additional} is {\it not} a necessary consequence of the
Galilean-relativistic limit (`c$\;\rightarrow\infty$') of Einstein's field
equations, but, instead, is an added restriction on the Newton-Cartan
structure\cite{Ehlers, 81}\cite{Malament}. Indeed, it is not possible to
recover the Poisson equation ${\Delta\Phi = 4\pi G\,\rho}$ of Newton's
theory from the usual Newton-Cartan field equations \Ep{three-cond},
\Ep{constraint}, and \Ep{Poisson} without any global
assumptions unless this extra condition prohibiting any rotational holonomy
is imposed on the curvature tensor\cite{Ehlers, 81}. It becomes redundant,
however, if non-intersecting spacelike hypersurfaces covering the Galilean
spacetime are required to asymptotically resemble Euclidean
space\cite{Kunzle, 72}\cite{Kunzle, 76}. Such a global boundary
condition of asymptotically flat spacetime --- which idealizes the
gravitating systems as isolated systems --- is, of course,
of great historical and physical importance not only in the case of
Newton's theory, but also in the case of Einstein's theory.
Nevertheless, any such global condition can only encompass special
physical situations (`island universes'), and, in general, the extra 
condition \Ep{additional} is inevitable to ensure smooth recovery of 
the Newtonian Poisson equation from the Newton-Cartan field equations.
Therefore, in what follows, we shall view equation \Ep{additional} as an
extraneously imposed but necessary field equation on the Newton-Cartan
structure.

For reasons that will become increasingly clear as we go on, it is convenient
to express the above intrinsic condition \Ep{additional} in terms of local
variables ${u^{\nu}}$ and ${A_{\mu}}$ in analogy with the equation
\Ep{one-Lagrangian} representing the condition \Ep{constraint}. The desired
expression --- which must hold at least locally --- is
$$
\nabla_{\!\lbrack\gamma}\nabla_{\!\delta\rbrack}\,A_{\mu}\;-\;
{\!\!\!\!\!\!\buildchar{\;\;\;\;h_{\mu\nu}}{u}{}}\,
\nabla_{\!\lbrack\gamma}\nabla_{\!\delta\rbrack}\,u^{\nu}\;=\;0\,.\EQN New-Can
$$
Contracting with ${h^{\alpha\mu}\,}$, replacing
${-\,v^{\alpha}}$ for ${h^{\alpha\mu}A_{\mu}\,-\,u^{\alpha}\,}$,
multiplying by ${-\,{{2}\over{A^2}}\,h^{\lambda\beta}A_{\beta}\,}$, and
using equation \Ep{path-in-u} reveals that this expression is equivalent to the
intrinsic form \Ep{additional}. It tells us that,
as long as the additional condition \Ep{additional} on the curvature
tensor is satisfied, path-independence of
parallel-transport around a small closed curve holds true not only for the
fiducial observer ${u\,}$ (cf. equation \Ep{path-in-u}), but also for any
observer ${v}$ in 4-rotation with ${u}$.

Finally, we end this subsection by observing the logical completeness of the
Newton-Cartan structure described in the last two subsections.
First of all, straightforward evaluations from equation \Ep{curvature} show
that the curvature tensor of the connection \Ep{gauge-depend} satisfies the
relations \Ep{constraint} and \Ep{additional}. Conversely, and more
importantly for our purposes,
Trautman\cite{Trautman} has shown that equations \Ep{constraint} and
\Ep{additional} together imply the existence of a scalar potential
${\buildchar{\Phi}{v}{}}$ and a flat connection ${\buildchar{\Gamma}{v}{}}$
such that the components of the `curved' connection ${\Gamma}$ satisfy
the relation \Ep{gauge-depend} and its curvature tensor satisfies equation
\Ep{curvature}. This necessary and sufficient equivalence of equation
\Ep{curvature} with the pair \Ep{constraint} and \Ep{additional} consistently
rounds off the geometrization of Newton's theory of gravity due to Cartan.
For convenience, let us rewrite the complete geometric set of gravitational
field-equations of the Newton-Cartan theory:
$$
\EQNalign{
h^{\alpha\beta}t_{\beta}\;=\;0\,,\;\;\;\;
\nabla_{\!\alpha}h^{\beta\gamma}\;&=\;0\,,\;\;\;\;
\nabla_{\!\alpha}t_{\beta}\;=\;0\,,\;\;\;\;
\partial_{\lbrack\alpha}\,t_{\beta\rbrack}\;=\;0\;,\EQN cset; a \cr
R^{\,\alpha\;\;\>\gamma}_{\,\;\;\>\beta\;\;\>\delta}
&= R^{\,\gamma\;\;\>\alpha}_{\,\;\;\>\delta\;\;\>\beta}\;,\EQN cset; b \cr
R^{\,\alpha\lambda\;\;\>}_{\;\;\>\;\;\>\gamma\delta}\;&=\;0\,,\EQN cset; c \cr
{\rm and}\;\;\;\;\;R_{\mu\nu}\;+\;\Lambda\,t_{\mu\nu} \;&=\;
4\pi G\,M_{\mu\nu}\;,\EQN cset; d \cr} 
$$
where the first four equations specify the degenerate `metric' structure
and a set of torsion-free connections on ${\M\,}$, the fifth one picks
out the Newton-Cartan connection from this set of generic possibilities,
the sixth one postulates the existence of absolute rotation, and the last one
relates spacetime geometry to matter in analogy with Einstein's field equation.

\subsection{Gauge and Lie-algebraic structures of Newton-Cartan spacetime}

\subsubsection{The gauge structure}

Since the `\ae ther-frame' with its four-velocity $u$ has been adapted as an
auxiliary structure into the description of Newton-Cartan framework delineated
in the previous subsections,
any physical theory constructed in accordance with this framework must be
invariant under changes in $u$ --- and, hence, under associated changes in the
`vector-potential' $A$ --- if the theory is to maintain general covariance.
Using a particular pair of potentials ${(u,\;A)}$ rather than a given
Newton-Cartan connection $\Gamma$ amounts to choosing a `Bargmann gauge' in the
fiber-bundle formulation of the Newtonian gravity studied by Duval and 
K\"unzle\cite{Duval and Kunzle}. They take Bargmann bundle $B(\M)$ to be a
principal bundle over the Galilean manifold $\M$ with the Bargmann group $B$
(a non-trivial central extension of the inhomogeneous Galilean group ${\cal G}$
by an abelian one-dimensional phase group U(1) appearing in the exact sequence
${1\rightarrow{\rm U(1)}\rightarrow B\rightarrow{\cal G}\rightarrow 1}$) as its
structure group. It is a U(1)-extension of a sub-bundle ${{\cal G}(\M)}$ of the
bundle of Galilean-relativistic affine frames of ${\M\,}$, with a surjective
principal bundle homomorphism ${B(\M)\rightarrow {\cal G}(\M)\,}$, constructed
as follows. Let ${Gl(\M)}$ be the principal bundle of linear frames over $\M$.
A {\it Galilean structure} is then a reduction of
${Gl(\M)}$ to the homogeneous subgroup ${{\cal G}_0 := SO(3)\,\semidirect\,
{\real}^3}$ of the Galilean group (where ${\semidirect}$ denotes a semidirect
product), and the sub-bundle ${{\cal G}(\M)}$ of the bundle of
Galilean-relativistic affine frames of $\M$ is the pull-back of 
${{\cal G}_0(\M)}$ by the canonical projection $T\M\rightarrow \M$, where
$T\M$ is the tangent bundle over $\M$. If ${i_o:{\cal G}_0(\M)\hookrightarrow
{\cal G}(\M)}$ denotes an embedding through the zero section of $T\M$, and
${B(\M)}$ is a U(1)-extension of ${{\cal G}(\M)}$ with structure group $B$,
then let ${B_0(\M)}$ be the pull-back of ${B(\M)}$ by ${i_o}$.
The quotient bundle ${P(\M):=B_0(\M)/{\cal G}_0}$ is
then a U(1)-principal bundle over $\M$. Now, it can be
shown\cite{Duval and Kunzle} that any compatible Newton-Cartan connection
defined by equation \Ep{constraint} above designates an entire class of
connections on ${P(\M)}$ which are in one-to-one correspondence with the
sections $u$ of the unit tangent bundle ${T\!\suba \M = {\cal G}_0(\M)/SO(3)}$
of the structure ${(\M;\,{\tau},\,{h})}$.
Clearly, the sections $u$ are nothing but the four-velocity
observer fields considered above with their gauge-dependent spatial projection
fields ${\!\!\!\!\!\!\buildchar{\;\;\;\;h_{\mu\nu}}{u}{}\,}$.
The gauge theory of Newtonian gravity we are
considering must, therefore, be covariant under changes of the sections $u$,
over and above the covariance under the automorphisms of 
${P(\M)}$. Duval and K\"unzle show that the Newton-Cartan connection defined by
the constraint \Ep{constraint} on the Galilean structure \Ep{three-cond}
is indeed invariant under the simultaneous changes
$$
\EQNalign{\chi&\mapsto{\widetilde \chi} = \chi + f \EQN vertical; a \cr
          u^{\alpha}&\mapsto{\widetilde u}^{\alpha} = u^{\alpha} + 
               h^{\alpha\sigma}{\rm w}_{\sigma} \EQN vertical; b \cr
          A_{\alpha}&\mapsto{\widetilde A}_{\alpha} = A_{\alpha} +
\partial_{\alpha}f + \,{\rm w}_{\alpha} - (u^{\sigma}{\rm w}_{\sigma} +
{\scriptstyle{1\over 2}}h^{\mu\nu}{\rm w}_{\mu}{\rm w}_{\nu})t_{\alpha}\;, 
\EQN vertical; c \cr}
$$
where ${\chi}$ is the U(1)-phase (treated as the only intrinsic group
coordinate), ${f\in C^{\infty}(\M,{\real})}$ is an arbitrary smooth map
${\M\rightarrow{\real}\,}$, ${\rm w}$ is a Galilean boost 1-form 
(defined only modulo ${t_{\alpha}}$) belonging to the space of
covector-fields ${\Omega^1(\M)}$
on $\M$, and ${A_{\alpha}}$ is the `vector-potential'
defined above. The transformations \Ep{vertical;a} and \Ep{vertical;b}
can be regarded as the `vertical automorphisms' of the unit tangent bundle
${B_0(\M)/SO(3)}$, and along with the diffeomorphisms
${\phi\in{\rm Diff}(\M)}$ of ${\M}$ they compose the complete automorphism
group
$$
{\cal A}ut(B(\M)) \,:=\, \Leftcases{21pt}
(\phi,{\rm w}, f)|\,\phi\in{\rm Diff}(\M),\,
{\rm w}\in \Omega^1(\M), \,f\in C^{\infty}(\M,{\real})
\Rightcases{21pt}\EQN gauge1
$$
of the Newton-Cartan theory of gravity. Conversely, the projections of the
fiber-preserving elements of the group ${{\cal A}ut(B(\M))}$ by the bundle
projection map ${\pi:B(\M)\rightarrow\M}$ are the smooth coordinate
transformations of ${\M}$ into itself which constitute the diffeomorphism
group ${{\rm Diff}(\M)\,}$; and, since the projection map ${\pi}$ is a group
homomorphism, the kernel of this map is the group
$$
{\cal V}(B(\M))\,:=\,[\Omega^1(\M)\times C^{\infty}(\M,{\real})]\EQN vert-group
$$
of vertical
gauge transformations defined by equation \Ep{vertical}. Thus, the group of
bundle automorphism \Ep{gauge1} encapsulates two classes of invariance which
must be respected by the action ${\cal I}$ of any gauge-invariant field theory
compatible with the Newton-Cartan structure ${(\M;\;{\tau},\;{h},\;\Gamma\,)}$.
Mathematically, the elements of one of these two classes correspond to
transforming the {\it fibers} of the bundle space ${B(\M)}$ by the action of
the normal (or invariant) subgroup ${{\cal V}(B(\M))}$ of the complete
automorphism group
${{\cal A}ut(B(\M))\,}$, whereas the elements of the other class correspond to
transforming the {\it base space} ${\M}$ of the bundle by the action of the
factor subgroup ${{{\rm Diff}(\M)}={{\cal A}ut(B(\M))}/{{\cal V}(B(\M))}\,}$.
This means that the automorphism group ${{\cal A}ut(B(\M))}$ has the structure
of a semidirect product ${{\cal V}(B(\M))\,\semidirect\,{\rm Diff}(\M)}$
indicating its status in the exact sequence
$$
1\;\longrightarrow\;{\cal V}(B(\M))\;\longrightarrow\;{\cal A}ut(B(\M))
\;\longrightarrow\;{\rm Diff}(\M)\;\longrightarrow\;1\;,\EQN
$$
and it acts on the set ${\{({h},{\tau}, u, A)\}}$ by
$$
(\phi,{\rm w}, f):\left(\matrix{{h}\cr\cr
                            {\tau}\cr\cr
                             u\cr\cr
                             A\cr}\right)\longmapsto \phi_*
\left(\matrix{{h}\hfill\cr\cr
              {\tau}\hfill\cr\cr
              {u + {h}({\rm w})}\hfill\cr\cr
              {A + df + {\rm w} - \{{\rm w}(u) + 
                  {\scriptstyle{1\over 2}}{h}
({\rm w},{\rm w})\}{\tau}}\cr}\right) \EQN gauge2
$$
giving the semidirect product group multiplication law:
$$
(\,\phi\suba ,{\rm w}\suba , f\suba\,)(\,
{\phi\subb},{{\rm w}\subb},{f\subb}\>)=
(\,\phi\suba{\phi\subb},\;{\phi\subb}^{\!\!*}\,{\rm w}\suba + 
{\rm w}\subb,\; 
{\phi\subb}^{\!\!*} f\suba + {f\subb}\>)\;,\EQN gauge3
$$
where ${\phi_*}$ is the push-forward map and ${\phi^*}$ is the pull-back
map corresponding to the diffeomorphism ${\phi\in{\rm Diff}(\M)\,}$.
More significantly, this complete gauge group ${{\cal A}ut(B(\M))}$ acts on the
Newton-Cartan structure ${(\M,\,h,\,\tau,\,\Gamma)}$ only via the quotient
subgroup ${{\rm Diff}(\M)\,}$,
$$
(\phi,{\rm w}, f):\left(\matrix{{h}\cr\cr
                            {\tau}\cr\cr
                            {\Gamma}\cr}\right)\longmapsto \phi_*
                  \left(\matrix{{h}\cr\cr
                            {\tau}\cr\cr
                            {\Gamma}\cr}\right)\;, \EQN gene-cove
$$
exhibiting that the principle of general covariance has
been quite consistently adapted from Einstein's theory of gravity to this
Newton-Cartan-Bargmann framework (cf. footnote ${\scriptstyle 2}$).

\subsubsection{The Lie-algebraic structure}

The degenerate `metric' structure of general Galilean spacetime permits some
plasticity in the Lie-algebraic structure ${T_e\,{\cal A}ut(B(\M))}$
of the automorphism group ${{\cal A}ut(B(\M))\,}$ (here ${e}$ is
the unit element of the group). 
At least {\it six} different nested Lie algebras of
infinitesimal `isometries' (which play a role analogous to that of the algebra
of Killing vector fields of Lorentzian spacetime) have been
shown\cite{Duval-93} to naturally arise as possible candidates for
Newton-Cartan symmetry algebras. These nested Lie algebras include two
extreme cases: (1) the {\it Coriolis algebra} --- i.e., the Lie algebra of the
infinite-dimensional Leibniz group\cite{Earman}
(the symmetry group of most general `metric' automorphisms
of the Galilean-relativistic spacetime) --- and (2) the all important
{\it Bargmann algebra} --- i.e., the Lie algebra of the Bargmann group (the
fundamental symmetry group of massive, non-interacting Galilean-relativistic
systems --- classical or quantal).

To see how these two Lie-algebraic structures arise as extreme cases, recall
that Newton-Cartan connection ${\Gamma}$ can be expressed in terms of the
vector fields ${u}$ and ${v\,}$, and the scalar potential
${\buildchar{\Phi}{v}{}}$ (cf. equations \Ep{gauge-depend},
\Ep{u-part} and \Ep{two-gauges}). In terms of these variables the action
\Ep{gauge2} of the group ${{\cal A}ut(B(\M))}$ on the full
Newton-Cartan-Bargmann
structure ${(\M;\,h,\,\tau,\,u,\,v,\,{\buildchar{\Phi}{v}{}}\,)}$
can be expressed as
$$
\;\;\;\;\left(\matrix{{h}\cr\cr
           {\tau}\cr\cr
                u\cr\cr
                v\cr\cr
{\buildchar{\Phi}{v}{}}\cr}\right)\longmapsto \phi_*\,
\left(\matrix{{h}\hfill\cr\cr
           {\tau}\hfill\cr\cr
 u + {h}({\rm w})\hfill\cr\cr
      v + {h}(df)\hfill\cr\cr
{\buildchar{\Phi}{v}{}} + v(df) + {\scriptstyle{1\over 2}}\,{h}(df,\,df)
                       \cr}\right)\;,\EQN
$$
whereas the corresponding infinitesimal action of this automorphism group on
the structure ${(\M;\,h,\,\tau,\,u,\,v,\,{\buildchar{\Phi}{v}{}}\,)}$ can be
seen as\cite{Duval-93}
$$
\delta\left(\matrix{{h}\cr\cr
                 {\tau}\cr\cr
                      u\cr\cr
                      v\cr\cr
{\buildchar{\Phi}{v}{}}\cr}\right)\;\;=\;\;
\left(\matrix{{\hbox{\it\char36}{\!}_{\rm x}}{h}\hfill\cr\cr
           {\hbox{\it\char36}{\!}_{\rm x}}{\tau}\hfill\cr\cr
{\hbox{\it\char36}{\!}_{\rm x}}u + {h}({\theta})\hfill\cr\cr
{\hbox{\it\char36}{\!}_{\rm x}}v + {h}(d{\rm g})\hfill\cr\cr
      {\rm x}({\buildchar{\Phi}{v}{}}) + v({\rm g})\cr}\right)\;,\;\;\;\;\EQN
$$
where ${\theta\in\Omega^1(\M)\,}$, ${{\rm g}\in C^{\infty}(\M)\,}$, and
${\hbox{\it\char36}{\!}_{\rm x}}$
denotes a Lie derivative with respect to the symmetry generating vector-fields
${{\rm x} = {\rm x}^{\alpha}\partial_{\alpha}\,}$ on ${\M\,}$. The associated
Lie-algebraic structure of this infinitesimal action works out to be
$$
[({\rm x},\,\theta,\,{\rm g}),\;({\rm x}',\,\theta',\,{\rm g}')]\;\;=\;\;
([{\rm x},\,{\rm x}'],\;{\hbox{\it\char36}{\!}_{\rm x}}{\theta'}\,-\,
{\hbox{\it\char36}{\!}_{{\rm x}'}}\theta,\;{\rm x}({\rm g}')\,-\,
{\rm x}'({\rm g})\,)\;.\EQN
$$
If we now set ${0=\delta h=\delta\tau=\delta u=\delta v=\delta
{\buildchar{\Phi}{v}{}}}$ and, thereby, look for the stabilizer of the
Newton-Cartan-Bargmann structure for the case of flat spacetime, we
recover the Lie algebra of the Bargmann group\cite{Duval-93}. Thus,
the isotropy subgroup (or the stabilizer) of the full automorphism group
${{\cal A}ut(B(\M))}$ corresponding to the immutable flat structure
${({h} = \delta^{ab}\partial_a\otimes\partial_b,\;{\tau}= dt,\; 
u=\partial_{\scriptscriptstyle 0}, \;A = 0)}$ is nothing but the Bargmann
group, as one would expect. By relaxing one or more of the restrictions
${0=\delta u=\delta v=\delta{\buildchar{\Phi}{v}{}}\,}$, and/or imposing
various different restrictions on the connection ${\Gamma\,}$, a variety of
intermediate algebraic structures associated with some physically interesting
special cases may be worked out\cite{Duval-93}. All of these intermediate
symmetry groups are necessarily wider than the stabilizing Bargmann group.
They form different subgroups of the Leibniz group --- the most general
infinite-dimensional `isometry' group of the Newton-Cartan structure.
The generators of Leibniz group are restricted only by the
conditions ${{\hbox{\it\char36}{\!}_{{\rm x}}}h = 0}$ and
${{\hbox{\it\char36}{\!}_{{\rm x}}}\tau = 0\,}$, and, in general,
do not Lie-transport
the connection: ${{\hbox{\it\char36}{\!}_{{\rm x}}}\Gamma\not= 0\,}$. 
Consequently, the infinite-dimensional Lie algebra --- the Coriolis algebra
--- corresponding to the Leibniz group preserves the absolute structure
${(\M;\,h,\,\tau\,)}$ of the Newton-Cartan spacetime, but leaves the connection
${\Gamma}$ completely malleable, or dynamical. The elements of this most
general symmetry group of the absolute structure represented in an arbitrary
rigid frame take the form
$$
\EQNalign{\;\;\;\;\; {\rm x}^{\,\4} &= x^{\4} + c^{\4} \equiv t + c^{\4}\>,\cr
          {\rm x}^{\,a} &= O^{a}_{\;\>b}(t)\>x^b + c^{a}\!(t)\>,
                \;\;\;\;\;\;\;\;({\scriptstyle{a,b\;=\;1,2,3}}),
\EQN Leibniz \cr}
$$
where ${O^{a}_{\;\>b}(t)\in {\rm SO}(3)}$
form an orthogonal rotation matrix for
each value of $t$, ${c^{a}(t)\in\real^3}$ are arbitrary functions of
${t\in\real}$, and
${c^{\4}\in\real}$ is an infinitesimal time translation. Physically, this
infinite-dimensional symmetry group correspond to transformations that connect
different Galilean observers in arbitrary (accelerating and rotating)
relative motion.

One physically and historically important subgroup of Leibniz
group is the Milne group\cite{Milne}, which results as a direct consequence of
the additional constraint \Ep{additional} on the curvature tensor. In the above
notation, it simply eliminates the time-dependence of the rotation matrix
${O^{a}_{\;\>b}}$ in equation \Ep{Leibniz}, and may also be characterized by
the restriction 
$$
\hbox{\it\char36}{\!}_{\rm x}(h^{\nu\sigma}
\Gamma_{\mu\;\>\sigma}^{\;\>\alpha})\,\equiv\,
\hbox{\it\char36}{\!}_{\rm x}\,\Gamma_{\mu}^{\,\alpha\,\nu}\,=\,0\EQN
$$
imposed directly on the connection\cite{Duval-93} in addition to the conditions
${{\hbox{\it\char36}{\!}_{{\rm x}}}h = 0}$ and
${{\hbox{\it\char36}{\!}_{{\rm x}}}\tau = 0}$ on the metrics. The
vector-fields ${\rm x}$ then constitute a Lie algebra corresponding to the
infinitesimal Milne transformations 
$$
\EQNalign{\;\;\;\;\; {\rm x}^{\,\4} &= t + c^{\4}\>,\cr
          {\rm x}^{\,a} &= O^{a}_{\;\>b}\>x^b + c^{a}\!(t)\>,
                \;\;\;\;\;\;\;\;({\scriptstyle{a,b\;=\;1,2,3}}),
\EQN Milne \cr}
$$
discussed in \Ref{Carter}.

\subsection{Derivation of the matter conservation laws}

As is well-known\cite{Hawking}\cite{Wald}, in general relativity the principle
of general covariance is sufficient for a derivation of the local differential
conservation law ${\nabla_{\!\mu}T^{\mu\nu} = 0}$ for relativistic matter and
non-gravitational fields from variation of an action functional with respect to
the Lorentzian metric ${g_{\mu\nu}\,}$. The principle of general covariance in
variational formulation of that theory is encapsulated in the mathematical
requirement of invariance of the
gravitation plus matter action ${{\cal I}\,[g_{\mu\nu},\,\Psi]
\,:=\, {\cal I}_g\,[g_{\mu\nu}] \,+\, {\cal I}_m\,[g_{\mu\nu},\,\Psi]}$  
under the diffeomorphisms of the Lorentzian spacetime ${(\M;\,g_{\mu\nu})\,}$;
i.e., in the requirement that if the map
${{}^{(s)}\!\phi:\M\rightarrow\M}$ defines a one-parameter family of
diffeomorphisms, then ${{\cal I}\,[{}^{(s)}\!\phi_{\ast}g_{\mu\nu},
\,{}^{(s)}\!\phi_{\ast}\Psi]\,=\,{\cal I}\,[g_{\mu\nu},\,\Psi]\,}$,
where ${{}^{(s)}\!\phi_{\ast}}$ is the
push-forward map corresponding to ${{}^{(s)}\!\phi \in {\rm Diff}(\M)\,}$,
and ${\Psi}$ represents
the matter fields. Moreover, general covariance demands that the matter action
${{\cal I}_m}$ by itself must be invariant under these diffeomorphisms if it
were to retain an unequivocal physical meaning. In other words, for such
diffeomorphic variations we must have
$$
0 \,\;=\;\, {d\over{ds}}\,{\cal I}_m[g_{\mu\nu},\,\Psi]\,\;\equiv\;\,
\delta{\cal I}_m[g_{\mu\nu},\,\Psi]\,\;=\;\,
\int {{\delta{\cal I}_m}\over{\delta g_{\mu\nu}}}\;\delta g_{\mu\nu}
\;+\,\int {{\delta{\cal I}_m}\over{\delta\Psi}}\;\delta\Psi\;.\EQN param
$$
If we now define
$$
{1\over 2}\,T^{(\mu\nu)} \;=\; {1\over 2}\,T^{\mu\nu}
\;:=\; {1\over{\sqrt{-g}}}\,
{{\delta{\cal I}_m}\over{\delta g_{\mu\nu}}}\;,\EQN Hil-stress
$$
and assume that the matter fields satisfy the Euler-Lagrange equations
${{{\delta{\cal I}_m}\over{\delta\Psi}}=0}$, then equation
\Ep{param} amounts to
$$
0 \;\,=\;\, \delta{\cal I}_m \;\,=\;\,\int_{\cal O}
{\scriptstyle{1\over 2}}
\,T^{\mu\nu}\delta g_{\mu\nu}\;d{\scriptstyle{V}}\;\EQN acti-func
$$
for some compact region ${{\scriptstyle{\cal O}}\subset\M}$ with a non-null
boundary ${\scriptstyle{\partial{\cal O}}\,}$, where ${d{\scriptstyle{V}}}$ 
is the Lorentzian 4-volume element on $\M\,$. Using the
well-known relation ${\delta g_{\mu\nu}\,=\,\hbox{\it\char36}{\!}_{{\!}_{X}}\,
g_{\mu\nu}\,=\,2\,\nabla_{\!\lparen\mu}{X}_{\nu\rparen}}$ for such variations
(where ${\hbox{\it\char36}{\!}_{{\!}_{X}}}$
denotes Lie derivative with respect to an
arbitrarily chosen smooth vector field ${X_{\mu}}$ on ${\M\,}$) the above
functional equation can be rewritten as
$$
0=\!\int_{\cal O}\!\!T^{\mu\nu}\nabla_{\!\mu}{X}_{\nu}\,d
{\scriptstyle{V}}
 =\!\int_{\cal O}\!\!\nabla_{\!\mu}(T^{\mu\nu}{X}_{\nu})\,d
{\scriptstyle{V}}
 -\!\int_{\cal O}\!\!(\nabla_{\!\mu}T^{\mu\nu}){X}_{\nu}\,d
{\scriptstyle{V}}
=-\!\int_{\cal O}\!\!(\nabla_{\!\mu}T^{\mu\nu}){X}_{\nu}\,d
{\scriptstyle{V}}.\EQN
$$
Here the last equality is obtained by converting the volume integral 
${\int_{\cal O}\nabla_{\!\mu}(T^{\mu\nu}{X}_{\nu})\,d
{\scriptstyle{V}}}$ to a surface
integral, which vanishes because ${{X}_{\mu}}$ vanishes on the boundary of the
volume by assumption. Finally, since the region
${\scriptstyle{\cal O}}$ and the vector field
${{X}_{\mu}}$ are arbitrarily chosen, the desired conservation law for the
stress-energy of matter and non-gravitational field,
${\nabla_{\!\mu}T^{\mu\nu} = 0\,}$, immediately follows from this
equation\cite{Hawking}\cite{Wald}.

Thus, in Lorentzian spacetime the stress-energy tensor ${T^{\mu\nu}}$ is seen
to be dual to the metric ${g_{\mu\nu}}$ in the sense that variations of the
matter action ${{\cal I}_m}$ with respect to the semi-Riemannian metric lead to
the conserved matter tensor ${T^{\mu\nu}}$. This state of affairs naturally
suggests that in our Galilean-relativistic spacetime there should be
{\it a pair} of quantities ${(S_{\mu\nu},\,C^{\mu})}$ dual to the pair of
metrics ${(h^{\mu\nu},\,t_{\mu})}$ playing a role analogous to that of the
relativistic stress-energy tensor ${T^{\mu\nu}\,}$. However, as shown by
K\"unzle and Duval\cite{Kunzle, 76}\cite{Duval and Kunzle, 78}, if we consider
an action functional analogous to \Ep{acti-func} defined on the Galilean
spacetime ${\M\,}$, say
$$
\delta{\cal I}_m\;=\;\int_{\cal O}\Leftcases{24 pt}
{\scriptstyle{1\over 2}}S_{\mu\nu}\delta h^{\mu\nu}\,+\,C^{\mu}\delta
t_{\mu}\Rightcases{24 pt}\;\wp\,d^4x\,,\EQN first-guess
$$
with ${\wp\,d^4x}$ being the Galilean 4-volume element (cf. equation
\Ep{vol}), then the equations following from it do not correctly correspond to
the well-known balance equations of energy and momentum (they lack an essential
acceleration term of the classical equations.) The culprit, of course, is the
degenerate or non-semi-Riemannian structure of the Galilean spacetime; as we
have often
noted, the degenerate pair of metrics ${(h^{\mu\nu},\,t_{\mu})}$ does not
determine the connection completely. Fortunately, the difficulty also suggests
its resolution. As we now know, introduction of the pair ${(u^{\alpha},\,
A_{\alpha})}$ of gauge fields as an auxiliary structure in the Galilean
spacetime does fix the connection uniquely, and, hence, all we need to do is
to enlarge the set of variables in the argument of the above action functional
by these two gauge variables. But then the principle of general covariance
demands that the variation of the resulting action,
${{\cal I}_m[h^{\mu\nu},\,t_{\nu},\,A_{\mu},\,u^{\nu},\,\Psi]\,}$,
must be invariant not just
under the diffeomorphisms of ${\M\,}$, but also under the vertical
transformations \Ep{vertical} of these gauge variables; i.e., the variations of
the matter action must be invariant under the entire gauge group
${{\cal A}ut(B(\M))}$ of the Newton-Cartan theory discussed in the previous
subsection. Accordingly, following Duval and
K\"unzle\cite{Duval and Kunzle, 78}\cite{Duval and Kunzle}, we augment
\Ep{first-guess} to be the functional
$$
\delta{\cal I}_m\;=\;\int_{\cal O}\Leftcases{24 pt}
{\scriptstyle{1\over 2}}S_{\mu\nu}\delta
h^{\mu\nu}\,+\,C^{\mu}\delta t_{\mu}\,+\,J^{\mu}\delta A_{\mu}\,+\,K_{\mu}
\delta u^{\mu}\Rightcases{24 pt}\;\wp\,d^4x\EQN second-guess
$$
of four different variations, ${\delta h^{\mu\nu}}$, ${\delta t_{\mu}}$,
${\delta A_{\mu}}$, and ${\delta u^{\mu}}$, and require it to be invariant
under the full gauge group ${{\cal A}ut(B(\M))\,}$.
Then, not only the equations
of motion for matter fields derived from the variations of this action are
gauge-invariant under ${{\cal A}ut(B(\M))}$ (see the following section for
an example), but the associated matter-current density ${J^{\mu}}$ and
`Hilbert' stress-energy tensor {\parindent 0.40cm\baselineskip 0.53cm
\footnote{$^{\scriptscriptstyle 4}$}{\ninepoint{\hang Following Duval and
K\"unzle we call this a `Hilbert' stress-energy tensor because it corresponds
to the general relativistic matter tensor \Ep{Hil-stress} obtained by
variations with respect to the Lorentzian metric. The canonical stress-energy
tensor implied by Noether's theorem is different and not
gauge-invariant.\par}}}
$$
N^{\mu}_{\,\;\;\nu} \;:=\; h^{\mu\sigma}S_{\sigma\nu} - C^{\mu}t_{\nu}
+ J^{\mu}(v_{\nu} - {\scriptstyle{1\over 2}}v^2t_{\nu}) + u^{\mu}K_{\nu}
\,,\EQN hilb-tens
$$
are also invariant under the action of that group,
where ${\rho:=J^{\sigma}t_{\sigma}}$, ${J^{\sigma}:=\rho v^{\sigma}}$,
${v_{\sigma}:={\!\!\!\!\!\!\buildchar{\;\;\;\;h_{\sigma\alpha}}{u}{}}
v^{\alpha}}$, and ${v^2:=v^{\sigma}v_{\sigma}}$. What is more, the
matter-current density and the `Hilbert' stress-energy tensor satisfy the
balance equations
$$
\EQNalign{\nabla_{\!\mu}J^{\mu} &\,=\, 0 \EQN balance; a \cr
{\rm and}\;\;\;\;\;\;\;\nabla_{\!\mu}N^{\mu}_{\,\;\;\nu} &\,=\,
\rho\,{\!\!\!\!\!\!\buildchar{\;\;\;\;h_{\nu\sigma}}{v}{}}\,v^{\alpha}
\nabla_{\!\alpha}v^{\sigma}\;\EQN balance; b \cr}
$$
(as meticulously shown by Duval and K\"unzle\cite{Duval and Kunzle, 78}),
which now correctly reduce to the classical expressions on flat
spacetime\cite{Kunzle, 76}.
Thus, the relativistic stress-energy tensor decouples into the pair
${(N^{\mu}_{\,\;\;\nu},\,J^{\mu})}$ in this nonrelativistic theory, describing
the stress and energy flow by the tensor ${N^{\mu}_{\,\;\;\nu}}$ and
the matter flow by the vector ${J^{\mu}\,}$. This distinct role of the concept
of mass-current from that of the stress-energy flow in the Newton-Cartan
theory is closely related to the well-known fact that mass plays distinctly
different roles in the Galilean and Poincar\'e invariant mechanics.

If we now define a mass-momentum-stress tensor (or matter tensor, for short) by
$$
M^{\mu\nu}\;\,:=\;\,-\,h^{\nu\sigma}
N^{\mu}_{\,\;\;\sigma}\;+\;J^{\mu}v^{\nu}\;,\EQN ma-mo-st-te
$$
then the balance equation \Ep{balance;b} immediately yields the nonrelativistic
matter conservation law
$$
\nabla_{\!\mu}M^{\mu\nu}\;=\;0\EQN cons-law
$$
analogous to the relativistic conservation law ${\nabla_{\!\mu}T^{\mu\nu}=0}$.
As a matter of fact, \Ep{cons-law} is precisely
the Galilean-relativistic limit-form of
the relativistic conservation law, as shown by K\"unzle\cite{Kunzle, 76}.
Substituting the expression \Ep{hilb-tens} into the definition
\Ep{ma-mo-st-te} of the matter tensor, and using the relation
${K_{\nu}\,=\,-\,{\!\!\!\!\!\!\buildchar{\;\;\;\;h_{\nu\sigma}}{u}{}}
J^{\sigma}}$ + ${(constant)\,t_{\nu}}$ (which has been derived by Duval and
K\"unzle in \Ref{Duval and Kunzle, 78}), we obtain the tensor
${M^{\mu\nu}}$ in a more transparent form:
$$
M^{\mu\nu}\;\,=\;\,\rho\,u^{\mu}u^{\nu}\,-\,2\rho\,u^{\lparen\mu}
h^{\nu\rparen\sigma}A_{\sigma}\,-\,
h^{\mu\sigma}h^{\nu\alpha}S_{\sigma\alpha}\;,\EQN matte-trans
$$
which is identical to the mass-momentum-stress tensor \Ep{mass-tens}
if we identify ${s^{\nu}\,\equiv\,\rho\,h^{\nu\sigma}A_{\sigma}\,}$.
Consequently, the matter conservation law \Ep{cons-law} is identical to the
condition \Ep{mass-cons}, which, at that stage, we had to impose independently.
Amiably enough, in the present variational approach it is a {\it derived}
result, following directly from the principle of general covariance, provided,
of course, the matter field equations
${{{\delta{\cal I}_m}\over{\delta\Psi}}=0}$ are satisfied.

\section{Quantum field theory on the curved Newton-Cartan spacetime}

\subsection{One-particle Schr\"odinger theory}

In the previous section we reviewed the classical Newton-Cartan
theory in some detail. The first systematic study of the {\it quantum}
theory of freely falling particles in (unquantized) Newton-Cartan
spacetime with illustrations of how the principle of equivalence works for 
such quantum systems has been carried out by Kucha\v r\cite{Kuchar}.
He arrives at a generally-covariant 
version of the Schr\"odinger's equation for a free quantum particle
in an arbitrary, noninertial frame in such a classical spacetime which, in
the special case of the Galilean frame with its distinguished
gravitational potential, reduces to the usual Galilean-relativistic
Schr\"odinger equation. Following the `parameterized' canonical
formalism\cite{Dirac},
Kucha\v r starts with an action integral, giving Newton-Cartan geodesics as
the extremal paths, as it appears to an arbitrary observer in an arbitrary
gauge. After `deparameterizing' it by labelling the worldlines with the
Newtonian absolute time $t\,$, 
he casts the action into the generalized Hamiltonian form to prepare
for the Dirac's constraint quantization. The transformation to the quantum
theory is then easily achieved in an unambiguous, coordinate independent
fashion by quantizing the motion of the particle with the use of the Dirac
method. The resulting quantum mechanical equation of motion, or the quantum
mechanical analog of the geodesic equation, turns out to be nothing but a
covariant version of the ordinary Schr\"odinger equation for the free
particle, with gravitational forces absorbed in the structure of spacetime
in which the particle is freely falling. Thus, in a nutshell, Kucha\v r has
successfully shown that quantum mechanics remains consistent in the presence
of the Newtonian connection-field viewed as an effect of the curving of
spacetime, and, due to the unique foliation possessed
by the Newton-Cartan spacetime, Galilean-relativistic quantum theory escapes
`the problem of time'\cite{Time}\cite{Kuchar, Boston}
usually encountered in the
attempts to canonically quantize parameterized dynamical systems in the
presence of Einsteinian connection-field. In other words, Kucha\v r has
shown that it is possible to unequivocally quantize Galilean-relativistic
parameterized systems by means of the Dirac method such that the evolution
of their quantum state ${\Psi\,:\,\M\rightarrow\comps\,}$,
unlike their general-relativistic counterpart,
does not depend on the choice of spacetime foliation (which, of course, is
uniquely given in the Newtonian case). Consequently, the corresponding
Hilbert-space inner-product ${\bra{\Psi_1}\Psi_2\rangle_t}$
remains the same for all such domains of simultaneity. 
It is worth noting here that in a subsequent work
De Bi\`evre\cite{De Bievre} has obtained essentially the same 
covariant Schr\"odinger-Kucha\v r equation (in the framework of the 
Bargmann bundle discussed in the previous section)
directly from an application of the
principle of equivalence rather than {\it a posteriori} exhibiting the 
compatibility of this principle with the correct quantum dynamics of a free
test particle in a gravitational field {\it \`a la} Kucha\v r. This further
justifies the validity of the principle of equivalence in the quantum domain,
and, to put more strongly, logically {\it demands} the generally-covariant
reformulation of the Galilean-relativistic quantum dynamics.

For our purposes in this paper, however, it is convenient to follow
the covariant framework of Duval and K\"unzle\cite{Duval and Kunzle}, who
show that a prescription of minimal (Newtonian) gravitational coupling
naturally leads to the four-dimensional spacetime-covariant
Schr\"odinger-Kucha\v r equation as a result of extremization of an action
of the form \Ep{second-guess} discussed in the previous section. Recall the 
U(1)-principle bundle $\M'$ from the subsection 2.4, 
and consider a vector bundle ${\rm E}$ with the Hilbert space
${L^2(\real^3)}$ of square integrable functions on ${\real^3}$
as the standard fiber associated with it. Under the gauge transformation
\Ep{gauge2} a section ${\Psi}$ of ${\rm E}$, or a wave-function, changes as
$$
(\phi,{\rm w},f): \Psi\>\longmapsto\>
{\phi_*}[\exp(i{m\over\hbar}f)\Psi]\;;\EQN functrans
$$
and, in analogy with the electromagnetic gauge theory, the covariant derivative
induced by the connection on $\M'$ acting on sections of ${\rm E}$ is
$$
{\rm D}_{\alpha}\Psi := (\partial_{\alpha} 
- i{m\over\hbar}A_{\alpha})
\Psi\;.\EQN
$$
Now, a Lagrangian density for the free one-particle Schr\"odinger equation
$$
{{\hbar}^2\over{2m}}\,\delta^{ab}\partial_a\partial_b
\Psi + i{\hbar}\,\partial_{\scriptscriptstyle 0}\Psi = 0 
\;\;\;\;\;\;\;({\scriptstyle a,\,b\,=\,1,\,2,\,3})\EQN Schr
$$
may be taken to be
$$
{\cal L}_{\scriptscriptstyle Schr} = 
{{\hbar}^2\over{2m}}\,\delta^{ab}\partial_a\Psi\partial_b{\overline\Psi}
+ i{\hbar\over 2}
(\Psi\partial_{\scriptscriptstyle 0}{\overline \Psi} 
- {\overline \Psi}\partial_{\scriptscriptstyle 0}\Psi)\;,\EQN
$$
which on the curved Newton-Cartan spacetime becomes
$$
{\cal L}_{\scriptscriptstyle Schr}\longrightarrow
{\cal L}_{\scriptscriptstyle Kuch} = \wp\,4\pi G\,\Leftcases{36pt}
{{\hbar}^2\over{2m}}\,h^{\alpha\beta}{\rm D}_{\alpha}\Psi
{\overline{{\rm D}_{\beta}\Psi}} 
\,+\, i{\hbar\over 2}\,u^{\alpha}\!(
\Psi{\overline{{\rm D}_{\alpha}\Psi}} - {\overline\Psi}{\rm D}_{\alpha}
\Psi)\!\!\Rightcases{36pt}\EQN NC
$$
if we use the contravariant 3-metric 
$h^{\alpha\beta}$ for the Laplacian in place of the Kronecker delta
${\delta^{ab}}$, write ${\wp\,d^4x}$ (cf. equation \Ep{vol})
for the spacetime volume element on ${\M}$, and
replace ${\partial_{\scriptscriptstyle 0}\Psi}$ by
${u^{\alpha}{\rm D}_{\alpha}\Psi\,}$ (the propagation covariant derivative
along the timelike vector field ${u^{\alpha}}$)
and ${\partial_{\alpha}\Psi}$ by ${{\rm D}_{\alpha}\Psi}$ in accordance with
the minimal interaction principle (the meaning of the multiplicative factor
${4\pi G}$ will become clear in the following section). This latter
Lagrangian density is manifestly covariant with respect to the
diffeomorphisms ${\phi\in{\rm Diff}(\M)}$ of $\M$ since ${h^{\alpha\beta}}$, 
$t_{\alpha}$, $A_{\alpha}$, $u^{\alpha}$, $\Psi$, and ${\overline\Psi}$ are
all tensor fields on the spacetime. What is more,
it is also invariant under the simultaneous vertical gauge transformations 
\Ep{vertical} of $A$, $u$, $\Psi$, and ${\overline\Psi}$\,. Consequently, we
expect the resulting
Schr\"odinger-Kucha\v r theory to be independent of the choice of a Galilean
observer represented by $u$ with the corresponding gravitational interaction
being uniquely described by
the Newton-Cartan structure $(\M;\,{h},\,{\tau},\,\nabla)$ alone. 
The Euler-Lagrange equations corresponding to the Lagrangian density
\Ep{NC} obtained by {\it independently}
extremizing the corresponding action functional ${{\cal I}_{Kuch}}$
with respect to ${\overline\Psi}$ and ${\Psi}$ are
$$
{\cal E}_{\scriptscriptstyle{Kuch}}[\Psi] = [
{\hbar^2\over{2m}}{\rm D}^{\alpha}{\rm D}_{\alpha} +
i{\hbar}u^{\alpha}{\rm D}_{\alpha} +
i{\hbar\over 2}{\nabla}_{\!\alpha}u^{\alpha}]\Psi = 0\EQN E-L
$$
and its complex conjugate, respectively; and once the gauge-covariant
derivatives ${{\rm D}_{\alpha}}$ in equation \Ep{E-L} are worked out, we indeed
obtain the desired covariant Schr\"odinger-Kucha\v r equation on the curved 
Newton-Cartan spacetime:
$$
[{\hbar^2\over{2m}}\nabla^{\alpha}\partial_{\alpha} +
i{\hbar}\!(u^{\alpha} - h^{\alpha\beta}A_{\beta})
\!\partial_{\alpha}
+ m\!(u^{\alpha}A_{\alpha} - {\scriptstyle{1\over 2}}h^{\alpha\beta}A_{\alpha}
A_{\beta})\!+ i{\hbar\over 2}\nabla_{\!\alpha}\!(u^{\alpha} -
h^{\alpha\beta}A_{\beta})\!]\Psi = 0.\EQN Schr-Kuch
$$
Among many interesting results in this
theory\cite{Duval and Kunzle}\cite{Kuchar}, 
the two we will need later are the expressions for the conserved
matter-current density (cf. equation \Ep{balance;a})
$$
J^{\alpha} \,:=\,{1\over{\wp\,4\pi G}}\,
{{\delta{\cal I}_{Kuch}}\over{\delta A_{\alpha}\;\;\;\;}}\,
=\,m\Psi{\overline\Psi}\>u^{\alpha} +\,
i{\hbar\over 2}\,h^{\alpha\beta}\!(
\Psi{\overline{{\rm D}_{\beta}\Psi}} - {\overline \Psi}{\rm D}_{\beta}
\Psi)\,,\;\;\;\;\;\;\;\nabla_{\alpha}J^{\alpha} = 0\,,\EQN J
$$
where ${\rho\equiv m\Psi{\overline\Psi}}$ is the invariant mass density, and,
more significantly, the conserved matter
tensor (cf. equation \Ep{matte-trans})
$$
M^{\mu\nu} \,=\, m\Psi{\overline\Psi}\,u^{\mu}u^{\nu}\,-\,2
\,m\Psi{\overline\Psi}
\,u^{\lparen\mu}h^{\nu\rparen\sigma}A_{\sigma}\,-\,S^{\mu\nu}\,,\;\;\;\;\;\;\;
\nabla_{\!\mu}M^{\mu\nu} = 0\,,\EQN M
$$
with
$$
S^{\mu\nu}\;=\;\Leftcases{24pt}{\scriptstyle{1\over 2}}h^{\sigma\alpha}
\Omega_{\alpha\sigma}\,+\,u^{\sigma}\Omega_{\sigma}\,-\,m\Psi{\overline\Psi}
\,u^{\sigma}\!A_{\sigma}\Rightcases{24pt}h^{\mu\nu}\,-\,2\,u^{\lparen\mu}
h^{\nu\rparen\sigma}\Omega_{\sigma}\,-\,h^{\mu\sigma}h^{\nu\alpha}
\Omega_{\sigma\alpha}\,,\EQN
$$
${\Omega_{\mu\nu}:={{\;\hbar^2}\over m}{\rm D}_{\lparen\mu}\Psi\,
{\overline{{\rm D}_{\nu\rparen}\Psi}}\,}$, and
${\Omega_{\mu}:={{i\hbar}\over 2}
(\Psi\partial_{\mu}{\overline{\Psi}}-{\overline{\Psi}}\partial_{\mu}\Psi)}$.
As the above covariant Schr\"odinger-Kucha\v r equation, both of these
conserved flows appear here as quantities {\it derived} from the action
functional, and their respective expressions are invariant under the 
combined diffeomorphisms and gauge transformations \Ep{gauge2} as expected.

\subsection{Galilean-relativistic quantum field theory}

In the previous subsection we have reviewed one-particle Schr\"odinger
mechanics on the curved Newton-Cartan spacetime. It is
well-known\cite{Schweber}\cite{Brown}\cite{Gross},
however, that such a mechanics on the usual flat
Galilean spacetime can be interpreted also as a second-quantized field theory
in which the wave-function ${\Psi}$ becomes an operator-valued distribution
in the Fock space of unspecified number of identical particles. Thanks to the
above described minimal coupling formulation of the Schr\"odinger-Kucha\v r
theory, it turns out that this Fock-space representation of many-particle
system can be quite straightforwardly generalized to our case of curved
background. Given the Lagrangian density \Ep{NC}, the first step towards
constructing the alleged generalization is to evaluate the momenta conjugate
to the classical c-number fields ${\Psi}$ and ${\overline\Psi}$:
$$
{\overline {\rm P}}\;:=\;{{\delta\,{\cal L}_{Kuch}}
\over{\delta(u^{\sigma}{\rm D}_{\sigma}
\Psi)}}\;=\;-\,2\,i\,\hbar\,\wp\,\pi\,G\,{\overline \Psi}\,,\EQN c-bum-mom
$$
and its complex conjugate. This relation suggests that we should not regard 
${\overline\Psi}$ as an independent variable, but rather as proportional to
the canonical conjugate of ${\Psi}$. Therefore, we introduce a new, more
appropriate set of canonical variables ${\{\psi,\,p\}\,}$,
$$
\EQNalign{
\psi\;&:=\;{1\over{\hbar\sqrt{8\pi G\,\wp}}}\,
(\,2\pi G\,\wp\,\hbar\,{\overline\Psi}\;+\;i\,{\overline{\rm P}}\,),\cr
p\;&:=\;{1\over{\sqrt{8\pi G\,\wp}}}\,
(\,{\rm P}\;+\;i\,2\pi G\,\wp\,\hbar\,\Psi\,),\EQN new-c\cr}
$$
so that equation \Ep{c-bum-mom} and its conjugate yield
$$
\psi\;=\;{\sqrt{2\pi G\,\wp}}\;
\,{\overline{\Psi}}\;\;\;\;\;\;\;\;{\rm and}
\;\;\;\;\;\;\;\;p\;=\;i\,\hbar\;{\overline\psi}\;.\EQN
$$ 
Given a spacelike hypersurface ${\Sigma_t}$ (cf. subsection 4.2 below), the
quantum theory can now be easily defined by the usual equal-time commutation
${(-)}$ or anticommutation ${(+)}$ relations,
$$
\EQNalign{
\lbrack\widehat{\psi}(x),\,\widehat{\psi}(x')\rbrack_{\mp}\;&=\;0\,,\;\;
\lbrack\widehat{\psi}^{\dagger}(x),\,\widehat{\psi}^{\dagger}(x')\rbrack_{\mp}
\;=\;0\,,\cr
\;\;{\rm and}\;\;\;\;\lbrack\widehat{\psi}(x),\,&\,\widehat{\psi}^{\dagger}(x')
\rbrack_{\mp}
\;=\;\widehat{\ident}\,\delta\!({\vec{\rm x}}-{\vec{\rm x}}\,')\,,
\EQN CCR \cr}
$$
corresponding to the bosonic ${(-)}$ or fermionic ${(+)}$ algebra,
where ${{\vec{\rm x}}\in \Sigma_t\,}$, ${\hat{\ident}}$ is the identity
operator in the Fock space, and, as always, we have replaced the complex
conjugate function ${\overline\psi}$ with the Hermitian conjugate operator
${\widehat{\psi}^{\dagger}\,}$. It is worth recalling the well-known
fact\cite{Brown} that {\it a Galilean-relativistic theory provides no
connection between spin and statistics}. Put differently, the Schr\"odinger
field can be quantized equally well by imposing either commutation or
anticommutation relations on the field operators as we have done.
In either case, the operator ${\widehat{\psi}^{\dagger}}$ acts as a creation
operator, whereas its Hermitian conjugate ${\widehat{\psi}}$ acts as an
annihilation operator, which is assumed to annihilate the Milne-invariant
(cf. equation \Ep{Milne}) vacuum state:
$$
\widehat{\psi}\,\ket{\,0\,}\;=\;0\,.\EQN vacuum
$$
All the operators
needed to describe the Schr\"odinger field theory on the curved Newton-Cartan
spacetime can now be constructed from these two operators. One of the most
important among these is the number density operator
${\widehat{\varrho}=\widehat{\psi}^{\dagger}\widehat{\psi}}$
which, as a consequence of the condition \Ep{vacuum}, annihilates the vacuum:
${\widehat{\varrho}\,\ket{\,0\,}=0\,}$.
A related operator relevant to us here is the mass operator
$$
{\widehat \bom}\;\;:=\;\;m
\int_{\Sigma_t}\widehat{\psi}^{\dagger}\widehat{\psi}\,
\;d^3x\;,\EQN operator-mass
$$
which implies
$$
[{\widehat\bom},\,\widehat{\psi}^{\dagger}(x)]_{\mp}\;=\;m\,
\widehat{\psi}^{\dagger}(x)
\;\;\;\;\;\;{\rm and}\;\;\;\;\;\;
[{\widehat\bom},\,\widehat{\psi}(x)]_{\mp}\;=\;-\,m\,\widehat{\psi}(x)
\,.\EQN
$$
As a result, we may also define a covariant mass-density operator
$$
{\widehat M}_{\mu\nu}\;:=\;
m\,\widehat{\psi}^{\dagger}\widehat{\psi}\,t_{\mu\nu}
\;,\EQN mass-density-operator
$$
which will be discussed further in the last section. Another important
operator of interest to us is the angular-momentum operator
$$
{\widehat{\bf J}}\;\;:=\;
\int_{\Sigma_t}\widehat{\psi}^{\dagger}\;{\bf s}\;\widehat{\psi}\,\;d^3x\;,
\EQN operator-angular
$$
giving
$$
[{\widehat{\bf J}},\,\widehat{\psi}^{\dagger}(x)]_{\mp}\;=\;{\bf s}\,
\widehat{\psi}^{\dagger}(x)\;\;\;\;\;\;{\rm and}\;\;\;\;\;\;
[{\widehat{\bf J}},\,\widehat{\psi}(x)]_{\mp}\;=\;-\,{\bf s}\,
\widehat{\psi}(x)\;.\EQN
$$
In particular, a vanishing commutation with ${{\widehat{\bf J}}}$ of a given
field operator is inferred to imply that the field corresponds to spinless
bosons.

As usual, the conjugate momentum \Ep{c-bum-mom} and its conjugate allows the
construction of Hamiltonian density operator:
$$
\EQNalign{
{\widehat{\rm H}}_{Kuch}
\;:=&\;\,\;\widehat{\rm P}^{\dagger}\,u^{\sigma}\!({\rm D}_{\sigma}
\widehat{\Psi})\;+\;u^{\sigma}\!({\rm D}_{\sigma}\widehat{\Psi})^{\!\dagger}\,
\widehat{\rm P}\;-\;{\widehat{\cal L}}_{Kuch}\cr
=&\;-\,4\pi G\,\wp\,
{{\hbar^2}\over{2m}}\,h^{\mu\nu}({\rm D}_{\mu}\widehat{\Psi})
({\rm D}_{\nu}\widehat{\Psi})^{\dagger}\cr
=&\;-\,{{\hbar^2}\over{2m}}\;h^{\mu\nu}
[{\rm D}_{\mu}({\sqrt{2}}\;\widehat{\psi})]^{\dagger}
[{\rm D}_{\nu}({\sqrt{2}}\;\widehat{\psi})]\;.\EQN particle-hami\cr}
$$
Here the ordering of noncommuting operators is, of course, important, since, as
it is obvious form equation \Ep{CCR},
${\widehat{\psi}}$ and ${\widehat{\psi}^{\dagger}}$ in this expression
are not genuine operators but rather operator-valued distributions in the Fock
space. We have made use of the usual normal-ordering prescription of
keeping ${\widehat{\psi}^{\dagger}}$ to the left of ${\widehat{\psi}\,}$.

This completes our field-theoretic generalization of the Schr\"odinger theory
on the curved Newton-Cartan spacetime. Of course, the equal-time
(anti-)commutation relations \Ep{CCR} break manifest covariance of
the theory. However, as we
shall see in the subsection 4.3 below, this is not inevitable. What is
physically more significant is the following observation. The theory we
have constructed here is a quantum theory of {\it free} particles on the
curved Newton-Cartan spacetime. If we now heuristically try
to substitute the quantum operator \Ep{mass-density-operator} 
in the right-hand side of
the field equation \Ep{cosmology-1}, we realize that what is needed for
consistency is a theory in which these
quantized Schr\"odinger particles {\it produce} the quantized Newton-Cartan
connection-field through which they interact. In other words, consistency
requires a generally-covariant quantum field theory of identical
Schr\"odinger particles interacting through their own quantized Newtonian
gravitational field. A construction of such an interacting quantum field
theory is the goal of the remaining of this paper.

\section{Newton-Cartan-Schr\"odinger theory from an action principle}

Our aim in this section is, firstly, to obtain the complete
Newton-Cartan-Schr\"odinger theory epitomized in equations \Ep{cset},
\Ep{balance; a}, \Ep{cons-law}, and \Ep{Schr-Kuch}
from extremizations of a {\it single} ${{\cal A}ut(B(\M))}$-invariant
action, and, then, to recast the theory into a {\it constraint-free}
Hamiltonian form in 3+1-dimensions, as well as into a manifestly covariant
reduced phase-space form in 4-dimensions, to pave the way for quantization
in the next section.

\subsection{Covariant Lagrangian formulation of the theory}

Let us first segregate the dynamical variables from the non-dynamical
variables. For this purpose, recall that the Newton-Cartan connection is
dynamical in general and not an invariant backdrop: as discussed in the
subsection 2.4 above, the most general group of symmetry transformations of the
Galilean-relativistic structure --- the Leibniz group --- does not leave the
connection invariant; the generators ${\rm x}$ of the Leibniz group forming an
infinite-dimensional Lie algebra are constrained only by the conditions
${\hbox{\it\char36}{\!}_{\rm x}t_{\mu\nu} = 0}$ and
${\hbox{\it\char36}{\!}_{\rm x}h^{\mu\nu} = 0\,}$, and, in general,
do not Lie-transport the Newton-Cartan connection,
${\hbox{\it\char36}{\!}_{\rm x}\,\Gamma_{\alpha\;\>\beta}^{\;\>\gamma}
\not= 0\,}$. In other words, the respective `isometries' 
${\hbox{\it\char36}{\!}_{\rm x}t_{\mu\nu} = 0}$ 
and ${\hbox{\it\char36}{\!}_{\rm x}h^{\mu\nu} = 0}$ of the tensor fields
${t_{\mu\nu}}$ and ${h^{\mu\nu}}$ dictate that these fields are simply parts
of the immutable background structure of the Newton-Cartan spacetime, and it
is only the connection ${\Gamma}$ of the full Newton-Cartan structure
${(\M;\,h,\,\tau,\,\Gamma)}$ which is left unrestrained by the Leibniz
group.{\parindent 0.40cm
\baselineskip 0.53cm\footnote{$^{\scriptscriptstyle 5}$}{\ninepoint{\hang
Continuing our discussion in footnote ${\scriptstyle 2}$ regarding
the meaning of general covariance, it is worth emphasising that the metric
fields ${h^{\mu\nu}}$ and ${t_{\mu\nu}}$ here defining (in Stachel's
terminology\cite{Stachel}) the chronogeometrical structure of the
Newton-Cartan gravity have been specified {\it independently} of the
inertio-gravitational field ${\Gamma_{\alpha\;\>\beta}^{\;\>\gamma}\,}$.
Now in general relativity, due to the dynamical nature of the metric field
${g_{\mu\nu}\,}$, the meaninglessness of an {\it a priori} labelling of
individual spacetime points is axiomatic: a point in the bare manifold ${\M}$
is not distinguishable form any other point --- and, indeed, does not even
become a point with physical meaning --- until the
metric field is dynamically determined. Unlike in general relativity, however,
where the affine structure of spacetime is inseparably (and rather
wholistically) identified with the
inertio-gravitational field, in Newton-Cartan theory it {\it is} possible to
individuate spacetime points as {\it entia per se} existing
independently of, and logically prior to, the inertio-gravitational field
${\Gamma_{\alpha\;\>\beta}^{\;\>\gamma}\,}$. This is because (again in
Stachel's language) the metric fields
${h^{\mu\nu}}$ and ${t_{\mu\nu}}$ serve as non-dynamical
{\it individuating fields}\cite{Stachel} specifying once and for all the
immutable chronogeometrical structure, which in Newton-Cartan theory happens
to be independent of the mutable inertio-gravitational structure.
See \Ref{Penrose, GR}, however, for a somewhat differing emphasis on the
difficulties in pointwise identification of two different Newton-Cartan
spacetimes.\par}}}
Consequently, it is sufficient to treat the non-metric connection ${\Gamma}$
as the {\it only} dynamical attribute determined by the matter distribution
via the field equation \Ep{cosmology-1}.
On the other hand, we recall that any Newton-Cartan
connection can be affinely decomposed
in terms of an \ae ther-field ${u}$ and an arbitrary vector-potential ${A}$
as ${\Gamma_{\alpha\;\>\beta}^{\;\>\gamma} \;= \!\!\!\!\!\!
{\buildchar{\>\;\;\;\;\Gamma_{\alpha\;\>\beta}^{\;\>\gamma}}{u}{}}\, + 
\!\!\!\!\!\!\buildchar{\>\;\;\;\;\Gamma_{\alpha\;\>\beta}^{\;\>\gamma}}{A}{}}$
(cf. equation \Ep{affinely}). Therefore, it is sufficient to take the fields
$A$ and $u$ of the full Newton-Cartan structure ${(\M;\,h,\,\tau,\,u,\,A)}$
as the only dynamical variables of the gravitational
field when we proceed next to formulate the Newton-Cartan-Schr\"odinger
theory in a Lagrangian form.

In what follows, however, it is convenient to view our theory as a
{\it parameterized field theory}\cite{Dirac}\cite{Torre} by introducing
supplementary kinematical variables on which the two metrics depend. To see
how this is done, consider an arbitrary covector-field
${\!\!\buildchar{\;A_{\mu}}{\circ}{}\,}$, an observer field
${\buildchar{\;u^{\nu}}{\circ}{}\,}$, the {\it classical}
Schr\"odinger-Kucha\v r complex scalar field ${\buildchar{\Psi}{\circ}{}}$,
and its complex conjugate ${\buildchar{{\overline{\Psi}}}{\circ}{}}$ as our
dynamical field variables defined on a generic background structure
${({\buildchar{\M\,}{\circ}{}};\,{\buildchar{h}{\circ}{}},\,
{\buildchar{\tau}{\circ}{}},\,{\buildchar{\nabla}{\circ}{}})}$ satisfying
${\hbox{\it\char36}{\!}_{\rm x}{\!\!\!\!\buildchar{\;\;\;t_{\mu\nu}}{\circ}{}}
=0=
\hbox{\it\char36}{\!}_{\rm x}{\!\!\!\!\buildchar{\;\;\;h^{\mu\nu}}{\circ}{}}}$
with an as-yet-unspecified connection ${\buildchar{\Gamma}{\circ}{}}$
given by equation \Ep{decomposed} ---
i.e., let us not restrain the connection {\it a priori} to be Newton-Cartan
by means of the constraint \Ep{constraint}; in other words,
at this juncture we do not impose the conditions \Ep{three-cond} on this
structure, and do not subject the arbitrary 1-form
${{\buildchar{A\,}{\circ}{}}={\!\!\buildchar{\;A_{\mu}}{\circ}{}}dx^{\mu}}$
on ${\buildchar{\M\,}{\circ}{}}$ to satisfy the Newton-Cartan
selection condition \Ep{2-form}. Since the metrics
${\!\!\!\!\buildchar{\;\;\;h^{\alpha\beta}}{\circ}{}}$ and
${\!\!\buildchar{\;\,t_{\beta}}{\circ}{}}$ are fixed, an action functional
defined on such a structure will not be invariant under all possible
diffeomorphisms of ${\M\,}$. Therefore, we enlarge the configuration space
of the theory by a new one-parameter family of kinematical variables
${{{}^{(s)}\!y}\in{\rm Diff}(\M)}$ constituting the map
$$
{{}^{(s)}\!y}\,:\,\M\longrightarrow{\buildchar{\M\,}{\circ}{}}\EQN map
$$
from a copy $\M$ of the manifold ${\buildchar{\M\,}{\circ}{}}$ to the manifold
${\buildchar{\M\,}{\circ}{}}$ itself, and think of ${{}^{(s)}\!y(x)}$ as a
field on $\M$ taking values in ${\buildchar{\M\,}{\circ}{}\,}$. As is
well-known,
this is nothing but the procedure of {\it parameterization}; i.e., the
procedure 
for obtaining a generally-covariant version of a field theory originally
defined on some non-dynamical, background spacetime. Using the map \Ep{map},
we can now pull-back the two metrics as well as the dynamical fields from the
`fixed' manifold ${\buildchar{\M\,}{\circ}{}}$ to the `parameterized' manifold
${\M\,}$:
$$
\EQNalign{h^{\mu\nu}(y)&\;:=\;(y_*)^{\mu}_{\;\;\alpha}(y_*)^{\nu}_{\;\;\beta}
    {\!\!\!\buildchar{\;\;\;h^{\alpha\beta}}{\circ}{}}\;,  \EQN pul-bak; a \cr
          t_{\mu}(y)   &\;:=\;(y^*)^{\alpha}_{\;\;\mu}
       {\!\!\buildchar{\;\,t_{\alpha}}{\circ}{}}\;,        \EQN pul-bak; b \cr}
$$
${A_{\mu}(y):=(y^*)^{\alpha}_{\;\;\mu}
{\!\buildchar{\;A_{\alpha}}{\circ}{}}\,}$,
${u^{\nu}(y):=(y_*)^{\nu}_{\;\;\mu}{\!\!\buildchar{\;\,u^{\mu}}{\circ}{}}\,}$,
${\Psi(y):=y^*{\buildchar{\Psi}{\circ}{}}\,}$, and
${{\overline{\Psi}}(y):=y^*{\buildchar{{\overline{\Psi}}}{\circ}{}}\,}$; where
${(y^*)^{\mu}_{\;\;\alpha}}$ denotes the induced map from the tangent space of
${x\in\M}$ to the tangent space of ${y(x)\in{\buildchar{\M\,}{\circ}{}}}$,
or, equivalently, the pull-back map from the cotangent space of ${y(x)}$ to the
cotangent space of ${x\,}$, and ${(y_*)^{\alpha}_{\;\;\mu}}$ denotes the
inverse of ${(y^*)^{\mu}_{\;\;\alpha}\,}$.

Having defined the dynamical variables ${A_{\mu}}$, ${u^{\nu}}$,
$\Psi$, and ${\overline\Psi}$ on the manifold ${\M}$ along with 
the supplementary kinematical variables ${{}^{(s)}\!y\,}$,
our next concern is to construct a meaningful classical phase-space for
the system using Lagrangian and Hamiltonian formalisms, and then define Poisson
brackets on this space in order to proceed with quantization in the next
section. Accordingly, we demand that extremizations of a stationary action
defined on $\M$ with respect to variations of these variables
lead to the complete set of gravitational field equations \Ep{three-cond},
\Ep{constraint}, and \Ep{cosmology-1}, the matter conservation law
\Ep{cons-law}, and the equation of motion \Ep{Schr-Kuch} for the classical
fields $\Psi$ and ${\overline\Psi}$ on the curved spacetime,
of the full Newton-Cartan-Schr\"odinger theory. A Lagrangian density which
fulfils this demand can then be used to construct a Hamiltonian density,
which would then lead us to the desired phase-space, and, subsequently,
quantization of the theory can be accomplished by interpreting the functions
on this phase-space as quantum mechanical operators.

Now, as discussed in the subsection 2.1, any Galilean spacetime
${(\M;\,{h},\,{\tau},\,\nabla)}$ is orientable since it possesses a canonical
4-volume element ${\wp\,d^4x}$ (cf. equation \Ep{vol}), which is,
conveniently, non-dynamical; in particular,
${\nabla_{\!\mu}\,{\scriptstyle{\cal E}}_{\alpha\beta\gamma\delta}=0\,}$.
Unlike the general relativistic 4-volume element, which is determined from the
dynamical gravitational field variables ${g_{\mu\nu}\,}$, here the volume
element is derived using only the {\it non-dynamical} metric fields
${h_{\mu\nu}}$ and ${t_{\mu\nu}\,}$, and, more importantly, is independent of
the dynamical field variables ${A_{\mu}}$ and ${u^{\nu}\,}$. Furthermore, even
in the absence of the conditions \Ep{three-cond} defining the Galilean
spacetime, one can begin with a measure form
${{\scriptstyle{\cal E}}_{\alpha\beta\gamma\delta}}$ defined on the generic
structure ${(\M;\,h^{\mu\nu},\,t_{\nu})}$ by equation \Ep{vol}. Therefore,
if ${Z}$ is the collection of all field configurations on the manifold
${\M\,}$, then we can form a tentative stationary action
${{\cal I}:\,{Z}\rightarrow\real}$ over some measurable region
${{\scriptstyle{\cal O}}\subset\M}$ with a non-null boundary
${\scriptstyle{\partial{\cal O}}}$ as follows:

$$
{\cal I}\,=\int_{\cal O}\!
d^4x\;{\cal L}\!(A_{\nu},\nabla_{\!\mu}A_{\nu},
\nabla_{\!\mu}\nabla_{\!\nu}A_{\alpha},\,u^{\nu},
\nabla_{\!\mu}u^{\nu},\nabla_{\!\mu}\nabla_{\!\nu}
u^{\alpha},\,\Psi,\,
\partial_{\mu}\Psi,\,{\overline \Psi},\,\partial_{\mu}{\overline{\Psi}}\,;\,
{{}^{(s)}\!y})\;\;\,\EQN action-int
$$
$$
\EQNalign{{\rm where}\;\;
{\cal L}\;\>&\;\equiv\;{\cal L}_{Grav}\;+\;{\cal L}_{Kuch}\;\equiv\;
              [{\cal L}_{\rm B}\,+\,{\cal L}_{\Gamma}\,+\,
  {\cal L}_{\rm N}\,+\,{\cal L}_{\Phi}\,+\,{\cal L}_{\Lambda}]\;+\;
            [{\cal L}_{\Psi}\,+\,{\cal L}_{\rm I}]\,,\;\;\;\;\,\EQN cl; a \cr
{\cal L}_{\rm B}&\;\equiv\; +\,\wp\,\Leftcases{26pt}
        \Upsilon_{\mu}\,h^{\mu\nu}t_{\nu}\;+\;\Upsilon^{\mu}_{\nu\sigma}\,
           \nabla_{\!\mu}h^{\nu\sigma}\;+\;\Upsilon^{\mu\nu}\,
          \nabla_{\!\mu}t_{\nu}\;+\;{\widetilde{\Upsilon}}^{\mu\nu}\,
   \partial_{\lbrack\mu}\,t_{\nu\rbrack}\Rightcases{26pt}\,,  \EQN cl; b \cr
{\cal L}_{\Gamma}&\;\equiv\; +\,\wp[\zeta\,\Leftcases{26pt}
      u^{\nu}t_{\nu}\,-\,1\Rightcases{26pt}\;+\;{\chi}^{\mu\nu}\Leftcases{32pt}
                           \nabla_{\!\lbrack\mu}\,A_{\nu\rbrack}\,+
                {\!\!\!\!\!\!\buildchar{\;\;\;\;\;h_{\sigma\lbrack\mu}}{u}{}}
                       \nabla_{\!\nu\rbrack}\,u^{\sigma}
                                        \Rightcases{32pt}]\,,  \EQN cl; c \cr
{\cal L}_{\rm N}&\;\equiv\; +\,\wp\;{\chi}^{\alpha\mu\nu}\Leftcases{30pt}
                  \nabla_{\!\lbrack\mu}\nabla_{\!\nu\rbrack}\,A_{\alpha}\,-
                    {\!\!\!\!\!\!\buildchar{\;\;\;\;h_{\alpha\sigma}}{u}{}}
                  \nabla_{\!\lbrack\mu}\nabla_{\!\nu\rbrack}\,u^{\sigma}
                                        \Rightcases{30pt}\,,  \EQN cl; d \cr
{\cal L}_{\Phi}&\;\equiv\; +\,\wp\;{{\kappa}\over 2}\Leftcases{26pt}
                               h^{\mu\nu}\,\nabla_{\!\mu}\Theta
                               \nabla_{\!\nu}\Theta\!\Rightcases{26pt}\,,
                             \;\;\;\;\;\;\Theta\!(A_{\alpha},\,u^{\sigma})
                             \,\equiv\,{1\over 2}\,h^{\mu\nu}A_{\mu}A_{\nu} -
  A_{\sigma}u^{\sigma}\,,\;\;\;\;\;\;\;\;\;\;\;\;\;\;\;\;\;\;   \EQN cl; e \cr
{\cal L}_{\Lambda}&\;\equiv\; +\,\wp\;\chi\,\Leftcases{26pt}
                    t_{\mu}\nabla_{\!\nu}{\chi}^{\mu\nu} \,-\, (1-{\kappa})
                           \nabla_{\!\sigma}\Theta^{\sigma}\,+\,
                            \lambda\,\Lambda\subn\,-\,\Lambda\subo\!
                           \Rightcases{26pt}\,,\;\;\;\Theta^{\sigma}\!:=
                      h^{\sigma\mu}\nabla_{\!\mu}\Theta\,,\;\;\;\EQN cl; f \cr
{\cal L}_{\Psi}&\;\equiv\; +\,\wp\;4\pi G\,\Leftcases{36pt}
                  {{\hbar}^2\over{2m}}\,h^{\alpha\beta}\partial_{\alpha}\Psi
         \partial_{\beta}{\overline{\Psi}}\;+\;i{\hbar\over 2}\,u^{\alpha}\!(
   \Psi\partial_{\alpha}{\overline{\Psi}}\,-\,{\overline \Psi}\partial_{\alpha}
                                 \Psi)\!\!\Rightcases{36pt}\,,  \EQN cl; g \cr
     {\cal L}_{\rm I}\;&\;\equiv\; -\,\wp\;4\pi G\int J^{\alpha}\,
                                  dA_{\alpha}\,,               \EQN cl; h \cr}
$$
${\Upsilon_{\mu}\,}$, ${\Upsilon^{\mu}_{\nu\sigma}\,}$,
${\Upsilon^{\mu\nu}\,}$, ${{\widetilde{\Upsilon}}^{\mu\nu}=
{\widetilde{\Upsilon}}^{\lbrack\mu\nu\rbrack}\,}$,
${\zeta\,}$, ${{\chi}^{\mu\nu}={\chi}^{\lbrack\mu\nu\rbrack}\,}$, and
${{\chi}^{\alpha\mu\nu}={\chi}^{\alpha\lbrack\mu\nu\rbrack}}$ are all
(independent) undetermined Lagrange multiplier fields (symmetric in
permutations of their indices unless indicated otherwise),
${\kappa}$ and ${\lambda}$ are {\it arbitrary} free parameters (with values
{\it including} ${0\,}$, ${1\,}$, and, in case of ${\lambda\,}$,
${{\Lambda\subo}\over{\Lambda\subn}}$), ${\chi}$
(also treated as a Lagrange multiplier field) will be
seen in the next subsection to be the arbitrary function of gauge
transformations (if ${\chi\mapsto\chi+f}$, then
${A_{\mu}\mapsto A_{\mu} + \partial_{\mu}f\,}$; cf. equation
\Ep{vertical}), the scalars ${\Lambda\subo}$ and
$$
\Lambda\subn\;:=\;
t_{\alpha}\nabla_{\!\sigma}\nabla_{\!\gamma}\,{\chi}^{\alpha\sigma\gamma}
\EQN Newton-Lambda
$$
will turn out to be parts of the cosmological constant, and 
$$
J^{\alpha} \;\equiv\; m\Psi{\overline\Psi}\,
\Leftcases{26pt}u^{\alpha}-h^{\alpha\beta}
A_{\beta}\Rightcases{26pt}\,+\,i{\hbar\over 2}\,h^{\alpha\beta}
\Leftcases{26pt}
\Psi\partial_{\beta}{\overline{\Psi}} - {\overline \Psi}\partial_{\beta}
\Psi\Rightcases{26pt}\,,\EQN J-un
$$
which is formally identical to the expression \Ep{J}. Note that we have
ensured the action ${\cal I}$ to have appropriate boundary terms by allowing
the Lagrangian ${\cal L}$ to contain pure divergences.
It is instructive to compare the Lagrangian density
${{\cal L}\equiv{\cal L}_{Grav}+{\cal L}_{\Psi}+{\cal L}_{\rm I}}$
with that for the Maxwell-Dirac system leading to
quantum electrodynamics with the corresponding action defined on
the background of Minkowski spacetime.
For the sake of convenience, in what follows we keep the notation of previous
sections, and continue to use the arbitrary observer field
${v^{\alpha}}$ as a difference field between the four-velocity vector field
${u^{\alpha}}$ and the arbitrary co-vector field ${A_{\alpha}}$ defined by
${v^{\alpha}:= u^{\alpha} - h^{\alpha\beta}A_{\beta}\,}$. Let us emphasize once
again that ${A=A_{\mu}dx^{\mu}}$ appearing in the action ${\cal I}$ is simply
an arbitrary 1-form on $\M\,$, as yet bearing no special relation to the 2-form
$F$ of equation \Ep{decomposed}.

Extremizations of the action ${\cal I}$ with respect to variations of the
multiplier fields ${\zeta\,}$, ${\Upsilon_{\mu}\,}$,
${\Upsilon^{\mu}_{\nu\sigma}\,}$, ${\Upsilon^{\mu\nu}\,}$, and
${{\widetilde{\Upsilon}}^{\mu\nu}}$ immediately yield the normalization
condition
$$
u^{\nu}t_{\nu}\;=\;1\EQN norm-cond
$$
for the timelike vector-field ${u}$ and the conditions
$$
h^{\mu\nu}t_{\nu}\;=\;0\,,\;\;\;\;
\nabla_{\!\mu}h^{\nu\sigma}\;=\;0\,,\;\;\;\;
\nabla_{\!\mu}t_{\nu}\;=\;0\,,\;\;\;\;{\rm and}\;\;\;\;
\partial_{\lbrack\mu}\,t_{\nu\rbrack}\;=\;0\;\EQN pre-cond
$$
specifying the Galilean structure (cf. equation \Ep{three-cond}).
Whereas its extremization with respect to variations of the tensor-field
${{\chi}^{\mu\nu}}$ yields the equation
\Ep{one-Lagrangian}:
$$
2\,\nabla_{\lbrack\mu}A_{\nu\rbrack} \;=\; -\,2
{\!\!\!\!\!\!\buildchar{\;\;\;\;\;h_{\sigma\lbrack\mu}}{u}{}}
\nabla_{\nu\rbrack}\,u^{\sigma}\,.\EQN Q-euler
$$
Consequently, once the last equation is compared with equation \Ep{useful},
we immediately obtain the condition
\Ep{2-form} entailing that the 2-form $F$ appearing in the connection
\Ep{decomposed} is closed. As discussed in subsection 2.2, the condition
\Ep{2-form} is equivalent to the Newton-Cartan field equation
$$
R^{\,\alpha\;\;\>\gamma}_{\,\;\;\>\beta\;\;\>\delta}
\,=\, R^{\,\gamma\;\;\>\alpha}_{\,\;\;\>\delta\;\;\>\beta}\;,\EQN int-cont
$$
which picks out the Newton-Cartan connection from the general Galilean
connections \Ep{decomposed}.
Thus, the extremization of ${\cal I}$ with respect
to ${{\chi}^{\mu\nu}\,}$ not only yields this Newton-Cartan field equation,
but, thereby, together with the relations \Ep{pre-cond} and \Ep{norm-cond},
also fixes the hitherto unspecified connection to be Newton-Cartan --- i.e.,
the one given by equations
\Ep{gauge-depend} and \Ep{affinely}. What is more, as a result of
extremizations of the action with respect to variations of the tensor-field
${{\chi}^{\alpha\mu\nu}}$ we also have the relation
$$
\nabla_{\!\lbrack\gamma}\nabla_{\!\delta\rbrack}\,A_{\mu}\;-\;
{\!\!\!\!\!\!\buildchar{\;\;\;\;h_{\mu\nu}}{u}{}}\,
\nabla_{\!\lbrack\gamma}\nabla_{\!\delta\rbrack}\,u^{\nu}\;=\;0\,,\EQN new-not
$$
or, equivalently (cf. equation \Ep{New-Can}),
$$
h^{\lambda\sigma}\,R^{\,\alpha}_{\;\;\;\sigma\,\gamma\,\delta}
\;\equiv\;R^{\,\alpha\lambda\;\;\>}_{\;\;\>\;\;\>\gamma\delta}
\;=\;0\,,\EQN add-constraints
$$
which, together with \Ep{int-cont}, allows us to
identify our spacetime structure (${\M;\,{h},\,{\tau},\,\nabla}$) as the
Newton-Cartan structure (cf. the last paragraph of subsection 2.3).
Consequently, we can now recognize the arbitrary covector-field ${A_{\mu}}$ as
the gravitational `vector-potential' defined by equation \Ep{2-form} and adopt
the entire body of formulae exclusive to the Newton-Cartan structure from
sections 2 and 3. In particular, we can now use the expression \Ep{curvature}
for the curvature tensor associated with the Newton-Cartan connection to
obtain the corresponding Ricci tensor
$$
R_{\mu\nu} \,=\, \Leftcases{31pt}
h^{\alpha\beta}\,{\!\!\buildchar{\,\;\nabla_{\!\alpha}}{v}{}}\,
{\!\!\!\buildchar{\,\;\nabla_{\!\beta}}{v}{}}\,
{\buildchar{\Phi}{v}{}}\Rightcases{31pt}t_{\mu\nu}\,,\EQN Ricci-old
$$
which, of course, is gauge-independent despite its appearance.
One way to see this gauge-independence is to note that --- thanks to the
tracelessness of the gravitational field tensor (${\!\!\!\!\!\!\!
{\buildchar{\>\;\;\;\;G_{\mu\;\>\alpha}^{\;\>\alpha}}{v}{}} = 0\,}$;
cf. equation \Ep{gravi-poten}) --- the
distinction between the `flat' covariant derivative
${{\!\!\buildchar{\,\;\nabla_{\!\alpha}}{v}{}}}$
and the `curved' covariant derivative ${\nabla_{\!\alpha}}$
disappears in the case of divergence. In particular,
${{\!\!\buildchar{\,\;\nabla_{\!\alpha}}{v}{}}
{\!\!\!\buildchar{\;\;\Phi^{\alpha}}{v}{}} =
\nabla_{\alpha}{\!\!\!\buildchar{\;\;\Phi^{\alpha}}{v}{}}\,}$, where
${\!\!{\buildchar{\;\;\Phi^{\alpha}}{v}{}}
:= h^{\alpha\beta}\,{\!\!\buildchar{\,\;
\nabla_{\!\beta}}{v}{}}\,{\buildchar{\Phi}{v}{}} =
h^{\alpha\beta}\,\partial_{\beta}{\buildchar{\Phi}{v}{}} =
h^{\alpha\beta}\,{\!\!\buildchar{\,\;
\nabla_{\!\beta}}{}{}}\,{\buildchar{\Phi}{v}{}}\,}$ because 
${\buildchar{\Phi}{v}{}}$ is a scalar. Therefore, as a direct consequence, the
above expression for the Ricci tensor can also be written in terms of the
{\it curved} covariant derivative operator ${\nabla_{\!\alpha}}$ as
$$
R_{\mu\nu} \,=\, \Leftcases{31pt}
\nabla_{\alpha}{\!\!\buildchar{\;\;\Phi^{\alpha}}{v}{}}
\Rightcases{31pt}t_{\mu\nu}
\,=\, \Leftcases{31pt}
h^{\alpha\beta}\,\nabla_{\!\alpha}\,\nabla_{\!\beta}\,
{\buildchar{\Phi}{v}{}}\Rightcases{31pt}t_{\mu\nu}\;.\EQN Ricci-new
$$

To extract further dynamical information from the action ${\cal I}$, we
next extremize it with respect to variations of the covector-field
${A_{\alpha}}$ and look for the corresponding Euler-Lagrange equations
$$
{{\delta{\cal L}\;}\over{\delta A_{\alpha}}}\;\equiv\;
{{\partial{\cal L}}\over{\partial A_{\alpha}}}\;-\;
\nabla_{\!\mu}\Leftcases{37pt}
{{\partial{\cal L}}\over{\partial(\nabla_{\!\mu}A_{\alpha})}}\Rightcases{37pt}
\;+\;\nabla_{\!\mu}\nabla_{\!\nu}\Leftcases{37pt}
{{\partial{\cal L}}\over{\partial(\nabla_{\!\mu}\nabla_{\!\nu}
A_{\alpha})}}\Rightcases{37pt}\;=\;0\,.\EQN
$$
Since the Newton-Cartan connection $\Gamma$ is invariant under variations of
the gauge variables $A$ and ${u\,}$ (cf. subsection 2.4), these equations give
the relations ${{{\delta{\cal L}_{\rm B}}\over{\delta{A_{\alpha}}}}=
{{\delta{\cal L}_{\Lambda}}\over{\delta{A_{\alpha}}}}=
{{\delta{\cal L}_{\Psi}}\over{\delta{A_{\alpha}}}}=0\,}$, and
$$
{{\delta{\cal L}\;}\over{\delta{A_{\alpha}}}}\;=\;
\wp\,\nabla_{\!\sigma}\,{\chi}^{\alpha\sigma}\;+\;
\wp\,\nabla_{\!\sigma}\nabla_{\!\gamma}\,{\chi}^{\alpha\sigma\gamma}\;+\;
\wp\,{\kappa}\,v^{\alpha}\,\nabla_{\!\sigma}\Theta^{\sigma}
\;-\; \wp\,4\pi G\,J^{\alpha}\;=\;0\,.\EQN gauge-euler
$$
As it stands, the last equation explicitly contains the
observer field ${v^{\alpha}\,}$, which can be eliminated by contracting both
sides of the equation with ${t_{\alpha}\,}$,
and using ${t_{\alpha}v^{\alpha}=1\,}$. Subsequently, after using \Ep{J-un}
with ${t_{\alpha}u^{\alpha}=1}$ in the resulting
equation and dividing it through by ${\wp\,}$, we obtain
$$
{\kappa}\,\nabla_{\!\sigma}\Theta^{\sigma}
\;+\;t_{\alpha}\nabla_{\!\sigma}\,{\chi}^{\alpha\sigma}\;+\;
t_{\alpha}\nabla_{\!\sigma}\nabla_{\!\gamma}\,{\chi}^{\alpha\sigma\gamma}
\;=\;4\pi G\,\rho\,,\EQN intern
$$
where ${\rho\equiv m\Psi{\overline\Psi}}$ as in equation \Ep{J}.

Amiably enough, it turns out that extremization of the action with
respect to the \ae ther-field ${u^{\alpha}}$ leads back to the same equation
\Ep{intern}; i.e., the dynamical pieces of
information extractable from the variations
of ${\cal I}$ with respect to ${A_{\alpha}}$ and ${u^{\alpha}}$ are identical.
The components of Euler-Lagrange equations in this twin case are:
${{{\delta{\cal L}_{\rm B}}\over{\delta{u^{\alpha}}}}=
{{\delta{\cal L}_{\Lambda}}\over{\delta{u^{\alpha}}}}=0\,}$,
$$
\EQNalign{{{\delta{\cal L}_{\Gamma}}\over{\delta{u^{\alpha}}}}\;&=\;
+\,\wp\,\Leftcases{26pt}\zeta\;t_{\alpha}\;-\;
{\!\!\!\!\!\!\buildchar{\;\;\;\;h_{\alpha\mu}}{u}{}}
\nabla_{\!\gamma}{\chi}^{\mu\gamma}\Rightcases{26pt}\,,    \EQN ae-u; a \cr
{{\delta{\cal L}_{\rm N}}\over{\delta{u^{\alpha}}}}\;&=\;
-\,\wp\,\Leftcases{26pt}{\!\!\!\!\!\!\buildchar{\;\;\;\;h_{\alpha\mu}}{u}{}}
\nabla_{\!\delta}\nabla_{\!\gamma}{\chi}^{\mu\delta\gamma}\Rightcases{26pt}
                                                         \,,\EQN ae-u; b \cr
          {{\delta{\cal L}_{\Phi}}\over{\delta{u^{\alpha}}}}\;&=\;
+\,\wp\;{\kappa}\,A_{\alpha}\,\Leftcases{26pt}
h^{\delta\gamma}\,\nabla_{\!\delta}\nabla_{\!\gamma}\,\Theta\!(A_{\mu},\,
u^{\nu})\Rightcases{26pt}\,,                               \EQN ae-u; c \cr
      {\,{\delta{\cal L}_{\Psi\,}}\over{\delta{u^{\alpha}}}}\;&=\;
+\,\wp\;4\pi G\;i\,{\hbar\over 2}\,\Leftcases{26pt}
\Psi\partial_{\alpha}{\overline{\Psi}} - {\overline \Psi}\partial_{\alpha}
\Psi\Rightcases{26pt}\,,                                   \EQN ae-u; d \cr
{\rm and}\;\;\;{\;{\delta{\cal L}_{{\rm I}\,}}\over{\delta{u^{\alpha}}}}\;\,&
=\;-\, \wp\;4\pi G\,\{m\Psi{\overline\Psi}\,\}\,A_{\alpha}\,.\EQN ae-u; e \cr}
$$
In evaluating the first two of the displayed equations we have used properties
\Ep{ink-rat}, \Ep{uniformity}, and \Ep{con-uniformity} of a geodesic observer.
Combining all of these components in the Euler-Lagrange equation
${{{\delta{\cal L}\;}\over{\delta u^{\alpha}}} = 0\,}$, dividing it through by
${\wp\,}$, contracting it with ${h^{\sigma\alpha}}$ and substituting 
${\delta^{\;\,\sigma}_{\mu}-t_{\mu}u^{\sigma}}$ in it for
${h^{\sigma\alpha}{\!\!\!\!\!\!\buildchar{\;\;\;\;h_{\alpha\mu}}{u}{}}\,}$,
using equation \Ep{gauge-euler} to substitute for
${\nabla_{\!\gamma}{\chi}^{\sigma\gamma}\,+\,
\nabla_{\!\delta}\nabla_{\!\gamma}{\chi}^{\sigma\delta\gamma}\,}$, 
and, finally, contracting the
result of all these operations (in this order) with ${t_{\sigma}}$ and using
${t_{\sigma}u^{\sigma}=1}$ yields equation \Ep{intern} as asserted. This
result is hardly surprising since, as we shall see in a moment, the
momenta canonically conjugate to the variables ${A_{\mu}}$ and ${u^{\nu}}$
are directly proportional to each other. 

Before we can analyze equation \Ep{intern} any further, however, we need to
either eliminate the undetermined multiplier tensors ${{\chi}^{\mu\nu}}$ and
${{\chi}^{\alpha\mu\nu}\,}$, or interpret them in
physical terms. A physical meaning of the multiplier field ${{\chi}^{\mu\nu}}$
is readily
revealed if we evaluate the four-momentum density canonically conjugate to the
gravitational field variable ${A_{\mu}}$ at each point ${x}$ on $\M\,$:
$$
\Pi^{\mu}\,\delta A_{\mu}\;:=\;{\cal J}^{\alpha}_{{}_{\!\!A}}\,t_{\alpha}
\;=\;\wp\,(t_{\alpha}{\chi}^{\alpha\mu})\,\delta A_{\mu}\;,\EQN four-moment
$$
where
$$
{\cal J}^{\mu}_{{}_{\!\!A}}\;:=\;\Leftcases{37pt}
{{\delta{\cal L}}\over{\delta(\nabla_{\!\mu}A_{\alpha})}}\Rightcases{37pt}
\,\delta A_{\alpha}\;+\;\Leftcases{37pt}
{{\delta{\cal L}}\over{\delta(\nabla_{\!\mu}\nabla_{\!\nu}
A_{\alpha})}}\Rightcases{37pt}\nabla_{\!\nu}\,\delta A_{\alpha}
\EQN symplectic-current
$$
is the {\it presymplectic potential current density}\cite{Lee}\cite{Barnich}, 
defined and discussed in the appendix A below (cf. equation \Ep{PPCD}). Here
we have dropped the 4-divergence term 
$$
\wp\,\nabla_{\!\nu}\,(t_{\mu}\,\chi^{\alpha\mu\nu}\,\delta A_{\alpha})
\EQN 4-term
$$
from the expression \Ep{four-moment}, arising from the ${{\cal L}_{\rm N}}$
component of the action, since, thanks to the vanishing of the boundary of a
boundary theorem, it does not contribute to the {\it presymplectic potential}
\Ep{pre-potential}. Note that
the antisymmetric nature of ${{\chi}^{\mu\nu}}$ requires ${\Pi^{\mu}}$
to be spacelike: ${t_{\mu}\Pi^{\mu}\equiv 0\,}$. It is also noteworthy that
the only non-zero contribution to the conjugate momentum density comes from
the ${{\cal L}_{\Gamma}}$ component of the action, which is also responsible
for the condition \Ep{constraint} specifying the Newton-Cartan
connection out of the generic possibilities \Ep{decomposed}. Thus, we now
understand the physical meaning of the multiplier field ${{\chi}^{\mu\nu}\,}$.
The physical meaning of the multiplier field ${{\chi}^{\alpha\mu\nu}\,}$, on
the other hand, has already been anticipated in the definition
\Ep{Newton-Lambda} of ${\Lambda\subn\,}$, which, in what follows, will be
viewed as a cosmological contribution.

The relations \Ep{four-moment} and \Ep{Newton-Lambda} allow us to rewrite
equation \Ep{intern} in
terms of the physical field variables ${\Psi}$, ${\overline\Psi}$, ${u^{\nu}}$,
${A_{\mu}\,}$, and the canonical conjugate field ${\Pi^{\mu}}$ of
${A_{\mu}\,}$:
$$
{\kappa}\,
\nabla_{\!\sigma}\Theta^{\sigma}\!(A_{\mu},\,u^{\nu})\;+\;{1\over{\wp}}\,
\nabla_{\!\sigma}\Pi^{\sigma}\;+\;\Lambda\subn\;=\;
4\pi G\,m\Psi{\overline\Psi}\;\equiv\;
4\pi G\,\rho\,,\EQN mod-intern
$$
where we have made use of ${\wp\,t_{\alpha}\nabla_{\!\sigma}\,
{\chi}^{\alpha\sigma}
\,=\,\wp\,\nabla_{\!\sigma}(t_{\alpha}{\chi}^{\alpha\sigma})\,=\,
\nabla_{\!\sigma}\Pi^{\sigma}\,}$. It is not surprising that only the
four-momentum density ${\Pi^{\mu}}$ conjugate to the field variable
${A_{\mu}}$ appears in this expression, because, as one may expect, the
four-momentum density ${\!\!{\buildchar{\;\;\Pi_{\nu}}{u}{}}}$ conjugate
to the observer field ${u^{\nu}}$ is just the `anti-dual' of ${\Pi^{\mu}\,}$,
$$
{\!\!{\buildchar{\;\;\Pi_{\nu}}{u}{}}}\,\delta u^{\nu}\;:=\;
{\cal J}^{\alpha}_u\,t_{\alpha}\;=\;(\,-\;
{\!\!\!\!\!\!\buildchar{\;\;\;\;h_{\nu\mu}}{u}{}}\,\Pi^{\mu})\,\delta u^{\nu}
\;,\EQN anti-dual
$$
(again, modulo the 4-divergence term
$$
\wp\,\nabla_{\!\nu}\,(t_{\mu}\,\chi^{\alpha\mu\nu}\,
{\!\!\!\!\!\!\buildchar{\;\;\;\;h_{\alpha\sigma}}{u}{}}\,\delta u^{\sigma})
\EQN 2nd4-term
$$
not affecting the symplectic structure)
which can be easily checked by explicitly evaluating ${{\cal J}^{\alpha}_u\,}$.
In fact, we have the relationship
$$
\Pi^{\mu}\;+\;h^{\mu\nu}{\!\!{\buildchar{\;\;\Pi_{\nu}}{u}{}}}\;\;\;=\;\;\;0
\;\;\;=\;\;\;{\!\!{\buildchar{\;\;\Pi_{\mu}}{u}{}}}\;+\;
{\!\!\!\!\!\!\buildchar{\;\;\;\;h_{\mu\nu}}{u}{}}\,\Pi^{\nu}\;,\EQN net-four-m
$$
which indicates that, strictly speaking, we should not view ${A_{\mu}}$ and
${u^{\nu}}$ as independent variables. In what follows, however, for the sake
of convenience, we shall continue treating them as if they were independent.
Eventually, in the next section, the constraint \Ep{net-four-m} will be taken
into account in a consistent manner.

Now, from the extremization of the action with respect to variations of the
scalar multiplier field ${\chi}$ we immediately infer that
$$
{1\over{\wp}}\,\nabla_{\!\sigma}\Pi^{\sigma}\;-\;(1-{\kappa})\,
\nabla_{\!\sigma}\Theta^{\sigma}\;+\;\lambda\,\Lambda\subn\;-\;
\Lambda\subo\;=\;0\;,\EQN chi-variation
$$
which, upon substitution into equation \Ep{mod-intern}, yields
$$
\nabla_{\!\sigma}\Theta^{\sigma}\;+\;\Lambda
\;=\;4\pi G\,\rho\,,\EQN tathe
$$
where
$$
\Lambda\;:=\;\Lambda\subo\,+\,(1-\lambda)\Lambda\subn\,.\EQN
$$
For the flat Galilean spacetime, the left-hand side of this equation 
is simply the Laplacian of the scalar ${\Theta(A_{\mu},\,u^{\nu})}$
plus the cosmological parameter ${{\Lambda}\,}$; and, hence,
it is just the Poisson equation of the classical
Newtonian theory of gravity provided we can interpret
${\Theta(A_{\mu},\,u^{\nu})}$ as the corresponding scalar gravitational
potential. But, of course, with ${A_{\mu}}$ recognized to be the gravitational
vector-potential as a result of equations \Ep{norm-cond}, \Ep{pre-cond}, and
\Ep{int-cont}, it is indeed
possible to identify the function ${\Theta(A_{\mu},\,u^{\nu})}$ in equation
\Ep{tathe} with the Newtonian gravitational scalar-potential
${{\buildchar{\Phi}{v}{}}(A_{\alpha},\,u^{\nu})}$
with respect to the observer-field 
${v^{\alpha}\equiv u^{\alpha}-h^{\alpha\beta}A_{\beta}\,}$:
$$
{\buildchar{\Phi}{v}{}}\!(A_{\alpha},\,u^{\nu})\;\equiv\;
\Theta\!(A_{\alpha},\,u^{\nu})\;=\;{1\over 2}\,h^{\mu\nu}
A_{\mu}A_{\nu} - A_{\sigma}u^{\sigma}\EQN equiv
$$
(cf. equation \Ep{two-gauges}). Using this identification in equation
\Ep{Ricci-new}, and multiplying equation \Ep{tathe} with ${t_{\mu\nu}}$,
it is now easy to obtain the last of the Newton-Cartan field equations,
$$
R_{\mu\nu}\;+\;\Lambda\,t_{\mu\nu} \;=\; 4\pi G\,M_{\mu\nu}\,,\EQN fil-eq
$$
as a generally-covariant generalization of the Newton-Poisson equation.
As noted before, an immediate inference one gains from this field equation is
that spacelike hypersurfaces of simultaneity ${\Sigma_t}$ embedded
in ${\M}$ are copies of flat, Euclidean three-spaces:
${h^{\mu\alpha}h^{\nu\sigma}R_{\alpha\sigma}\,=\,0\,}$. 
Unlike the general relativistic case, however, here the nonrelativistic
contracted Bianchi identities \Ep{bianchi} by themselves do not render the
scalar ${\Lambda}$ to be a spacetime constant. Nevertheless,
given the condition
\Ep{int-cont} on ${\M\,}$, a detailed analysis\cite{Dixon} of the tensor
representations of the Galilean group provided by both the physically sensible
matter tensor ${M^{\mu\nu}}$ and the curvature tensor
${R^{\,\alpha}_{\;\;\>\mu\,\sigma\,\nu}}$ of the general Galilean
spacetime, together with the
constraints due to nonrelativistic contracted Bianchi identities \Ep{bianchi},
reveals that ${\Lambda}$ in the above field equation at the most could be a
spacetime constant.

Note that neither ${\Lambda\subo}$ nor ${\Lambda\subn}$ are individually
required to be spacetime constants, only the net ${\Lambda}$ is.  
Further, there is nothing sacrosanct about the interpretation we have given
to ${\Lambda\subn\,}$. This contribution to ${\Lambda}$ arises from the
${{\cal L}_{\rm N}}$ term in the action. This is the term which gives rise to
the additional constraint \Ep{additional} on the curvature tensor. But,
as discussed in subsection 2.3, in the asymptotic limit this constraint is
automatically satisfied, and, hence, the ${{\cal L}_{\rm N}}$ term can be
dropped from the action; i.e, in such a case, the variation of the multiplier
tensor ${\chi^{\alpha\mu\nu}}$ itself must be set equal to zero. Moreover, in
this limit any cosmological contribution is also generally ruled out. And,
indeed, these physical requirements all come out consistently in our
interpretation of ${\Lambda\subn}$ if we set
${\lambda\,=\,{{\Lambda\subo}\over{\Lambda\subn}}}$ in the Lagrangian
${{\cal L}_{\Lambda}}$ giving ${\Lambda\,\equiv\,\Lambda\subn}$. However,
if one wishes, one can easily set ${\lambda\,=\,1}$ to avoid such a strong
link between the `Newtonian restriction' \Ep{additional} and the cosmological
constant. As long as the controversy over cosmological contributions is
unsettled, a choice of this part of the action is simply a matter of taste.

Turning now to the matter part of the action,
we recognize the Lagrangian density ${{\cal L}_{\Psi}+{\cal L}_{\rm I}}$
as nothing but the Schr\"odinger-Kucha\v r Lagrangian density
${{\cal L}_{Kuch}}$ expressed by equation \Ep{NC} of the previous section.
Since ${{\cal L}_{Grav}}$ is independent of the variables
${\Psi}$, ${\partial_{\mu}\Psi}$, and their conjugates, extremizations of the
action ${\cal I}$ with respect to variations of ${\Psi}$ and ${\overline\Psi}$
are identical to those of an action with Lagrangian density
${{\cal L}_{Kuch}\equiv{\cal L}_{\Psi}+{\cal L}_{\rm I}}$ which we have already
discussed in that section. The resulting Euler-Lagrange equations
${{{\delta{\cal L}}\over{\delta{\overline\Psi}}} = 0}$ and
${{{\delta{\cal L}}\over{\delta\Psi}} = 0}$ yield, respectively,
the Schr\"odinger-Kucha\v r equation \Ep{Schr-Kuch},
$$
[{\hbar^2\over{2m}}\nabla^{\alpha}\partial_{\alpha} +
i{\hbar}\!(u^{\alpha} - h^{\alpha\beta}A_{\beta})
\!\partial_{\alpha}
+ m\!(u^{\alpha}A_{\alpha} - {\scriptstyle{1\over 2}}h^{\alpha\beta}A_{\alpha}
A_{\beta})\!+ i{\hbar\over 2}\nabla_{\!\alpha}\!(u^{\alpha} -
h^{\alpha\beta}A_{\beta})\!]\Psi = 0,\EQN 4.23
$$
and its conjugate, describing the motion of a classical
Galilean-relativistic Schr\"odinger-Kucha\v r field ${\Psi}$
on the curved Newton-Cartan manifold achieved by minimally coupling it to the
gravitational `vector-potential' ${A_{\mu}\,}$.

Finally, let us not forget the remaining kinematical variables
${{}^{(s)}\!y:\M\rightarrow{\buildchar{\M\,}{\circ}{}}\,}$.
Extremization of the
action under variations of these auxiliary variables leads to
$$
0 \,\;=\;\, {d\over{ds}}\,{\cal I}\,\;\equiv\;\,
\delta{\cal I}\,\;=\;\,
\int {{\delta{\cal I}_{Grav}}\over{\delta {{}^{(s)}\!y^{\alpha}}\;\;\;}}\;
\delta {{}^{(s)}\!y^{\alpha}}\,\;+\;\,
\int {{\delta{\cal I}_{Kuch}}\over{\delta {{}^{(s)}\!y^{\alpha}}\;\;\;}}\;
\delta {{}^{(s)}\!y^{\alpha}}\;\,.\EQN kine-action
$$
As we already saw in the subsection 2.5 and the section 3, given the
matter field equations \Ep{4.23} and its complex conjugate, one of the
consequences of the covariance of the matter action ${{\cal I}_{Kuch}}$ is the
matter conservation laws
$$
\nabla_{\!\mu}J^{\mu}\,=\,0\;\;\;\;\;\;{\rm and}\;\;\;\;\;\;
\nabla_{\!\mu}M^{\mu\nu}\,=\,0\EQN remains-cons
$$
(cf. equations \Ep{balance; a} and \Ep{cons-law}), which correspond to the
relativistic conservation law ${\nabla_{\!\mu}T^{\mu\nu}=0\,}$.
In addition, the covariance of the matter action reduces equation
\Ep{kine-action} to
$$
0\,=\,
\int {{\delta{\cal I}_{Grav}}\over{\delta {{}^{(s)}\!y^{\alpha}}\;\;\;}}\,
\delta {{}^{(s)}\!y^{\alpha}}\,=\,
\int {{\delta{\cal I}_{Grav}}\over{\delta {h^{\mu\nu}}\;\;}}\,
\delta {h^{\mu\nu}}\,+\,
\int {{\delta{\cal I}_{Grav}}\over{\delta {t_{\mu}}\;\;\;}}\,
\delta {t_{\mu}}\,+\,
\int {{\delta{\cal I}_{Grav}}\over{\delta {A_{\mu}}\;\;\;}}\,
\delta {A_{\mu}}\,+\,
\int {{\delta{\cal I}_{Grav}}\over{\delta {u^{\nu}}\;\;\;}}\,
\delta {u^{\nu}}\,,\EQN kine-action
$$
since the variations of the action ${{\cal I}_{Grav}}$ with respect to all of
the multiplier fields have been required to vanish.
But now we can parallel the reasoning of the
subsection 2.5 with ${{\cal I}_m}$ replaced by ${{\cal I}_{Grav}}$ and obtain
the condition
$$
\nabla_{\!\mu}{\cal G}^{\mu\nu}\;=\;0\EQN grav-cons
$$
analogous to the equation \Ep{remains-cons} above, where
$$
{\cal G}^{\mu\nu}\,\;:=\;\,\{\nabla_{\!\sigma}\Theta^{\sigma}\,+\,\Lambda\}\,
u^{\mu}u^{\nu}\,-\,2\,\{\nabla_{\!\sigma}\Theta^{\sigma}\,+\,\Lambda\}\,
u^{\lparen\mu}h^{\nu\rparen\sigma}A_{\sigma}\,-\,h^{\mu\sigma}h^{\nu\alpha}
(S_{Grav})_{\sigma\alpha}\EQN
$$
(cf. equations \Ep{matte-trans} and \Ep{tathe}). Comparing this latter
expression with the general definition \Ep{upper-R} for ${R^{\mu\nu}}$ we
immediately see that it is equivalent to
$$
{\cal G}^{\mu\nu}\;\,\equiv\,\;R^{\mu\nu}\;+\;\Lambda\,v^{\mu}v^{\nu}\;.\EQN
$$
Substituting this expression into equation \Ep{grav-cons} and using
${\nabla_{\!\sigma}v^{\sigma}=0=v^{\mu}\nabla_{\!\mu}v^{\alpha}}$ for the
geodetic observer ${v^{\alpha}}$ gives
$$
\nabla_{\!\mu}R^{\mu\nu}\;=\;0\;,\EQN contbian-R
$$
which is nothing but the contracted Bianchi identities \Ep{bianchi} because
${h^{\mu\nu}R_{\mu\nu}=:R=0}$ according to equation \Ep{fil-eq}. Thus,
in close analogy with the Einstein-Hilbert theory, extremization
of the total action with respect to variations of the supplementary variables
${{}^{(s)}\!y}$ yields both the matter conservation law and the contracted
Bianchi identities.

It is remarkable that we have been able to obtain {\it all} of the field
equations of the Newton-Cartan theory, \Ep{cset}, the matter conservation laws
\Ep{balance; a} and \Ep{cons-law},
as well as the equation of motion \Ep{Schr-Kuch} for the
classical Schr\"odinger-Kucha\v r field ${\Psi}$ on the curved Newton-Cartan
spacetime, from a {\it single} action principle.{\parindent 0.40cm
\baselineskip 0.53cm\footnote{$^{\scriptscriptstyle 6}$}{\ninepoint{\hang All
previous
attempts to satisfactorily reformulate Newton-Cartan theory in four-dimensions
using variational principles have met insurmountable difficulties due to the
non-semi-Riemannian nature of the Galilean spacetime. For a partially
successful attempt, see reference\cite{Goenner}. For a review
of a {\it five-dimensional} formulation overcoming at least
some of the difficulties,
see reference\cite{remark}.\par}}} What is more, the resultant theory we have
obtained is covariant under the complete gauge group ${{\cal A}ut(B(\M))}$
discussed in the subsection 2.4. The invariance of ${{\cal L}_{Kuch}}$ under
this gauge group has already been emphasized in the previous section. Among
the four components of the Lagrangian
density associated with the connection-field, it can be explicitly checked that
${{\cal L}_{\rm B}}$, ${{\cal L}_{\Gamma}}$, ${{\cal L}_{\rm N}}$,
and ${{\cal L}_{\Lambda}}$
are all invariant under the complete
automorphism group ${{\cal A}ut(B(\M))}$ (the factor
${\{h^{\sigma\gamma}\,\nabla_{\!\sigma}\nabla_{\!\gamma}\,\Theta\}}$, of
course, does not change under the full gauge transformation; cf. equations
\Ep{Ricci-old} and \Ep{Ricci-new}). 
The third one, ${{\cal L}_{\Phi}}$, on the other
hand, is invariant only under the diffeomorphism subgroup,
${{\rm Diff}(\M)}$, and the `boost' subgroup,
$$
\EQNalign{u^{\alpha}&\mapsto u^{\alpha} + 
                                h^{\alpha\sigma}{\rm w}_{\sigma}\,, \cr
          A_{\alpha}&\mapsto A_{\alpha} +\,
                      {\rm w}_{\alpha} - (u^{\sigma}{\rm w}_{\sigma} +
{\scriptstyle{1\over 2}}h^{\mu\nu}{\rm w}_{\mu}{\rm w}_{\nu})t_{\alpha}\,,
\EQN boost \cr}
$$
of the full group ${{\cal A}ut(B(\M))}$. It is not invariant, in particular,
under the transformations ${A_{\mu}\mapsto A_{\mu} + \partial_{\mu}f}$.
As a result, the Euler-Lagrange equation \Ep{gauge-euler} depends on the choice
of this internal gauge;
for, under these transformations, ${v^{\mu}\mapsto v^{\mu}
- h^{\mu\nu}\partial_{\nu}f\,}$. However, this gauge-dependence is projected
out in the actual field equation \Ep{intern} --- since the gauge-dependent
part is spacelike: ${t_{\alpha}\,\partial^{\alpha}{\!f} = 0}$ ---
making the whole theory invariant under the complete Newton-Cartan
gauge group ${{\cal A}ut(B(\M))}$. Better still, if one wishes --- say, for
aesthetic reasons --- to eliminate the gauge-dependent part ${{\cal L}_{\Phi}}$
of the total Lagrangian density to make the Lagrangian prescription
{\it manifestly} generally-covariant, one is completely free to do so by
simply choosing the arbitrary constant ${\kappa=0\,}$.
Therefore, our variational
reformulation of the classical Newton-Cartan-Schr\"odinger theory is
no less generally-covariant than Einstein's theory of gravity (modulo, of
course, the philosophical caveat made in the footnote ${\scriptstyle 2}$). 
In fact, in close analogy with variational formulations of Einstein's theory,
all we have assumed here {\it a priori} is an ${{\cal A}ut(B(\M))}$-invariant
action functional --- dependent only on local degrees of freedom --- defined on
some measurable region of a sufficiently smooth, real, differentiable
Hausdorff 4-manifold ${\M\,}$. The rest follows squarely from extremizations
of this action functional.

\subsection{Constraint-free Hamiltonian formulation in 3+1 dimensions}

Unlike the spacetime covariant Lagrangian formulation of any field theory, the
conventional Hamiltonian formulation of such a theory requires a blatantly
non-covariant Aristotelian 3+1 decomposition of spacetime into
3-spaces ${\Sigma_t}$ at instants of time ${t\in\real\,}$.
For such a foliation of spacetime to be possible, it is necessary to assume
that the manifold $\M$ is of the globally hyperbolic
form: ${\M=\real\times\Sigma\,}$, where ${\Sigma}$ is taken to be an embedded
(cf. footnote ${\scriptstyle 3}$) achronal{\parindent 0.39cm
\baselineskip 0.53cm\footnote{$^{\scriptscriptstyle 7}$}{\ninepoint{\hang
A set is achronal if none of its elements can be joined by a timelike
curve. For a definition of `the domain of dependence' and a general discussion
on the construction of regular Cauchy surfaces,
see reference\cite{Wald}.\par}}}
closed submanifold of $\M$ such that its domain of dependence
${{\cal D}(\Sigma)=\M\,}$. In the general relativistic case such an {\it a
priori} topological constraint on spacetime is obviously too severe, and
therefore canonical approaches to quantize Einstein's gravity are sometimes
criticized for being overly restrictive if not completely misguided. However,
in Newtonian physics time plays a very privileged role. Therefore, for
a Galilean spacetime the breakup ${\M=\real\times\Sigma}$ is not only natural
but, in fact, a part of the intrinsic structure determined by the non-dynamical
metrics ${h^{\mu\nu}}$ and ${t_{\mu\nu}\,}$. This fact, of course, does not
compensate for the loss of covariance of any field theory based on such a
breakup of spacetime into 3-spaces at times. However, as we shall see in the
next subsection, the apparent loss of covariance in the Hamiltonian formulation
of our theory is only an aesthetic loss, and can be rectified using the
relatively less-popular symplectic approach yielding a manifestly
covariant description of canonical
formalism\cite{Lee}\cite{Barnich}\cite{Woodhouse}\cite{Witten}\cite{Ashtekar}.

To see in detail the intuitively obvious fact that Newton-Cartan structure is
naturally globally hyperbolic, first recall that the structure
${(\M,\,t_{\mu\nu})}$ is
time-orientable; i.e., there exists a globally defined smooth vector field
${t_{\mu}}$ on $\M$ inducing the temporal metric ${t_{\mu\nu}=t_{\mu}t_{\nu}}$
and determining a time-orientation. Further, since $\M$ is contractible by
definition, the compatibility condition ${\nabla_{\!\mu}t_{\nu}=0}$ together
with the Poincar\'e lemma allows one to define the absolute time globally by a
map ${t:\M\rightarrow\real\,}$, foliating the spacetime {\it uniquely} into
one-parameter family of smooth Cauchy surfaces ${\Sigma_t\,}$ --- the domains
of simultaneity. Here, one can change
the scalar function $t$ into ${t'=t'(t)\,}$, but the regular foliation
${\M=\cup\,\{\Sigma_t\}}$ with ${\Sigma_t=\{x\in\M\,|\,t(x)=c\,,\,
c\in\real\}}$ remains a part of the intrinsic structure of spacetime. Moreover,
these Cauchy surfaces have the property that all curves with images confined
to them are spacelike; consequently, they may also be defined directly
by the 1-form ${\tau=t_{\alpha}dx^{\alpha}}$ as the 3-dimensional
subspaces\cite{Kunzle, 76}
$$
\Sigma_x \;\;:=\;\; \Leftcases{21pt}
\eta\in {\rm T}_x\M\;|\;\eta\cont\tau=0\Rightcases{21pt}\EQN
$$
of tangent spaces ${{\rm T}_x\M}$ at points $x$ on ${\M\,}$. Since $\tau$
is closed, this differential system of hypersurfaces ${\Sigma_x}$ is
completely integrable, and defines a foliation of regular Cauchy surfaces on
$\M$ which are given by ${\Sigma_t}$ with a locally defined scalar function
${t(x)\in C^{\infty}(\M)}$ satisfying ${dt=\tau\,}$. What is more, these smooth
3-surfaces ${\Sigma_t}$ are orientable since the 4-manifold ${\M}$ is
orientable; given the 4-volume measure
${{\scriptstyle{\cal E}}_{\mu\nu\rho\sigma}}$ on
${\M}$ (cf. equation \Ep{vol}), the corresponding contravariant tensor
${{\scriptstyle{\cal E}}^{\mu\nu\rho\sigma}}$ can be used\cite{Carter} to
obtain a 3-volume measure ${{\scriptstyle{\cal E}}^{[\mu\nu\rho]} =
{\scriptstyle{\cal E}}^{\mu\nu\rho}:={\scriptstyle{\cal E}}^{\mu\nu\rho\sigma}
t_{\sigma}}$ on the Cauchy surfaces ${\Sigma_t\,}$. In the special case of
Newton-Cartan spacetime these Cauchy surfaces are flat Riemannian 3-surfaces
of spacelike vectors, carrying
a non-degenerate Euclidean metric induced by projecting ${h^{\mu\nu}}$ with
respect to any unit timelike vector field --- say the Galilean observer field
${u^{\alpha}\,}$. That is,
${(\Sigma_t,\,{\!\!\!\!\!\!\buildchar{\;\;\;\;h_{\mu\nu}}{u}{}})\,}$, with
${\!\!\!\!\!\!\buildchar{\;\;\;\;h_{\mu\nu}}{u}{}}$ as the induced metric field
on ${\Sigma_t}$ corresponding to the vector field ${u^{\alpha}\,}$, is a
copy of Euclidean 3-space, and, hence, $\M$ is homeomorphic to ${\real^4\,}$.
Since ${\!\!\!\!\!\!\buildchar{\;\;\;\;h_{\mu\nu}}{u}{}}$ has a spatial
inverse, namely ${h^{\mu\nu}\,}$ (cf. equation \Ep{inverse}), it follows
that\cite{Malament} there exists a unique 3-dimensional derivative operator
${^{(3)}\nabla_{\!\mu}}$ on ${\Sigma_t}$ compatible with
${{\!\!\!\!\!\!\buildchar{\;\;\;\;h_{\mu\nu}}{u}{}}:{^{(3)}\nabla_{\!\sigma}}
{\!\!\!\!\!\!\buildchar{\;\;\;\;h_{\mu\nu}}{u}{}}=0\,}$. It can be
characterized in terms of the 4-dimensional operator ${\nabla_{\!\mu}\,}$; for
example, for a tensor field ${V^{\alpha}_{\mu\nu}\,}$,
$$
{^{(3)}\nabla_{\!\sigma}}V^{\alpha}_{\mu\nu} \;=\;
{\!\!\!\buildchar{\>\;\;\delta_{\sigma}^{\;\;\beta}}{u}{}}
{\!\!\!\buildchar{\>\;\;\delta_{\mu}^{\;\;\gamma}}{u}{}}
{\!\!\!\buildchar{\>\;\;\delta_{\nu}^{\;\;\lambda}}{u}{}}
\nabla_{\!\beta}V^{\alpha}_{\gamma\lambda}\;.\EQN
$$
Note that there is no need to
project the contravariant indices because they remain spacelike even after the
application of ${\nabla_{\!\mu}}$ because of the compatibility condition
${\nabla_{\!\mu}t_{\nu}=0\,}$, and that ${^{(3)}\nabla_{\!\sigma}}$ satisfies
$$
{^{(3)}\nabla_{\!\sigma}}{\!\!\!\buildchar{\>\;\;\delta_{\mu}^{\;\;\nu}}{u}{}}
\;\;=\;\;
{^{(3)}\nabla_{\!\sigma}}{\!\!\!\!\!\!\buildchar{\;\;\;\;h_{\mu\nu}}{u}{}}
\;\;=\;\;
{^{(3)}\nabla_{\!\sigma}}h^{\mu\nu}\;\;=\;\;0\EQN
$$
over and above all of the defining conditions for a derivative
operator. The two operators
${\nabla_{\!\sigma}}$ and ${^{(3)}\nabla_{\!\sigma}}$ induce the same parallel
transport condition for spacelike vectors on ${\Sigma_t\,}$: 
${U^{\alpha}({^{(3)}\nabla_{\!\alpha}})V^{\mu}
=U^{\alpha}{\!\!\!\buildchar{\>\;\;\delta_{\alpha}^{\;\;\gamma}}{u}{}}
{\nabla_{\!\gamma}}V^{\mu}=U^{\gamma}{\nabla_{\!\gamma}}
V^{\mu}\,}$, for all spacelike fields ${U^{\mu}}$ and ${V^{\mu}}$
on ${\Sigma_t\,}$.

The vector field
${u^{\alpha}}$ may be viewed as describing the flow of time in $\M$ and can
be used to identify each Cauchy surface ${\Sigma_t}$ with the initial surface
${\Sigma_{\scriptscriptstyle 0}\,}$. Given such an observer field
${u^{\alpha}}$ and its relative spatial projection field
${{\!\!\!\buildchar{\>\;\;\delta_{\mu}^{\;\;\nu}}{u}{}}\equiv
{\!\!\!\!\!\!\buildchar{\;\;\;\;h_{\mu\sigma}}{u}{}}h^{\sigma\nu}\,}$, one may
decompose any spacetime quantity into its projections normal and tangential to
${\Sigma_t}$. For example, using the relations 
$$
t_{\alpha}{\!\!\!\buildchar{\>\;\;\delta_{\mu}^{\;\;\alpha}}{u}{}}\;\,=\;\,0
\;\;\;\;\;\; {\rm and} \;\;\;\;\;\; u^{\alpha}
{\!\!\!\buildchar{\>\;\;\delta_{\alpha}^{\;\;\mu}}{u}{}}\;\,
=\;\,0\;,\EQN diff-maps
$$
a contravector ${V^{\alpha}}$ on $\M$ may be decomposed as
$$
V^{\alpha}\;\;=\;\;{^{\bot}}V\,u^{\alpha}\;+\;{^{\|}}V^{\alpha}\EQN cont-split
$$
with ${{^{\bot}}V:=t_{\mu}V^{\mu}}$ and ${{^{\|}}V^{\alpha}:=
{\!\!\!\buildchar{\>\;\;\delta_{\mu}^{\;\;\alpha}}{u}{}}V^{\mu}\,}$, whereas
a covector ${V_{\alpha}}$ on $\M$ can be split as
$$
V_{\alpha}\;\;=\;\;{_{\bot}}V\,t_{\alpha}\;+\;{_{\|}}V_{\alpha}\EQN co-split
$$
with ${{_{\bot}}V:=u^{\mu}V_{\mu}}$ and ${{_{\|}}V_{\alpha}:=
{\!\!\!\buildchar{\>\;\;\delta_{\alpha}^{\;\;\mu}}{u}{}}V_{\mu}\,}$. Note
that the distinction between covectors and contravectors made conspicuous
by these decompositions has much greater significance here than in the
general relativistic spacetimes. 

With the above rather elaborate delineation of the convenient fact that
Newton-Cartan spacetime is naturally globally hyperbolic, we are ready to cast
our theory in a Hamiltonian form. The next two steps for which are, firstly, 
to construct a
configuration space for the field variables on $\M$ by specifying instantaneous
configuration fields corresponding to these variables on a chosen Cauchy
surface ${\Sigma_t\,}$, and then, secondly, to evaluate conjugate momentum
densities of these instantaneous fields on the chosen surface. As we shall see,
the ensuing canonical variables constitute the proper Cauchy data to be
propagated from one Cauchy surface to another. Let us begin by
concentrating on the
pair ${(A_{\mu},\,u^{\nu})}$ of pure gravitational field variables on $\M$ and
tentatively take ${A_{\mu}}$ and ${u^{\nu}}$ evaluated on ${\Sigma_t}$ as our
instantaneous configuration variables. Using the defining equations
\Ep{co-split} and \Ep{cont-split}, these variables may be decomposed into their
components normal and tangential to ${\Sigma_t}$ as
$$
A_{\alpha}\;\;=\;\;{_{\bot}}A\,t_{\alpha}\;+\;{_{\|}}A_{\alpha}\EQN A-split
$$
with ${{_{\bot}}A:=u^{\mu}A_{\mu}}$ and ${{_{\|}}A_{\alpha}:=
{\!\!\!\buildchar{\>\;\;\delta_{\alpha}^{\;\;\mu}}{u}{}}A_{\mu}\,}$, and
$$
u^{\alpha}\;\;=\;\;{^{\bot}}u\,u^{\alpha}\;+\;{^{\|}}u^{\alpha}\EQN u-split
$$
with ${{^{\bot}}u:=t_{\mu}u^{\mu}=1}$ and ${{^{\|}}u^{\alpha}:=
{\!\!\!\buildchar{\>\;\;\delta_{\mu}^{\;\;\alpha}}{u}{}}u^{\mu}= 0\,}$.
The Lagrangian density, expression \Ep{cl}, can now be rewritten
in terms of these decompositions and normal and tangential parts of the
momentum densities conjugate to the fields ${A_{\mu}}$ and ${u^{\nu}}$
specified on ${\Sigma_t}$ can be evaluated from it. It is easier, however, to
use the already evaluated four-momentum densities \Ep{four-moment} and
\Ep{anti-dual} and decompose them according to equations \Ep{cont-split} and
\Ep{co-split}. Whichever procedure is used, the outcomes are:
$$
{^{\bot}}\Pi\;=\;t_{\mu}\Pi^{\mu}\;=\;0\;\;\;\;\;\;\;\;{\rm and}\;\;\;\;\;\;
\;\;{^{\|}}\Pi^{\alpha}\;=\;
{\!\!\!\buildchar{\>\;\;\delta_{\mu}^{\;\;\alpha}}{u}{}}\,\Pi^{\mu}\;=\;
\Pi^{\alpha}\;,\EQN A-mome
$$
whereas
$$
{_{\bot}}{\buildchar{\Pi}{u}{}}\;=\;u^{\nu}
{\!\!{\buildchar{\;\;\Pi_{\nu}}{u}{}}}\;=\;0\;\;\;\;\;\;\;\;{\rm and}
\;\;\;\;\;\;\;\;
{_{\|}}\!{\!\!{\buildchar{\;\;\Pi_{\alpha}}{u}{}}}\;=\;
{\!\!\!\buildchar{\>\;\;\delta_{\alpha}^{\;\;\mu}}{u}{}}
{\!\!{\buildchar{\;\;\Pi_{\mu}}{u}{}}}\;=\;-\,
{\!\!\!\!\!\!\buildchar{\;\;\;\;h_{\alpha\sigma}}{u}{}}\Pi^{\sigma}\;.
\EQN u-mome
$$
The vanishing of the canonical momentum densities ${{^{\bot}}\Pi}$ and
${{_{\bot}}{\buildchar{\Pi}{u}{}}}$ implies that the conjugate momenta are
spacelike, and, more importantly for our purposes, suggests that we should not
view their conjugate variables ${{_{\bot}}A}$ and ${{^{\bot}}u}$ as dynamical
variables because they do not constitute suitable Cauchy data
(their presence as dynamical variables would be an obstacle to the Legendre
transformation required to specify a Hamiltonian functional on ${\Sigma_t}$).
Since ${{{^{\bot}}u\equiv 1}}$ and ${{^{\|}}u^{\alpha}\equiv 0\,}$, we shall
see that taking the observer field
${u^{\nu}}$ as one of our configuration variables does not cause any serious
problems. However, we must take only the tangential component
${{_{\|}}A_{\mu}}$ of ${A_{\mu}}$ as our gravitational configuration
variable along with the observer field ${u^{\nu}}$ if we are to maintain
substantive uniqueness of the propagation from Cauchy data. In addition to
these gravitational field variables, we also have the matter
configuration variables ${\Psi}$ and ${\overline \Psi}$ specified on
${\Sigma_t\,}$, and, together with them, we also must add their respective
conjugate momentum densities
$$
{\overline{\rm P}}\;:=\;{{\delta\,{\cal L}_{Kuch}}
\over{\delta(u^{\sigma}\partial_{\sigma}
\Psi)}}\;=\;-\,2\,i\,\hbar\,\wp\,\pi\,G\,{\overline \Psi}\;\;\;\;\;\;
{\rm and}\;\;\;\;\;\;
{\rm P}\;:=\;{{\delta\,{\cal L}_{Kuch}}\over{\delta(u^{\sigma}\partial_{\sigma}
{\overline \Psi}\,)}}\;=\;+\,2\,i\,\hbar\,\wp\,\pi\,G\,\Psi\EQN mat-can-mom
$$
to our list of canonical phase-space variables, where
${u^{\sigma}\nabla_{\!\sigma}}$ is a propagation covariant derivative (i.e.,
a time derivative) along the unit timelike vector field ${u^{\sigma}\,}$.

Next, in order to remain in close contact with the parameterized
formulation\cite{Dirac}\cite{Torre}
used in the previous subsection, we view the foliation $e$ of ${\M\,}$,
$$
e\,:\,{\real}\times\Sigma\;\longrightarrow\;\M\;,\EQN
$$
as the map which allows us to pass to the Hamiltonian formulation of the
theory. Then, for each ${t\in\real\,}$, $e$ becomes an embedding
$$
{{}^{(t)}\!e}\,:\,\Sigma\;\longrightarrow\;\M\EQN
$$
(cf. footnote ${\scriptstyle 3}$).
In order to make the embedded hypersurfaces ${\Sigma_t}$
(i.e., the images of ${\Sigma}$ under the above embedding)
spacelike, we require the differential ${{{}^{(t)}\!e}_{i}^{\;\;\mu}}$ of the
map ${{{}^{(t)}\!e}}$ to satisfy
$$
{{}^{(t)}\!e}_{i}^{\;\;\mu}\,t_{\mu}\;=\;0\;\;\;\;\;\;{\rm and}\;\;\;\;\;\;
{{}^{(t)}\!e}^{i}_{\;\;\nu}\,u^{\nu}\;=\;0\;,\EQN
$$
where ${{{}^{(t)}\!e}^{i}_{\;\;\nu}}$ is the inverse of the differential map
${{{}^{(t)}\!e}_{i}^{\;\;\nu}\,}$, and in these expressions (and only in these
expressions) the tensors evaluated on the closed 3-surfaces ${\Sigma_t}$
are represented via Latin indices for clarity. Comparing these conditions with
the relations \Ep{diff-maps}, we immediately see that
$$
{{}^{(t)}\!e}_{i}^{\;\;\mu}\;\equiv\;
{\!\!\!\buildchar{\>\;\;\delta_{i}^{\;\;\mu}}{u}{}}
\;\;\;\;\;\;{\rm and}\;\;\;\;\;\;
{{}^{(t)}\!e}^{i}_{\;\;\nu}\;\equiv\;
{\!\!\!\buildchar{\>\;\;\delta_{\nu}^{\;\;i}}{u}{}}
\;.\EQN
$$
In analogy with what we have done in the previous subsection, we can now
pull-back all the fields from ${\buildchar{\M\,}{\circ}{}}$ to the 3-manifold
${\Sigma\,}$, and enlarge the configuration space of these fields
by spacelike embedding variables
$$
{{}^{(o)}\!y}\circ{{}^{(t)}\!e}\;=:\;\vartheta\,:\,\Sigma\;\longrightarrow\;
{\buildchar{\M\,}{\circ}{}}\;,\EQN
$$
with ${{}^{(o)}\!y\in \{{}^{(s)}\!y\}}$ being any fixed diffeomorphism defined
by equation \Ep{map} of the previous subsection.
In terms of this map the traditional `deformation vector' 
${{\dot\vartheta}^{\sigma}:={u^{\sigma}\nabla_{\!\sigma}\vartheta^{\sigma}}
={{\partial\vartheta^{\sigma}}\over{\partial t\;\,}}}$ dictating
transition from one leaf of foliation to another neighboring one,
for example, may be split into its `lapse' and `shift' coefficients as
$$
{\dot\vartheta}^{\sigma}\;\,=\;\,
{^{\bot}}{\dot\vartheta}\,u^{\sigma}\;+\;
{^{\|}}{\dot\vartheta}^{\sigma}\;,\EQN
$$
where ${{^{\bot}}{\dot\vartheta}:={\dot\vartheta}^{\sigma}t_{\sigma}}$
and ${{^{\|}}{\dot\vartheta}^{\sigma}:=
{\!\!\!\buildchar{\>\;\;\delta_{\mu}^{\;\;\sigma}}{u}{}}
{\dot\vartheta}^{\mu}\,}$.

These considerations finally allow reconstruction of the action functional
${\cal I}$ of the previous subsection in the Hamiltonian form:
$$
{\cal I}_{\rm{I\!H}}\;=\int_{\rm{I\!R}}dt\int_{\Sigma_t}d^3x\;
[{^{\|}}\Pi^{\sigma}\,{_{\|}}{\dot{A}}_{\sigma}\;+\;
{\!\!{\buildchar{\;\;\Pi_{\sigma}}{u}{}}}\,{\dot u}^{\sigma}\;+\;
{\overline{\rm P}}\,{\dot\Psi}\;+\;{\rm P}
\,{\dot{\overline\Psi}}\;+\;{\tt \Pi}_{\sigma}\,{\dot\vartheta}^{\sigma}\;-\;
\aleph^{\sigma}H_{\sigma}]\;,\EQN hami-action
$$
where `${\,\cdot\,}$' indicates the time derivative
`${\,u^{\sigma}\nabla_{\!\sigma}}$' with respect to the parameter ${t}$ of the
one-parameter family of embeddings, ${{\tt \Pi}_{\sigma}}$ are the kinematical
momentum densities
conjugate to the supplementary embedding variables ${\vartheta^{\sigma}\,}$,
${{\dot\vartheta}^{\sigma}}$ being geometrically a vector field on
${\buildchar{\M\,}{\circ}{}}$ makes ${{\tt \Pi}_{\sigma}}$ a 1-form density of
weight one on ${{\buildchar{\M\,}{\circ}{}}\,}$, ${\aleph^{\sigma}}$ is a
Lagrange multiplier field, ${\aleph^{\sigma}H_{\sigma}}$ defined by
$$
\aleph^{\sigma}[{\tt \Pi}_{\sigma}\,+\,{\!\!\buildchar{\;H_{\sigma}}{\circ}{}}]
\;=:\;\aleph^{\sigma}H_{\sigma}\;\equiv\;H
\;:=\;{^{\|}}\Pi^{\sigma}\,{_{\|}}{\dot{A}}_{\sigma}\;+\;
{\!\!{\buildchar{\;\;\Pi_{\sigma}}{u}{}}}\,{\dot u}^{\sigma}\;+\;
{\overline{\rm P}}\,{\dot\Psi}\;+\;{\rm P}
\,{\dot{\overline\Psi}}\;+\;{\tt \Pi}_{\sigma}\,{\dot\vartheta}^{\sigma}
\;-\;{\cal L}\;\,\EQN hamiltonian
$$
(with ${\!\!\buildchar{\;H_{\alpha}}{\circ}{}}$ being a functional of the
original `un-parameterized' canonical data as well as ${\vartheta^{\sigma}}$
representing the combined gravitational and material
`energy-momentum' flux through the surface element ${\wp\,u^{\alpha}}$ of the
embedding) is the Hamiltonian density giving the Hamiltonian functional,
$$
\boh\;\;=\;\;\int_{\Sigma_t}\;\aleph^{\sigma}H_{\sigma}
\;\,d^3x\;,\EQN ham-functional
$$
which governs the dynamical evolution of the system, the Lagrangian density
${\cal L}$ is given by equation \Ep{cl} with the fields evaluated on
${\Sigma_t\,}$, and ${\wp\,d^3x}$ is the volume element in ${\Sigma_t\,}$.
As always, the generalized velocities appearing both
explicitly and implicitly in these expressions are viewed as functions of
the phase-space variables only. The tensor field
${{\chi}^{\mu\nu}\equiv{^{\|}}{\chi}^{\mu\nu}}$ appearing in ${{\cal L}\,}$,
however, is viewed {\it not} as a multiplier field, but simply as a function of
the canonical variables ${{^{\|}}\Pi^{\mu}}$ (cf. equation \Ep{four-moment}).
The variables ${\Upsilon_{\mu}\,}$, ${\Upsilon^{\mu}_{\nu\sigma}\,}$,
${\Upsilon^{\mu\nu}\,}$, ${{\widetilde{\Upsilon}}^{\mu\nu}\,}$,
${\aleph^{\sigma}\,}$, ${{_{\bot}}A\,}$, ${\zeta\,}$,
${\chi^{\alpha\mu\nu}\,}$, and ${\chi\,}$, on the other hand,
{\it are} viewed as non-dynamical multiplier
variables, and, hence, the Hamiltonian equations of motion,
$$
\;{{\delta\,H}\over{\delta\,{\tt \Pi}_{\alpha}}}\;=\;+\,u^{\sigma}
\nabla_{\!\sigma}\,\vartheta^{\alpha}\;,\;\;\;\;\;\;\;\;\;\;\;\;
{{\delta\,H}\over{\delta\,\vartheta^{\alpha}}}\;=\;-\,u^{\sigma}
\nabla_{\!\sigma}\,{\tt \Pi}_{\alpha}\;,\EQN 0-hami
$$
$$
{{\delta\,H}\over{\delta\,
{_{\|}}A_{\alpha}}}\;=\;-\,u^{\sigma}
\nabla_{\!\sigma}\,{^{\|}}\Pi^{\alpha}\;,\;\;\;\;\;\;\;\;\;\;
{{\delta\,H}\over{\delta\,u^{\alpha}}}\;=\;-\,u^{\sigma}
\nabla_{\!\sigma}\,{\!\!{\buildchar{\;\;\Pi_{\alpha}}{u}{}}}\;,\EQN 1st-hami
$$
$$
{{\delta\,H}\over{\delta\,
{^{\|}}\Pi^{\alpha}}}\;=\;+\,u^{\sigma}
\nabla_{\!\sigma}\,{_{\|}}A_{\alpha}\;,\;\;\;\;\;\;\;\;\;
{{\delta\,H}\over{\delta
{\!\!{\buildchar{\;\;\Pi_{\alpha}}{u}{}}}}}\;=\;+\,u^{\sigma}
\nabla_{\!\sigma}\,u^{\alpha}\;,\;\EQN 2nd-hami
$$
$$
{{\delta\,H}\over{\delta\,{\overline\Psi}}}\;=\;-\,u^{\sigma}
\nabla_{\!\sigma}\,{\rm P}\;,\;\;\;\;\;\;\;\;\;\;\;\;\;\;
{{\delta\,H}\over{\delta\,{\rm P}}}\;=\;+\,u^{\sigma}
\nabla_{\!\sigma}\,{\overline\Psi}\;,\EQN 3rd-hami
$$
and complex conjugates of the latter two, must be supplemented by the
equations
$$
{{\delta\,H\,}\over{\delta\,\Upsilon_{\mu}}}\;=\;0,\;\;\;\;\;\;
{{\delta\,H\;}\over{\delta\,\Upsilon^{\mu}_{\nu\sigma}}}\;=\;0,\;\;\;\;\;\;
{{\delta\,H\;}\over{\delta\,\Upsilon^{\mu\nu}}}\;=\;0,\;\;\;\;\;\;
{{\delta\,H\;}\over{\delta\,{{\widetilde{\Upsilon}}^{\mu\nu}}}}\;=
\;0,\;\;\;\;\;\;\;\;\;\EQN backer-const
$$
$$
{{\delta\,H}\over{\delta\,\aleph^{\sigma}}}\;=\;0,\;\;\;\;\;
{{\delta\,H}\over{\delta{_{\bot}}A}}\;=\;0,\;\;\;\;\;
{{\delta\,H}\over{\delta\,\zeta}}\;=\;0,\;\;\;\;\;
{{\delta\,H\;\;}\over{\delta\,\chi^{\alpha\mu\nu}}}\;=\;0,\;\;\;\;\;
{\rm and}\;\;\;\;\;{{\delta\,H}\over{\delta\,\chi}}\;=\;0,\;\,\EQN gauss-const
$$
giving rise to constraints on the phase-space.
Note that, since ${\delta\,{^{\bot}}u\equiv 0\,}$, no additional constraint
corresponding to ${{^{\bot}}u}$ need be appended to the set of the Hamiltonian
equations of motion. On the other hand, we do have to add the linear
constraints
$$
\EQNalign{
\Pi^{\mu}\;+\;h^{\mu\nu}{\!\!{\buildchar{\;\;\Pi_{\nu}}{u}{}}}\;&=\;0\cr
{\rm or}\;\;\;\;\;\;\;{\!\!{\buildchar{\;\;\Pi_{\mu}}{u}{}}}\;+\;
{\!\!\!\!\!\!\buildchar{\;\;\;\;h_{\mu\nu}}{u}{}}\,\Pi^{\nu}\;&=\;0
\EQN gazer-1\cr}
$$
$$
\EQNalign{{\rm and}\;\;\;\;\;\,
{\overline{\rm P}}\;+\;2\,i\,\hbar\,\wp\,\pi\,G\,{\overline \Psi}\;&=\;0\cr
{\rm or}\;\;\;\;\;\;\;{\rm P}\;-\;2\,i\,\hbar\,\wp\,\pi\,G\,\Psi\;&=\;0
\EQN gazer-2\cr}
$$
to the above list (cf. equations \Ep{net-four-m} and \Ep{mat-can-mom}).
Here we do not bother to classify these constraints {\it \`a la}
Dirac\cite{Dirac} because our eventual goal is to get rid of {\it all} of
them in order to obtain a {\it constraint-free} phase-space.

The first four of these eleven constraint equations, \Ep{backer-const},
immediately yield the
conditions \Ep{pre-cond} specifying the Galilean structure. Among the
remaining equations, the fifth,
${{{\delta H\,}\over{\delta\aleph^{\sigma}}}=0\,}$,
imparts the kinematical constraint
$$
H_{\sigma}\;\equiv\;
{\tt \Pi}_{\sigma}\;+\;{\!\!\buildchar{\;H_{\sigma}}{\circ}{}}
\;=\;0\EQN diff-const
$$
due to the parameterization process,
the sixth, ${{{\delta\,H}\over{\delta{_{\bot}}A}}=0\,}$, gives the
equation \Ep{mod-intern}, the seventh,
${{{\delta\,H}\over{\delta\,\zeta}}=0\,}$,
reproduces the normalization condition ${u^{\nu}t_{\nu}=1}$ defining the
foliation, the eighth,
${{{\delta\,H\;\;\;\,}\over{\delta\,\chi^{\alpha\mu\nu}}}=0\,}$,
leads to the constraint \Ep{add-constraints} prohibiting rotational
holonomy, and the ninth, ${{{\delta\,H}\over{\delta\,\chi}}=0\,}$,
asserts the constraint \Ep{chi-variation}.

Among the equations of motion,
the first one of the equations \Ep{0-hami}
tells us that the multiplier field ${\aleph^{\sigma}}$ is nothing
but the generalized velocity field ${{\dot\vartheta}^{\sigma}\,}$,
$$
\aleph^{\sigma}\;=\;\,{\dot\vartheta}^{\sigma}\;,\EQN time-choice
$$
whereas the second one, after using this identification,
leads to the condition ${{\dot H}_{\sigma} = 0\,}$, which guarantees that the
constraint \Ep{diff-const} is preserved in time.
Each of the equations of motion \Ep{1st-hami} leads to the field equation
\Ep{mod-intern} again (where the
second of these derivations, reminiscent of the derivation leading to equation
\Ep{ae-u} of the previous subsection, requires a little more work compared to
the first one), and the combined effect of the equations of motion
\Ep{2nd-hami} is nothing but the field equation \Ep{Q-euler}. Finally, both of
the remaining two equations of motion \Ep{3rd-hami}, as well as their complex
conjugates, give the Schr\"odinger-Kucha\v r equation \Ep{4.23}, and its
complex conjugate, respectively.
As we elaborated in the previous subsection, given
the relation \Ep{chi-variation} implied by the last of the five equations
\Ep{gauss-const}, equations \Ep{Q-euler} and \Ep{mod-intern} implied by
equations \Ep{2nd-hami} and \Ep{1st-hami} are equivalent to the
Newton-Cartan field equations \Ep{int-cont} and \Ep{fil-eq},
respectively. Furthermore, the constraint \Ep{mod-intern} on the
Cauchy data implied by the second of
the five equations \Ep{gauss-const} is no constraint at all, because it is
automatically satisfied by both of the Hamiltonian equations of motion
\Ep{1st-hami}; the constraint ${u^{\nu}t_{\nu}=1}$ implied by the
third of these equations, on the other hand, is trivially satisfied
if we choose ${u^{\nu}}$ to be a unit timelike vector-field, as we have done.

The presence of the remaining constraints, of course, indicate that our
phase-space is still too large and we have not isolated the `true' dynamical
degrees of freedom in our choice of configuration space. Note, however, that,
among the remaining constraints, the set \Ep{pre-cond} implied by equations
\Ep{backer-const} consists of purely
kinematical constraints in that its elements do not
impose any relations among the genuinely dynamical degrees of freedom.
Each of the constraint functions ${h^{\mu\nu}t_{\nu}\,}$,
${\nabla_{\!\mu}h^{\nu\sigma}\,}$, ${\nabla_{\!\mu}t_{\nu}\,}$, and
${\partial_{\lbrack\mu}\,t_{\nu\rbrack}}$ has vanishing variations with
respect to the dynamical variables ${A_{\mu}\,}$, ${u^{\nu}\,}$, ${\Psi\,}$,
and ${\overline\Psi}$ (recall that the Newton-Cartan connection appearing
in the second and third of these functions is invariant under the changes of
gauge variables ${A_{\mu}}$ and ${u^{\nu}}$). This means that these are
simply constant functions on the phase-space. In fact, we could have taken them
as part of the background structure on the `un-parameterized' manifold
${\buildchar{\M\,}{\circ}{}}$ before pulling them back with the fields by the
map ${\vartheta\,:\,\Sigma\rightarrow{\buildchar{\M\,}{\circ}{}}}$.
Consequently, if we ignore for the moment the kinematical constraints
\Ep{diff-const} and the linear constraints \Ep{gazer-1} and \Ep{gazer-2}, 
what we have in hand is a Hamiltonian
formulation of the classical Newton-Cartan-Schr\"odinger theory with the
constraints \Ep{new-not} and \Ep{chi-variation} on the Cauchy data due to the
last two of the constraint equations \Ep{gauss-const}. If we use the field
equation
\Ep{mod-intern} implied by the Hamiltonian equations of motion \Ep{1st-hami}
to substitute for ${\nabla_{\!\sigma}\Theta^{\sigma}}$ in the
constraint \Ep{chi-variation}, then it takes the form
$$
{\cal C}(\kappa)\;:=\;
{1\over{\wp}}\,\nabla_{\!\sigma}\Pi^{\sigma}
\;+\;\Lambda\subn\;-\;(1-{\kappa})\,
4\pi G\,\rho\;=\;{\kappa}\,\Lambda\;,\EQN only-left
$$
where recall that ${\kappa}$ is an arbitrary free parameter.

Now, we have seen that
Newton-Cartan connection is invariant under internal gauge transformations
${A_{\mu}\mapsto A_{\mu} + \partial_{\mu}f}$ and boost transformations
\Ep{boost} over and above the diffeomorphisms ${\phi\in{\rm Diff}(\M)}$
(cf. equation \Ep{vertical}). Physically this implies that the pairs
${\{A_{\mu},\,u^{\nu}\}}$ and ${\{(A_{\mu}+\partial_{\mu}f+{\rm w}_{\mu}-
(u^{\sigma}{\rm w}_{\sigma}+{\scriptstyle{1\over 2}}\,h^{\alpha\sigma}
{\rm w}_{\alpha}{\rm w}_{\sigma})\,t_{\mu}),\,(u^{\nu}+h^{\nu\sigma}
{\rm w}_{\sigma})\}}$ of gravitational variables represent the same physical
configuration of the gravitational field. On the other hand, as we noted in
section 3, under the
vertical transformations \Ep{vertical} the Schr\"odinger-Kucha\v r field
transforms as ${\Psi\mapsto\exp(i{\scriptstyle{m\over\hbar}}f)\,\Psi\,}$, and,
hence, the pair ${\{\Psi,\,{\overline\Psi}\,\}}$ representing matter fields is
physically equivalent to the pair ${\{(\exp(i{\scriptstyle{m\over\hbar}}f)
\,\Psi),\,(\exp(-i{\scriptstyle{m\over\hbar}}f)\,{\overline\Psi}\,)\}\,}$.
Therefore, we must replace the configuration space ${Z}$
of our composite physical system by the space ${\widetilde Z}$
of {\it equivalence classes}
of representatives of the physical fields, where all representatives of the
fields are defined to be equivalent if they differ only by
the interconnecting vertical gauge transformations \Ep{vertical}. Now recall
that in the canonical formalism, quite generally, the space of possible momenta
of a physical system at a given configuration point $z$ is the cotangent space
${T^*_z{Z}}$ of the configuration space at that point, and, consequently,
the corresponding phase-space is a cotangent-bundle ${T^*{Z}}$ over the
configuration space endowed with a natural presymplectic
structure (cf. appendix A);
where, given a configuration manifold ${Z}$ and a point $z$ on that
manifold, a cotangent vector ${\cal P}$ at the point is defined to be a real
linear map, ${{\cal P}:\,T_z{Z}\rightarrow\real}$, from the tangent space
at that point to the
set of real numbers. For example, in the particular case under consideration,
the momentum density ${\Pi^{\mu}}$ is a cotangent vector at the point
${A_{\mu}}$ of the configuration space which maps the tangent vector
${\delta{A_{\mu}}}$ at ${A_{\mu}}$ into ${\real}$ via ${\delta{A_{\mu}}
\rightarrow\int_{\Sigma_t}{\Pi^{\mu}}\,\delta{A_{\mu}}\,}$. If we now take our
configuration space to be the space ${\widetilde Z}$
of {\it equivalence classes} of representatives of the physical
fields as defined above, then the cotangent space at a configuration point
would be the space of real linear functions of variations of the physical
fields which are independent of the vertical gauge transformations
\Ep{vertical}. Consequently, the momentum densities would be represented by
those vector fields which leave the integral
$$
\int_{\Sigma_t}\!d^3x\,({\Pi^{\mu}}\,\delta{A_{\mu}}\;+\;
{\!\!{\buildchar{\;\;\Pi_{\nu}}{u}{}}}\,\delta{u^{\nu}}\;+\;
{\overline{\rm P}}\,\delta\Psi\;+\;{\rm P}\,\delta{\overline\Psi}
\;+\;{\tt \Pi}_{\sigma}\,\delta\vartheta^{\sigma})\EQN bund-in
$$
invariant under the transformations \Ep{vertical}, where we freely use
the convenient fact that
${{^{\|}}\Pi^{\sigma}\,\delta {{_{\|}}A_{\sigma}} \equiv
\Pi^{\sigma}\,\delta A_{\sigma}\,}$. However, it is easy to see by direct
substitutions for the expressions in the integrand (and using equation
\Ep{rel-trans}) that this integral is left
invariant under \Ep{vertical}, or, effectively, under
$$
\EQNalign{A_{\mu}\;&\longrightarrow\;A_{\mu}\;+\;\nabla_{\!\mu}f\cr
      \Psi\;\;&\longrightarrow\;\exp(i{m\over\hbar}f)\,\Psi\,,\EQN effect \cr}
$$
if and only if the momentum densities satisfy
the condition
$$
{1\over{\wp}}\,\nabla_{\!\sigma}\Pi^{\sigma}\;-\;
4\pi G\,\rho\;=\;0\;.\EQN lightof
$$
Next, we wish to further restrict our configuration space and admit only those
configuration variables which satisfy the constraint \Ep{new-not} implied by
the fourth of the constraint equations \Ep{gauss-const}. Now, as
discussed in the previous subsection (cf. expressions \Ep{4-term} and
\Ep{2nd4-term}), this restriction amounts to an addition of a couple of
4-divergence terms in the integral \Ep{bund-in}, which, however, remains
unaffected by them because of the vanishing of the boundary of a boundary
theorem. Nevertheless, the ensuing condition \Ep{lightof} {\it is} modified
under the transformations \Ep{effect}, and becomes
$$
{1\over{\wp}}\,\nabla_{\!\sigma}\Pi^{\sigma}\;+\;\Lambda\subn\;-\;
4\pi G\,\rho\;=\;0\;,\EQN lightof-modified
$$
with the ${\Lambda\subn}$ term arising from the addition of the 4-divergence
\Ep{4-term} in the integral \Ep{bund-in}.

But this is just the constraint
\Ep{only-left} in the limit ${\kappa \rightarrow 0\,}$:
$$
\lim_{\kappa\to0}\,{\cal C}(\kappa)\;=\;
{1\over{\wp}}\,\nabla_{\!\sigma}\Pi^{\sigma}\;+\;\Lambda\subn\;-\;
4\pi G\,\rho\;=\;0\EQN k-to-0-limit
$$
(recall that, as far as the field equations and their derivations are
concerned, we are free to choose any desired value for ${\kappa}$ without
loss of generality). Further, in this harmless limit, not only the field
equation \Ep{mod-intern} and the constraint equation \Ep{only-left} become
identical, but also the Lagrangian density \Ep{cl} becomes invariant under
the full automorphism group \Ep{gauge1}. In summary, by eliminating the
spurious gauge-arbitrariness in the configuration space by working rather
with equivalence classes of field variables, we have been able to liberate
our phase-space from the constraints \Ep{new-not} and \Ep{only-left}. In
fact we have arrived at the following set of results: in the light of equation
\Ep{lightof-modified} being a direct consequence of the invariance of the
integral \Ep{bund-in} under the vertical gauge transformations \Ep{vertical},
the limit ${{\kappa}\rightarrow 0\,}$, (1), eliminates the
unwanted constraint \Ep{only-left} on the Cauchy data, (2), dictates that
the constituents of the configuration space of the system are automatically
entire classes of representatives of the physical fields which, in addition,
satisfy the constraint \Ep{new-not}, (3), makes the Lagrangian density
invariant under the full group ${{\cal A}ut(B(\M))\,}$, and, by virtue of (2),
(3) and ${{^{\|}}\Pi^{\sigma}\,{{_{\|}}{\dot{A}}_{\sigma}}
\equiv\Pi^{\sigma}\,{\dot{A}}_{\sigma}\,}$, (4), renders the Hamiltonian
functional \Ep{ham-functional} manifestly invariant under the vertical gauge
transformations \Ep{vertical}. In the remaining of this paper the limit
${\kappa\rightarrow 0}$ will be understood to have been taken.

Let us now turn our attention to the remaining linear constraints
\Ep{gazer-1} and \Ep{gazer-2}. To eliminate these constraints, we further
reduce our phase-space by defining a new, more appropriate set of canonical
variables ${\{v^{\alpha},\,\pi_{\alpha}\,;
\,\psi,\,p\,;\,{\vartheta}^{\sigma},\,{\tt \Pi}_{\sigma}\}\,}$, where
$$
\EQNalign{
v^{\alpha}\,&:=\;u^{\alpha}\;-\;h^{\alpha\sigma}A_{\sigma}\;=\;
u^{\alpha}\;-\;h^{\alpha\sigma}{{_{\|}}A_{\sigma}}\;,\cr
\pi_{\alpha}\,&:=\;{_{\|}}\!{\!\!{\buildchar{\;\;\Pi_{\alpha}}{u}{}}}\;=\;
{\!\!\!\buildchar{\>\;\;\delta_{\alpha}^{\;\;\mu}}{u}{}}
{\!\!{\buildchar{\;\;\Pi_{\mu}}{u}{}}}\;=\;
{\!\!{\buildchar{\;\;\Pi_{\mu}}{u}{}}}\;=\;-\,
{\!\!\!\!\!\!\buildchar{\;\;\;\;h_{\alpha\sigma}}{u}{}}\Pi^{\sigma}\;;\cr
\psi\,\;&:=\;{1\over{\hbar\sqrt{8\pi G\,\wp}}}\,
(\,2\pi G\,\wp\,\hbar\,{\overline\Psi}\;+\;i\,{\overline{\rm P}}\,),\cr
{\rm and}\;\;\;\;\;\;\;p\,\;&:=\;{1\over{\sqrt{8\pi G\,\wp}}}\,
(\,{\rm P}\;+\;i\,2\pi G\,\wp\,\hbar\,\Psi\,),\EQN whole-new-c\cr}
$$
so that equations \Ep{gazer-2} yield
$$
\psi\;=\;{\sqrt{2\pi G\,\wp}}\;
\,{\overline{\Psi}}\;\;\;\;\;\;\;\;{\rm and}
\;\;\;\;\;\;\;\;p\;=\;i\,\hbar\;{\overline\psi}\;\EQN
$$ 
(cf. equations \Ep{A-split}, \Ep{u-mome}, and \Ep{new-c}). In terms of these
new canonical variables the Hamiltonian density \Ep{hamiltonian} translates to
be
$$
\aleph^{\sigma}H_{\sigma}\;\equiv\;H
\;:=\;\pi_{\sigma}\,{\dot v}^{\sigma}\;+\;2\,p\,{\dot\psi}
\;+\;{\tt \Pi}_{\sigma}\,{\dot\vartheta}^{\sigma}
\;-\;{\cal L}\!(v^{\alpha},\,\pi_{\alpha}\,;\,\psi,\,p\,;\,
{\vartheta}^{\sigma},\,{\tt \Pi}_{\sigma})\;,\EQN new-hamiltonian
$$
which can be rewritten in the familiar form as
$$
\aleph^{\sigma}H_{\sigma}\;\equiv\;H
\;:=\;\pi_{\sigma}\,{\dot v}^{\sigma}\;+\;p\,{\dot\psi}
\;+\;{\tt \Pi}_{\sigma}\,{\dot\vartheta}^{\sigma}
\;-\;{\cal L}^{\rm c}\!(v^{\alpha},\,\pi_{\alpha}\,;\,\psi,\,p\,;\,
{\vartheta}^{\sigma},\,{\tt \Pi}_{\sigma})\;,\EQN familiar-Hamiltonian
$$
with
$$
{\cal L}^{\rm c}\;:=\;{\cal L}\,-\,p\,{\dot\psi}\EQN correct-Lagrangian
$$
being the `correct' (or `constraint-free') Lagrangian density. (Note that
${{\overline{\rm P}}\,{\dot\Psi}\,+\,{\rm P}\,{\dot{\overline\Psi}}}$
translates into ${2\,p\,{\dot\psi}}$ only upto a total time derivative, which,
of course, does not affect the action.) The `correct' Hamiltonian form of the
parameterized action functional now reads
$$
{\cal I}^{\rm c}_{\rm{I\!H}}\;=\int_{\rm{I\!R}}dt\int_{\Sigma_t}d^3x\;
[\pi_{\sigma}\,{\dot v}^{\sigma}\;+\;p\,{\dot\psi}
\;+\;{\tt \Pi}_{\sigma}\,{\dot\vartheta}^{\sigma}\;-\;
\aleph^{\sigma}H_{\sigma}]\;,\EQN new-hami-action
$$
with the Hamiltonian equations of motion
$$
\EQNalign{
\;{{\delta\,H}\over{\delta\,{\tt \Pi}_{\alpha}}}\;&=\;+\,u^{\sigma}
\nabla_{\!\sigma}\,\vartheta^{\alpha}\;,\;\;\;\;\;\;\;\;\;\;\;\;
{{\delta\,H}\over{\delta\,\vartheta^{\alpha}}}\;=\;-\,u^{\sigma}
\nabla_{\!\sigma}\,{\tt \Pi}_{\alpha}\;,\EQN news-hami; a \cr
\;{{\delta\,H}\over{\delta\,\pi_{\alpha}}}\;&=\;+\,u^{\sigma}
\nabla_{\!\sigma}\,v^{\alpha}\;,\;\;\;\;\;\;\;\;\;\;\;\;
{{\delta\,H}\over{\delta\,v^{\alpha}}}\;=\;-\,u^{\sigma}
\nabla_{\!\sigma}\,\pi_{\alpha}\;,\EQN news-hami; b \cr
{\rm and}\;\;\;\;\;
\;{{\delta\,H}\over{\delta\,p\,}}\,\;&=\;+\,u^{\sigma}
\nabla_{\!\sigma}\,\psi\;,\;\;\;\;\;\;\;\;\;\;\;\;\;\;
{{\delta\,H}\over{\delta\,\psi\,}}\;=\;-\,u^{\sigma}
\nabla_{\!\sigma}\,p\;.\EQN news-hami; c \cr}
$$
It is easy to check that this set of equations of motion is equivalent to the
previous set with the {\it same} Hamiltonian density, but, of course, expressed
in terms of the old set of somewhat redundant canonical variables. The
Hamiltonian density \Ep{familiar-Hamiltonian} clearly suggests the `correct'
Lagrangian form of the action functional,
$$
{\cal I}^{\rm c}\,=\int_{\cal O}\!
d^4x\;{\cal L}^{\rm c}\!(v^{\sigma},
\nabla_{\!\mu}v^{\sigma},\nabla_{\!\mu}\nabla_{\!\nu}
v^{\sigma},\,\psi,\,\partial_{\mu}\psi\,;\,
{{}^{(s)}\!y})\,,\EQN correct-action
$$
which, {\it a priori}, might be taken to indicate that we would have been
better off taking from the beginning the alternative form
$$
{\cal L}^{\rm c}_{\Psi}\;=\;+\,\wp\;4\pi G\,\Leftcases{36pt}
          {{\hbar}^2\over{2m}}\,h^{\alpha\beta}\partial_{\alpha}\Psi
     \partial_{\beta}{\overline{\Psi}}\;-\;i{\hbar\over 2}\,u^{\alpha}\!(
    {\overline \Psi}\partial_{\alpha}\Psi)\!\!\Rightcases{36pt}\EQN in-end
$$
of the component ${{\cal L}_{\Psi}}$ in the action integral \Ep{action-int}.
However, this form of the Lagrangian density is not
Hermitian and breaks the symmetry which naturally exists between ${\Psi}$ and
${{\overline\Psi}\,}$. Moreover, it contains an incorrect factor of
${1\over 2}$ in the second term. Nevertheless, in the end, it is this form
\Ep{in-end} which turns out to be the most appropriate one as far as the new
set of canonical variables are concerned.

Finally, what about the last remaining constraint --- the diffeomorphism
constraint \Ep{diff-const} due to the parameterization process --- which we
have left on hold so far? It turns out that this constraint \Ep{diff-const} can
also be readily untangled and, unlike the parallel case in the Hamiltonian
formulation of the `already parameterized' Einstein's theory of gravity, here
the process of `deparameterization'\cite{Dirac} can be easily carried out.
This is because in our case the kinematical momentum densities
${{\tt \Pi}_{\sigma}}$ conjugate to the supplementary embedding variables
${\vartheta^{\sigma}}$ appear {\it linearly} in the expression of the
constraint, neatly segregating themselves from the set of genuinely dynamical
variables. Consequently, all one has to do to recover the `true' Hamiltonian
form of the physical action is to solve the constraint for these momentum
densities and get ${{\tt \Pi}_{\sigma}\,=\,-\,
{\!\!\buildchar{\;H_{\sigma}}{\circ}{}}\,}$, substitute this result into the
Hamiltonian action functional \Ep{new-hami-action}, and then impose the
relation \Ep{time-choice} (obtained from one of the equations of motion) on
the resulting expression to complete the deparameterization.

Thus, we have been able to eliminate {\it all} redundancies from the
configuration space by redefining it to consist only true dynamical degrees of
freedom, and, as a result, succeeded in constructing a meaningful
(constraint-free) phase-space for our Newton-Cartan-Schr\"odinger system.
Let us now recapitulate the main features of this phase-space.
First, and foremost, as there are no topological complications
(${\Sigma_t\cong\real^3}$), the phase-space is naturally a
{\it cotangent-bundle} ${T^*{\widetilde Z}}$ over the infinite-dimensional
complex configuration space ${\widetilde Z}$ of equivalence classes of fields
${\{{\widetilde v}^{\sigma}(x);\,{\widetilde \psi(x)}\}}$ evaluated on
the 3-submanifold ${\Sigma}$ of ${\M\,}$:   
$$
{\widetilde Z}\,:=\,\{{v^{\sigma};\psi\;|\;
t_{\sigma}v^{\sigma}=1,\,\nabla_{\!\lbrack\mu}\nabla_{\!\nu\rbrack}
v^{\sigma}=0,\,v^{\sigma}\sim v^{\sigma}-h^{\sigma\mu}\nabla_{\!\mu}f\,;\;
\psi\sim\exp(-\,i{\scriptstyle{m\over\hbar}}f)\psi}\}\,,
\EQN configuration-space
$$
where ${\nabla_{\!\lbrack\mu}\nabla_{\!\nu\rbrack}v^{\sigma}\,=\,0}$ is
equivalent to the condition \Ep{new-not}, and
${v^{\sigma}\sim v^{\sigma}-h^{\sigma\mu}\nabla_{\!\mu}f}$ results form the
equivalence \Ep{effect}. What is more, this cotangent-bundle possesses a
well-defined, {\it non-degenerate} symplectic structure, which is just its
natural symplectic structure
$$
{\widetilde{\omega}}\;\,=\,\;\int_{\Sigma_t}d^3x\,
[\,d_z\,{\widetilde \pi}_{\mu}\,\wedge\,d_z\,{\widetilde v}^{\mu}\;+\;
d_z\,{\widetilde p}\,\wedge\,d_z\,{\widetilde \psi}\,]\EQN non-two-form
$$
with ${{\widetilde v}^{\alpha},\;{\widetilde \pi}_{\alpha},\;{\widetilde \psi},
\;{\rm and}\;{\widetilde p}}$ representing entire gauge-equivalence classes of
fields as discussed above. Note that, although ${\widetilde Z}$ has a natural
vector-space structure over the {\it complex} field ${\comps}$, the 2-form
${\widetilde{\omega}}$ is {\it real-valued} since
${p=i\times{\overline\psi}\,}$.

As it stands, however, it is not immediately clear whether this description of
symplectic structure is a generally-covariant one. In fact it appears not
to be a covariant one on two counts. First of all, in the present subsection
we have been working strictly within a non-covariant 3+1 decomposition of
spacetime. Secondly, the symplectic structure \Ep{non-two-form} we have arrived
at in the end corresponds to the {\it deparameterized} action --- i.e., an
action effectively corresponding to the original unparameterized theory on a
fixed background ${(\M;\,h,\,\tau)\,}$. As we shall see in the next subsection,
however, the expression \Ep{non-two-form} in fact provides a truly
generally-covariant description of the phase-space, provided it is viewed more
appropriately.

\subsection{Manifestly covariant description of the canonical formulation}

In the previous subsection we have constructed a canonical phase-space for the
Newton-Cartan-Schr\"odinger system by working within a 3+1 decomposition of
spacetime into 3-spaces at instants of time. Such a blatantly non-covariant
breakup of spacetime is admittedly somewhat natural for a Newtonian theory,
but it undermines the efforts of Cartan and followers to give a full
spacetime-covariant meaning to Newtonian gravity by violating manifest
covariance of even the old-fashioned Galilean kind. Therefore, at least for the
sake of Cartan's legacy if not for anything else, it is desirable to seek a
manifestly covariant version of the canonical phase-space constructed in the
previous subsection. Fortunately, it has long been
recognized --- dating all the way back to Lagrange\cite{Ashtekar} --- that the
phase-space
of a physical system is better viewed as the space of entire dynamical
histories of the system, without reference to a particular instant of time.
With this view of phase-space, the core of the canonical formalism can be
developed in a manner that manifestly preserves all relevant symmetries of a
given classical
system\cite{Lee}\cite{Barnich}\cite{Woodhouse}\cite{Witten}\cite{Ashtekar}.
Since this `covariant phase-space formalism' is relatively less-popular, and
since we shall be applying it to our Newton-Cartan-Schr\"odinger system in the
present subsection as well as in the subsection 5.2 below when we quantize the
system, we have briefly reviewed its main concepts in the appendix A below
setting our notational conventions. We urge the reader not familiar with these
concepts to carefully study the appendix before reading any further in order to
appreciate the elegance of the underlying ideas used in what follows. 

The essence of the covariant phase-space --- the space ${\cal Z}$ of solutions
of the equations of motion of a theory --- is most succinctly
encapsulated in a closed, non-degenerate symplectic 2-form ${\widetilde\omega}$
defined on the quotient space ${{\widetilde{\cal Z}}:={\cal Z}/{\cal K}\,}$,
where ${\cal K}$ is the characteristic distribution defined by equation
\Ep{distribution} of the appendix. As a first step towards evaluating
${\widetilde\omega}$ for our Newton-Cartan-Schr\"odinger system, we work out
the presymplectic potential current density \Ep{PPCD} for the
generally-covariant action functional \Ep{correct-action},
$$
{\cal J}^{\mu}_{\!{\scriptscriptstyle{\cal Q}}}\;\,=\,\;
{\cal J}^{\mu}_{v}\;+\;{\cal J}^{\mu}_{\!\psi}\;+\;
{\cal J}^{\mu}_{\!{{}^{(s)}\!y}}\;,\EQN
$$
giving
$$
{\cal J}^{\alpha}_{\!{\scriptscriptstyle{\cal Q}}}\,t_{\alpha}\;\,=\,\;
{\pi_{\mu}}\,\delta{v^{\mu}}\;+\;p\,\delta\psi\;+\;
{{}^{(s)}\!{\tt \Pi}_{\sigma}}\,\delta\,{{}^{(s)}\!y^{\sigma}}\;.\EQN
$$
If we assume that ${{\cal J}^{\alpha}_{\!{\scriptscriptstyle{\cal Q}}}
\rightarrow 0}$ at spatial infinity (or work with a compact ${\Sigma_t}$),
then the corresponding {\it presymplectic} 2-form ${\omega}$
is given by the equation \Ep{omega-defi}, with ${\omega^{\mu}=
d_z\,{\cal J}^{\alpha}_{\!{\scriptscriptstyle{\cal Q}}}}$ (cf. equation
\Ep{j-to-w}):
$$
\omega\;=\,\int_{\Sigma_t}d^3x\,[\omega^{\mu}t_{\mu}]\;=
\,\int_{\Sigma_t}d^3x\,[d_z\,\pi_{\mu}\,\wedge\,d_z\,{v^{\mu}}\;+\;
d_z\,p\,\wedge\,d_z\,\psi\;+\;
d_z{{}^{(s)}\!{\tt \Pi}_{\sigma}}\,\wedge\,d_z{{}^{(s)}\!y^{\sigma}}]\,.
\EQN 4.89
$$
Now, in accordance with the discussion around equation \Ep{distribution} of the
appendix, this presymplectic structure
has a degenerate direction for each infinitesimal gauge
transformation of the theory stemming from the action of the automorphism group
${{\cal A}ut(B(\M))\,}$. Consequently, we seek the non-degenerate projection
${\widetilde{\omega}}$ of ${\omega}$ on the physically relevant {\it reduced
phase-space} ${{\widetilde{\cal Z}}={\cal Z}/{\cal A}ut(B(\M))\,}$, which is
nothing but the space of orbits of the group ${{\cal A}ut(B(\M))}$ in the
solution-space ${{\cal Z}\,}$. Fortunately, the complete automorphism group
has the structure of a semidirect product: ${{\cal A}ut(B(\M))=
{\cal V}(B(\M))\,\semidirect\,{\rm Diff}(\M)\,}$, where ${{\cal V}(B(\M))}$ is
the group of vertical gauge transformations given by \Ep{vert-group}. This
makes it possible to discuss effects of the two subgroups successively, since
the isomorphism
$$
{\cal Z}/{\cal A}ut(B(\M))\;\cong\;[{\cal Z}/{\cal V}(B(\M))]/{\rm Diff}(\M)
\EQN
$$
holds. First note that, in analogy with the discussion in the previous
subsection, the
quotient space ${{\cal Z}/{\cal V}(B(\M))}$ is the space of solutions modulo
vertical gauge directions in which the integral ${\int_{\Sigma_t}\!d^3x\,
{\cal J}^{\alpha}_{\!{\scriptscriptstyle{\cal Q}}}t_{\alpha}}$ is rendered
gauge-invariant (cf. equation \Ep{bund-in}). In terms of the projection map
${\proj\,:\,{\cal Z}\rightarrow{\widetilde{\cal Z}}\,}$, this implies
$$
\proj(v)\;=\;{\widetilde v}\,,\;\;\;\;\;\;
\proj(\pi)\;=\;{\widetilde \pi}\,,\;\;\;\;\;\;
\proj(\psi)\;=\;{\widetilde \psi}\,,\;\;\;\;\;\;{\rm and}\;\;\;\;\;\;
\proj(p)\;=\;{\widetilde p}\,.\EQN
$$
Next, a globally valid gauge representing the moduli space
${[{\cal Z}/{\cal V}(B(\M))]/{\rm Diff}(\M)}$ can be easily constructed by
simply taking ${{}^{(s)}\!y\,:\,\M\,\rightarrow\,
{\buildchar{\M\,}{\circ}{}}}$ to be a fixed diffeomorphism ${{}^{(o)}\!y}$
(e.g., ${{}^{(o)}\!y\,=\,identity}$) and then specifying the values of the
dynamical variables on ${\Sigma_t}$ with respect to this choice. This is
possible because, as in a parameterized scalar field theory in Minkowski
spacetime\cite{Lee}\cite{Torre}, any arbitrary variation of the diffeomorphism
${{}^{(s)}\!y}$ sweeps out the entire degeneracy submanifold of ${\omega\,}$,
allowing us to uniquely characterize each degeneracy submanifold by a fixed
diffeomorphism ${{}^{(o)}\!y\,}$. But this immediately renders
${{\cal J}^{\mu}_{\!{{}^{(o)}\!y}}=0}$ (cf. equation \Ep{PPCD}).
Therefore, the net result of projecting the presymplectic 2-form ${\omega}$ on
the reduced phase-space ${{\widetilde{\cal Z}}={\cal Z}/{\cal A}ut(B(\M))}$ by
the projection map ${\proj\,:\,{\cal Z}\rightarrow{\widetilde{\cal Z}}}$ is the
desired non-degenerate 2-form

$$\boxit{
{\vbox{\hsize 253pt\strut
$$
\;\;{\widetilde{\omega}}\;\,=\,\;\int_{\Sigma_t}d^3x\,
[\,d_z\,{{\widetilde \pi}_{\mu}}\,\wedge\,d_z\,{{\widetilde v}^{\mu}}\;+\;
d_z\,{\widetilde p}\,\wedge\,d_z\,{\widetilde \psi}\,].
$$
\strut}}}\EQN expofw
$$
It is easy to see that ${\omega}$ in expression \Ep{4.89} is a unique pull-back
${\proj^*({\widetilde{\omega}})}$ of ${\widetilde{\omega}}$ form
${\widetilde{\cal Z}}$ to ${{\cal Z}\,}$. In other words, we have
$$
{{\tt Y}\!\!\subA}\cont{\widetilde{\omega}}\;=\;0\EQN
$$
(cf. equation \Ep{gauge-directions}),
where ${{\tt Y}\!\!\subA}$ are the gauge directions corresponding to the
complete
automorphism group ${{\cal A}ut(B(\M))\,}$. Again in close analogy with the
covariant phase-space description of a parameterized scalar field
theory in Minkowski spacetime\cite{Lee}\cite{Torre}, this non-degenerate
symplectic 2-form ${\widetilde{\omega}}$ describes precisely the phase-space
of the original unparameterized theory in the fixed, non-dynamical spacetime
${(\M;\,h^{\mu\nu},\,t_{\mu})\,}$. Thus, the reduction procedure in the
present case has the effect of `deparameterizing' the parameterized theory,
just as in the case of a scalar field theory. However, unlike in the previous
subsection, here our description is manifestly covariant. 
Further, the resulting {\it constraint submanifold}\cite{Lee}
${{\widetilde{\cal Z}}}$ of the dynamically possible states of the system
permeates {\it all} of the space ${\widetilde Z}$ of kinematically possible
states because the constraints have been eliminated (cf. equation
\Ep{Monique}). In fact, as comparison of equations \Ep{non-two-form}
and \Ep{expofw} immediately suggests, the covariant phase-space
${{\widetilde{\cal Z}}}$ of the present subsection is symplectically
diffeomorphic to the constraint-free canonical phase-space
${T^*{\widetilde Z}}$ of the previous subsection:
$$
{\widetilde{\cal Z}}\;\simeq\;T^*{\widetilde Z}\,.\;\;\,\EQN
$$

It is worth recalling here that the symplectic structure ${\widetilde{\omega}}$
is independent of the choice of a hypersurface ${\Sigma_t\,}$, and as such, in
particular, it is Milne-invariant (cf. equation \Ep{Leibniz}).
More generally,
${\widetilde{\omega}}$ is rather trivially invariant under the action of
${{\rm Diff}(\M)}$ on ${\cal Z}$ since all ingredients in its expression
above transform homogeneously, like tensors. In short,
we have been able to successfully crystallize
the diffeomorphism-invariant canonical essence of
Newton-Cartan-Schr\"odinger system in the expression \Ep{expofw}.

\section{Quantization of the Newton-Cartan-Schr\"odinger system}

So far we have considered only classical, external gravitational field.
However, as noted before, in Newton-Cartan theory the connection-field
is a dynamical object: it is not just a part of the immutable background
structure, but depends crucially on the distribution of matter-sources via the
field equation ${R_{\mu\nu} = 4\pi G\, M_{\mu\nu}\,}$. Since this equation
dictates the coupling of spacetime curvature to {\it quantum mechanically}
treated matter, the Newton-Cartan connection cannot have an {\it a priori}
definite value and must itself be treated in a quantum mechanical fashion.
Thus, a consistent
account of physical phenomena even at a Galilean-relativistic
level {\it necessitates} the construction of a quantum theory of gravity in
which the superposition principle holds not only for the states of matter, but
also for the states of the Newton-Cartan connection-field. In what follows we
construct such a generally-covariant, Milne-relativistic quantum theory of
gravity in which quantized Schr\"odinger particles produce the {\it quantized}
Newton-Cartan connection-field through which they interact.

\subsection{Covariant phase-space quantization}

Having successfully identified the constraint-free phase-space for our
classical Newton-Cartan-Schr\"odinger system in the previous section, the
desired quantum theory can be easily constructed using a manifestly covariant
approach to the usual canonical quantization method. An accessible reference on
the general procedure is \Ref{Wald-94}. Recall that classical
observables, say ${{\tt O}}$, are maps from the phase-space
${\widetilde{\cal Z}}$ to ${\real\,}$, and the non-degenerate symplectic 2-form
${\widetilde\omega}$ on ${\widetilde{\cal Z}}$ determines a set of Poisson
brackets inducing a Lie-algebra structure on the space of these observables. To
quantized such a system, we are supposed to replace the classical observables
with operators and Poisson brackets with commutators providing a corresponding
algebraic structure on the space of
these operators. More precisely, we are to seek a correspondence map,
${\,{\widehat{}}\,:\,{\tt O}\mapsto{\widehat{\tt O}}\,}$, and look for an
irreducible representation of the Lie-algebra of classical observables as an
algebra of operators acting on elements of some separable Hilbert space
${\cal H}$. It is well-known, however, that, as stated, this programme cannot
be carried out in general. As early as in 1951 van Hove demonstrated that, for
theories in which the position and momentum operators are represented in the
standard manner, no such correspondence map can provide an
irreducible representation of the full Poisson algebra of classical
observables\cite{Chernoff}. Fortunately, in our case the phase-space is
naturally isomorphic to a cotangent-bundle,
${{\widetilde{\cal Z}}\cong T^*{\widetilde Z}\,}$, for which the van Hove
obstruction is neutralised. Consequently, for us, it will turn out to be
possible to choose a Hilbert space ${\cal H}$ and a correspondence map
${\;{\widehat{}}\,:\,{\tt O}\mapsto{\widehat{\tt O}}}$ such that the
non-vanishing commutators satisfy{\parindent 0.40cm
\baselineskip 0.53cm\footnote{$^{\scriptscriptstyle 8}$}{\ninepoint{\hang
For definiteness and simplicity, we shall only follow the bosonic case here.
In this Galilean-relativistic context, where there is no connection between
spin and statistics, a parallel discussion with a fermionic field is
relatively straightforward (see subsection 4.7 of \Ref{Wald-94} for a
fermionic treatment in the relativistic case).\par}}}
$$
[\,{\widehat{\widetilde v}}^{\mu}(x),\;{\widehat{\widetilde\pi}}_{\nu}(x')]
\;=\;i\hbar\,\widehat{\ident}\,\delta^{\mu}_{\;\;\nu}\,
\delta\!({\vec{\rm x}}-{\vec{\rm x}}\,')
\;\;\;\;\;\;{\rm and}\;\;\;\;\;\;
[\,\widehat{\widetilde\psi}(x),\;\widehat{\widetilde\psi}^{\dagger}(x')]
\;=\;\widehat{\ident}\,\delta\!({\vec{\rm x}}-{\vec{\rm x}}\,')\EQN TCCR
$$
at equal-times (cf. equation \Ep{CCR}). The appearance of the Dirac
delta-`function' here necessitates that
${{\widehat{\widetilde v}}^{\mu}}$, ${{\widehat{\widetilde\pi}}_{\nu}}$,
${\widehat{\widetilde\psi}}$, and ${\widehat{\widetilde\psi}^{\dagger}}$ must
all be viewed as operator-valued distributions, and indicates
that only the fields
smeared with appropriate test-functions have physical meaning as observables
in accordance with the well-known analysis of measurements of field-observables
developed by Bohr and Rosenfeld\cite{Bohr}. As they stand, however, these
commutation relations are clearly not expressed in a manifestly covariant
manner. This deficiency can be quite easily removed in our case, partly
by exploiting the natural vector-space structure of the phase-space
${\widetilde{\cal Z}}$ noted in the previous section. On account of this
vector-space structure of ${\widetilde{\cal Z}}$, we may identify the tangent
space at any point ${z\in{\widetilde{\cal Z}}}$ with ${\widetilde{\cal Z}}$
itself. Furthermore, as discussed in the appendix, under this identification
the symplectic form ${\widetilde\omega}$ becomes an antisymmetric bilinear
function, ${{\widetilde\omega}\,:\,{\widetilde{\cal Z}}\times
{\widetilde{\cal Z}}\rightarrow\real\,}$, on the resultant symplectic
vector-space ${\widetilde{\cal Z}}$. Thus, we may rewrite the commutation
relations \Ep{TCCR} in terms of the operators
${\widehat{\widetilde\omega}({\cal Q}\,,\;\cdot\;)}$ corresponding to the
functions ${{\widetilde\omega}({\cal Q}\,,\;\cdot\;)}$ as the single commutator
$$
[\,\widehat{\widetilde\omega}({\cal Q}\suba,\;\cdot\;),\;\,
\widehat{\widetilde\omega}({\cal Q}\subb,\;\cdot\;)]\;\,=\,\;
-\,i\hbar\,\widehat{\ident}\;{\widetilde\omega}
({\cal Q}\suba,\,{\cal Q}\subb)\;,\EQN
$$
where ${{\cal Q},\,{\cal Q}\suba,\,{\cal Q}\subb\,\in\,{\widetilde{\cal Z}}}$
(see \Ref{Wald-94} for further details on such translations).
Since the self-adjoint operators appearing in this expression are unbounded,
and, hence, only densely defined, it is convenient to work with the equivalent
but better behaved {\it Weyl relations},
$$
\widehat{\rm W}^{\dagger}({\cal Q})\;\,=\,\;\widehat{\rm W}(-\,{\cal Q})
$$
$$
{\rm and}\;\;\;\;\;
\widehat{\rm W}({\cal Q}\suba)\,\widehat{\rm W}({\cal Q}\subb)\;\,=\,\;
\exp[{\scriptstyle{i\over 2}}\hbar\,{\widetilde\omega}
({\cal Q}\suba,\,{\cal Q}\subb)]\,
\widehat{\rm W}({\cal Q}\suba+{\cal Q}\subb)\;,\;\;\;\;\;\;\;\;\;\EQN
$$
where
$$
{\rm W}({\cal Q})\;:=\;\exp[\,i\,{\widetilde\omega}({\cal Q}\,,\;\cdot\;)]\EQN
$$
is unitary and varies with ${\cal Q}$ in the `strong operator
topology'\cite{Haag}.

\subsection{The GNS-construction and the choice of a vacuum state}

It is well-known that the set, ${\cal B(H)}$, of all bounded linear maps
on ${\cal H}$ has the natural structure of a C${^*}$-algebra with the
`${*}$-operation' corresponding to taking adjoints. The subalgebra ${\cal A}$
of ${\cal B(H)}$ generated by
${\{\,{\widehat{\rm W}}({\cal Q})\,|\,{\cal Q}\in
{\widetilde{\cal Z}}\,\}}$ satisfying the above relations is called
the {\it Weyl algebra} over the symplectic vector-space ${\widetilde{\cal Z}}$.
Each normalized, positive algebraic state
${\zeta\,:\,{\cal A}\rightarrow\comps}$ over the Weyl algebra ${\cal A}$ ---
viewed as an abstract C${^*}$-algebra --- determines a Hilbert space
${{\cal H}_{\zeta}}$ and a representation
${{\cal R}_{\zeta}\,:\,{\cal A}\rightarrow{\cal B}({\cal H}_{\zeta})}$ 
of ${\cal A}$ by bounded linear operators acting in ${{\cal H}_{\zeta}}$,
and thereby defines an Hermitian scalar product on ${\cal A}$ by
$$
\zeta(a^*b)\;:=\;\bra{a}b\rangle\;\;\;\;\;\;\forall\;a,b\,\in\,{\cal A}\,.\EQN
$$
Conversely, each choice of a measure
${\mu\,:\,{\cal H}_{\zeta}\times{\cal H}_{\zeta}\rightarrow\real}$ generates
a state ${\zeta}$ on the algebra ${\cal A}$. Consequently, the positivity
and normalization conditions on ${\zeta}$ can be expressed as
$$
\zeta(a^*a)\;\geq\;0\;\;\;\;\;\;\forall\;a\in{\cal A}\,,\;\;\;\;\;\;
{\rm and}\;\;\;\;\;\;\zeta(\ident)\;=\;1\,,\EQN
$$
where ${\ident}$ denotes the identity element of ${\cal A}$. Such a
construction of a representation of the Weyl algebra ${\cal A}$
as an algebra of operators on a Hilbert space ${{\cal H}_{\zeta}}$ is nothing
but the celebrated GNS-construction\cite{Haag}, which,
given a {\it cyclic} vector ${\ket{\xi^o}\in{\cal H}_{\zeta}}$, guarantees the
existence of a representation ${{\cal R}_{\zeta}\,:\,{\cal A}\rightarrow
{\cal B}({\cal H}_{\zeta})}$ such that
$$
\zeta_{\xi^o}(a)\;=\;\bra{\xi^o}\,{\cal R}_{\zeta}(a)\ket{\xi^o}\;\;\;\;\;\;
\forall\;a\,\in\,{\cal A}\,.\EQN
$$
The vector ${\xi^o}$ is called a cyclic vector because the set
${\{\,{\cal R}_{\zeta}(a)\ket{\xi^o}\,|\,a\in{\cal A}\,\}}$ constitutes a
{\it dense} subspace of ${{\cal H}_{\zeta}}$. Upto unitary equivalence, the
triplet ${({\cal H}_{\zeta}\,,\;{\cal R}_{\zeta}\,,\;\ket{\xi^o})}$ is uniquely
determined by these properties, with the cyclic vector
${\ket{\xi^o}\,\in\,{\cal H}_{\zeta}}$ corresponding to the identity element of
${\cal A}$.

Thus, an algebraic state ${\zeta}$ in ${\cal A}$ can be easily
represented in the Hilbert space ${{\cal H}_{\zeta}}$ by the state-vector
${\ket{\xi^o}}$. Note, however, that the states over an abstract
C${^*}$-algebra like ${\cal A}$ come in families: any vector ${\ket{\xi^i}}$
from a collection
${\{\,\ket{\xi^i}\,|\;\ket{\xi^i}\in{\cal H}_{\zeta}\,\}}$ represents the
family of algebraic state ${\zeta}$ by
$$
\zeta_{\xi^i}\!(a)\;=\;\bra{\xi^i}\,{\cal R}_{\zeta}(a)\ket{\xi^i}\,.\EQN
$$
(More generally, we may consider the states
${\zeta\subd(a)\,=\,\Tr[\,{\cal D}\,{\cal R}_{\zeta}(a)\,]\,}$, with
${\cal D}$ being a positive trace-class operator in
${{\cal B}({\cal H}_{\zeta})\,}$).
This is due to the fact that ${\zeta_{\xi^i}(a)}$ can be approximated as
closely as desired by ${\zeta_{\xi^o}(b^*a^{}b)}$,
for any ${b\in{\cal A}}$, since, thanks to the cyclicity of ${\ket{\xi^o}}$,
${\ket{\xi^i}}$ can be approximated as closely as desired by
${{\cal R}_{\zeta}(b)\ket{\xi^o}}$:
$$
\zeta_{\xi^i}\!(a)\;=\;\bra{\xi^i}\,{\cal R}_{\zeta}(a)\ket{\xi^i}\;\approx\;
\bra{\xi^o}\,{\cal R}^*_{\zeta}(b){\cal R}_{\zeta}(a)
{\cal R}_{\zeta}(b)\ket{\xi^o}\;=\;\zeta_{\xi^o}\!(b^*a^{}b)\,.\EQN
$$
Consequently, a GNS-representation crucially depends on the generating state
${\zeta}$ of the system. In particular, with changes in the state ${\zeta}$,
the measure ${\mu}$ --- with respect to which the inner-product of the Hilbert
space ${{\cal H}_{\zeta}}$ has been defined --- changes. But this is a bad
news: measures with different null-sets would in general lead to operators with
different norms in the corresponding representations, and the kernel (i.e.,
the set of those operators which are mapped onto zero) of the representations
will correspondingly differ. Consequently, such representations (uncountably
many of them) will be {\it unitarily inequivalent} in general since the states
which generate them could fail to determine quasi-equivalent measures --- i.e.,
measures with the same null-sets. This, of course, is a well-known problem for
quantum systems with infinitely many degrees of freedom, since the Stone-von
Neumann uniqueness theorem is inapplicable for such systems\cite{Wald-94}.

Fortunately, if we are willing to let go a bit of the mathematical elegance
maintained so far and find our way back to physics, there are two strategies
at our disposal to tackle this problem. The first obvious strategy is to select
a privileged cyclic state-vector using some physical criterion (some rule
external to the quantum theory proper), and thereby obtain an equivalence
class of
representations which contains this distinguished state. The standard choice in
Minkowskian quantum field theories is, of course, the {\it vacuum} state,
${\ket{\xi^o}\equiv\ket{0}}$, which is required to remain invariant under the
action of the Poincar\'e group --- the isometry group of the flat Minkowski
spacetime. Then, through GNS-construction, the vacuum expectation values of all
operators provide the Hilbert space equipped with a natural inner-product.
This choice of a GNS-representation is then simply the Fock-representation. In
our case it is most natural to use the Milne group defined by
$$
0\;=\;\hbox{\it\char36}{\!}_{{\rm x}}\,h^{\mu\nu}\;=\;
\hbox{\it\char36}{\!}_{{\rm x}}\,t_{\mu}\;=\;
\hbox{\it\char36}{\!}_{\rm x}\,\Gamma_{\mu}^{\,\alpha\,\nu}\;\;\;\EQN
$$
(cf. equation \Ep{Milne}) in place of the Poincar\'e group in the above
procedure, and take the Milne-invariant no-particle state as our privileged
state. The Hilbert space we thereby obtain is a Milne-relativistic
Fock-representation of the Weyl algebra for our Newton-Cartan-Schr\"odinger
system.

The second strategy to deal with the problem of inequivalent representations is
to follow the general `operational' philosophy historically motivated by a
result of Fell\cite{Fell}, and adopt his criterion of `physical equivalence'
as opposed to the strict mathematical equivalence. This too is a well-known
strategy, a good accessible account of which can be found in \Ref{Wald-94}
(see, especially, his Theorem 4.5.2). The idea is to first acknowledge the
finiteness
of accuracy and number of possible realistic measurements, and then realize
that, due to these limitations, it is physically impossible to distinguish
between the `inequivalent' representations of the C${^*}$-algebra ${\cal A}$.
Consequently, a choice form the myriad of inequivalent representations is
physically irrelevant, and the choice of Milne-invariant
Fock-representation made above is as good as any.

\subsection{The Hamiltonian operator: in general and in an inertial frame}

Having settled the problem of inequivalent representations does not, of course,
guarantee that every important observable needed to unambiguously define a
quantum theory is contained in the algebra ${\cal A}$. Therefore, let us
verify that at least the most important observables of our theory are
well-defined. We begin with one of the simplest operator: the covariant
mass-density operator
$$
{\widehat M}_{\mu\nu}\;:=\;m\,{\widehat{\widetilde\psi}}^{\dag}
{\widehat{\widetilde\psi}}\,t_{\mu\nu}\EQN MDO
$$
(cf. equation \Ep{mass-density-operator}). We have been careful to choose an
appropriate ordering in defining this operator,
and, as a result, it is manifestly well-defined. Similarly, the operator
corresponding to the Riemann tensor \Ep{curvature}
is also well-defined provided the operators corresponding to the gravitational
field appearing in its polynomial expression are properly normal-ordered.
On the other hand, the all important Hamiltonian-density operator
$$
{\widehat H}\;:=\;H({\widehat{\widetilde v}}^{\alpha},\,
{\widehat{\widetilde \pi}}_{\alpha}\,;\,{\widehat{\widetilde \psi}},\,
{\widehat{\widetilde \psi}}^{\dag})\EQN
$$
in our interacting theory
involves both matter and gravitational field variables. Hence, {\it a priori},
one might fear existence of potential operator ordering ambiguities in its
expression. However, recall that all of the matter variables commute with all
of the gravitational variables, removing any danger of intractable ordering
ambiguity. Consequently, the Hamiltonian operator
$$
{\widehat\boh}\;:=\;\int_{\Sigma_t}:{\widehat H}:\;d^3x\EQN
$$
is also well-defined, where matter and gravitational operators in
${\widehat H}$ are taken to be independently normal-ordered.

The vanishing commutation between matter and gravitational variables imply,
in particular, that there is no gravitational self-interaction in this linear,
Newtonian quantum gravity; i.e., the Schr\"odinger-Fock particles of the
system do not gravitationally self-interact, though they interact among
themselves. This result becomes most conspicuous from the perspective of an
observer confined to a {\it local} inertial (i.e., Galilean) frame of reference
equipped with a Cartesian
coordinate system. In a local inertial frame the inertial and gravitational
parts of the Newton-Cartan connection-field may be unambiguously (but, of
course, non-covariantly) separated as in the equation \Ep{gauge-depend} above,
and a linear coordinate system may be introduced such that the two metric
fields assume their canonical forms:
${h=\delta^{ab}\partial_a\otimes\partial_b}$ and ${\tau=dt\,}$. Furthermore,
the connection-field in such a frame corresponds to a gauge-choice
${u={{\partial\;}\over{\partial t}}}$ and ${A=-\Phi\tau}$ in equation
\Ep{affinely}, with ${\Phi}$ viewed as the usual Newtonian gravitational
potential. With a further gauge-choice of ${\chi=-{1\over 2}\Phi}$, and setting
the rest of the multiplier fields to zero (without any loss of information, of
course), the action functional \Ep{action-int} in the inertial frame becomes
$$
{\cal I}\;=\,\int dt\int d\x\;[{1\over{8\pi G}}\,\Phi\,\Delta\,
\Phi\;+\;{{\,\hbar^2}\over{2m}}\,\delta^{ab}\,\partial_a\Psi\,\partial_b\Psi\;
+\;i{\hbar\over 2}(\Psi\,\partial_t{\overline\Psi}\,-\,{\overline\Psi}\,
\partial_t\Psi)\;-\;m\,\Psi{\overline\Psi}\,\Phi],\EQN
$$
where we recall that ${\kappa=0}$, and for simplicity we also take
${\Lambda\subo=0\,}$. Extremization of this functional with respect to
variations of the scalar potential ${\Phi}$ immediately yields the
Newton-Poisson equation
$$
\Delta\Phi\;=\;{{4\pi G}\over{\bra{\psi}\psi\rangle}}\,m\,\psi\,
{\overline\psi}\,,\EQN Newton-N
$$
where ${\psi:=\sqrt{2\pi G\wp}\;{\overline\Psi}}$, and we set
${\bra{\psi}\psi\rangle :=\int d\x\;{\overline\psi}\,\psi=1\,}$.
On the other hand, extremization of the action
with respect to variations of the matter field ${\Psi}$ leads to the familiar
Schr\"odinger equation in the presence of an external gravitational field:
$$
i\hbar{{\partial\;}\over{\partial t}}\,\psi\;=\;[-\,
{{\,\hbar^2}\over{2m}}\Delta\;
+\;m\,\Phi]\psi\,.\EQN Newton-M
$$
The last two equations may be interpreted as describing a {\it single}
Galilean-relativistic particle gravitationally interacting with its own
Newtonian field. As such, the coupled equations \Ep{Newton-N} and \Ep{Newton-M}
constitute a {\it nonlinear} system, which can be easily seen as such by
first formally solving equation \Ep{Newton-N} for the gravitational potential
giving
$$
\Phi(\x)\;=\;-\,Gm\,\int d\x\,'\;
{{{\overline\psi}(\x\,')\,\psi(\x\,')}\over{|\x\,'-\x|}}\,,\EQN
$$
and then --- by substituting this solution into equation \Ep{Newton-M} ---
obtaining the nonlinear integro-differential equation
$$
i\hbar{{\partial\;}\over{\partial t}}\,\psi(\x,\,t)\;=\;
-\,{{\,\hbar^2}\over{2m}}
\Delta\,\psi(\x,\,t)\;-\;Gm^2\,\int d\x\,'\;
{{{\overline\psi}(\x\,',\,t)\,\psi(\x\,',\,t)}\over{|\x\,'-\x|}}\,\psi(\x,\,t)
\,.\EQN int-diff
$$
However, when ${\psi}$ is promoted to a `second-quantized' field operator
${\widehat\psi}$ satisfying
$$
[\,{\widehat\psi}(\x),\;{\widehat\psi}^{\dagger}(\x')]
\,=\;\widehat{\ident}\,\delta\!(\x-\x\,')\,,\EQN
$$
this equation describes a system of many identical particles in the Heisenberg
picture, with ${\widehat\psi}$ acting as an annihilation operator in the
corresponding Fock
space, analogous to the covariantly described `free' system discussed in the
subsection 3.2 above. In particular, the normal-ordered Hamiltonian operator
for the system now reads
$$
\EQNalign{{\widehat\boh}\;&=\;{\widehat\boh}\subo\;+\;{\widehat\boh}_{\rm I}\cr
{\rm with}\;\;\;\;\;\;\,{\widehat\boh}\subo &:=\,\int d\x\;\;
{\widehat\psi}^{\dag}(\x)[-{{\hbar^2}\over{2m}}\Delta]{\widehat\psi}(\x)\,,\cr
{\rm and}\;\;\;\;\;\;\;{\widehat\boh}_{\rm I}\,&:=\;-\,{1\over2}\,Gm^2\,
\int d\x\int d\x\,'\;
{{{\widehat\psi}^{\dag}(\x\,')\,{\widehat\psi}^{\dag}(\x)\,
{\widehat\psi}(\x)\,{\widehat\psi}(\x\,')}\over{|\x\,'-\x|}}\,,\EQN\cr}
$$
which, upon substitution into the Heisenberg equation of motion
$$
i\hbar{{\partial\;}\over{\partial t}}\,{\widehat\psi}(\x,\,t)\;=\;
[{\widehat\psi}(\x,\,t),\,{\widehat\boh}]\EQN
$$
yields an operator equation corresponding to the equation \Ep{int-diff}. It
is easy to show\cite{Brown} that the action of the Hamiltonian operator
${\widehat\boh}$ on a multi-particle state is given by
$$
\bra{\x\suba\,\x\subb\,\dots\,\x_n}{\widehat\boh}\ket{\,\xi\,}\;=\;
[{-\,{{\,\hbar^2}\over{2m}}}\,\sum_{a={\scriptscriptstyle 1}}^n\Delta_a\;-\;
Gm^2\,\sum_{a\,<\,b}
{1\over{|\x_a\,-\,\x_b|}}]\,
\bra{\x\suba\,\x\subb\,\dots\,\x_n}\,\xi\,\rangle\,,\EQN
$$
which is consistent with the classical multi-particle Hamiltonian with
gravitational pair-interactions.
Evidently, the interaction Hamiltonian annihilates a single particle state
${\ket{\,\x\,}:={\widehat\psi}^{\dag}(\x)\ket{\,0\,}\,}$,
$$
{\widehat\boh}_{\rm I}\,\ket{\,\x\,}\;=\;0\,,\EQN
$$
implying that, thanks to the appropriate normal-ordering of the operators in
${\widehat\boh\,}$, the matter particles do not gravitationally self-interact.
Moreover, the number operator, defined by
${{\widehat{\rm N}}:=\int d\x\,\;{\widehat\psi}^{\dag}(\x)\,{\widehat\psi}(\x)
\,}$,
commutes with the total Hamiltonian operator. In other words, the number of
particles in the theory under consideration is a constant of motion: as
expected, our Galilean-relativistic interaction does
not lead to particle production.

\subsection{The intuitive physical picture}

It is well-known that Newtonian gravitational field does not possess any
dynamical degrees of freedom of its own --- they `remain frozen' in the
`${c\rightarrow\infty}$' limit. On the other hand, Einsteinian gravitational
radiation propagates in vacuum with the speed of light $c$. Considering this
in the light of
the local quantum field theory of Newton-Cartan gravity we have constructed
here, it is rather convenient to maintain that Newtonian gravitational field
also possesses propagating degrees of freedom, but it so happens that such
gravitational disturbances travel instantaneously --- i.e., with the
Galilean-relativistic speed of light `${c=\infty}$'. Indeed, it is
possible to view Newton-Poisson equation ${\,\Delta\Phi\,=\,4\pi\,G\,\rho\,}$
as a wave-equation for the Newtonian gravitational waves
propagating with infinite speed:
$$
\lim_{c\rightarrow\infty}[\Delta\Phi\;-\;{1\over{c^2}}\,
{{\partial^2\Phi}\over{\partial t^2}}\;=\;4\pi\,G\,\rho]\;\longrightarrow\;
[\Delta\Phi\;=\;4\pi\,G\,\rho]\,.\EQN limit-relation
$$
Moreover, unlike the usual weak-field approach to Einstein's
gravity (`${c\not=\infty}$') where such {\it longitudinal} degrees of freedom
of the gravitational field are `gauged-away' and ignored (transverse-traceless
gauge), in the theory constructed here we have been able to avoid any
gauge-fixing procedure. Therefore, as far as Newton-Cartan theory is viewed as
a Galilean-relativistic limit-form of Einstein's theory of gravity, the limit
relation
\Ep{limit-relation} is a better interpretation of the Newton-Poisson equation
than the usual one in which one insists that there are no gravitational waves
in a Newtonian theory.

Now, consider the quantized gravitational degrees of freedom. According to the
above view, these also propagate with infinite speed. Moreover, if we take
expressions \Ep{operator-mass} and \Ep{operator-angular} in our theory as the
mass and angular-momentum operators, respectively (which are indeed the
correct operators for an inertial observer), then
$$
[{\widehat\bom},\,{\widehat{\widetilde v}}^{\alpha}]\;=\;0\;\;\;\;\;\;{\rm and}
\;\;\;\;\;\;[{\widehat{\bf J}},\,{\widehat{\widetilde v}}^{\alpha}]\;=\;0\EQN
$$
imply that Newton-Cartan gravitons are the massless and spinless mediating
particles between the matter particles.
In other words, the theory constructed here
may be viewed, in a particle interpretation, as a theory of
non-self-interacting quantized Schr\"odinger particles producing the
{\it longitudinal} Newton-Cartan gravitons --- the massless, spin-0 exchange
bosons propagating with infinite speed --- and interacting through them.

\section{Conclusion}

If Einstein's gravity is viewed as a result of two physical principles, (1)
strong equivalence of gravitational and inertial masses, and (2) relativization
of time, then it is the second principle
which prevents it from any straightforward subjugation to the otherwise
well-corroborated rules of quantization. As demonstrated here, the first one
by itself is completely unproblematic as far as the canonical quantization of
gravity is concerned. On the other hand, the invocation of relativization of
time in addition to the first principle induces the so far intractable
`problem of time' via the Hamiltonian constraint in general relativity --- as
is quite well-known\cite{Time}. Because of the presence of preferred foliation
--- whose covariant normal constitutes the kernel of the spatial metric
endowing spacetime with an absolute, observer-independent notion of distant
simultaneity --- no such intractable constraint arises in the Hamiltonian
formulation of the classical Newton-Cartan-Schr\"odinger theory. Consequently,
we have been able to successfully and unambiguously quantize this
{\it interacting} Galilean-relativistic field theory as an {\it unconstrained}
Hamiltonian system in a {\it manifestly covariant} manner. What is more, as
discussed in the Introduction, this exercise opens up a completely novel
direction of research in quantum gravity: the program of the
{\it special-relativization} of the quantum theory of Newton-Cartan gravity.

\nosechead{\bf Note Added to Proof}

Several years after I published this paper I learned from Roger Penrose --- who,
in turn, had learned from John Stachel --- that in the 1930s a remarkable Soviet
physicist called Bronstein had already represented the fundamental theories of
physics in a map very much like the one in my Fig. 1 above. His two-dimensional
map, however, neglected the Newtonian theory of gravity, which was eventually
included by Zelmanov in 1967, thereby transforming Bronstein's
two-dimensional picture into a full three-dimensional cube. Neither Bronstein nor
Zelmanov, however, made the crucial distinction between Newton's original theory
of gravity and Cartan's spacetime reformulation of it, let alone made any indication
of quantization of the latter theory. In particular, Zelmanov's version of the cube
(followed up by Okun in 1991\cite{Okun-1991}) wrongly represents Newton's theory of
gravity as a Galilean-relativistic limit-form of Einstein's theory of gravity. In
fact, as we saw above, the true limit-form of Einstein's theory is the Newton-Cartan
theory with its mutable connection field. To appreciate the importance of this crucial
distinction in the contemporary context, see, for example, section 14.4.2 of
Ref.\cite{Christian-2001} and references therein. For an excellent historical account
of Bronstein's remarkable life and work in physics --- including the Soviet history
of ``the cube of theories'' --- see Ref.\cite{Gorelik-1994} and references therein.

\bigskip

\nosechead{\bf Acknowledgements}

I am deeply indebted to Jeremy Butterfield, Abner Shimony, Jeeva Anandan,
Paul Busch, Ashwin Srinivasan, Andreas Fuentes, Wolfram Latsch, Rui Camacho,
Savitaben Christian,
and The Mrs L.D. Rope Third Charitable Settlement for their generous and timely
support without which this paper would have never materialized. I am also
grateful to an anonymous referee for the \Ref{Christian} for inadvertently
setting me off to embark upon the present project. Finally, without the
generous interest shown by Roger Penrose on my precursory ideas, I would have
never been
able to overcome my stifling doubts on the significance of such a Newtonian
quantum gravity.

\bigskip
\bigskip

\appendix{A}{${\!\!\!\!\!\!}$ppendix: ~Review of the covariant
phase-space formalism}

\smallskip

Recall that the essence of the canonical formalism for a classical system with
finitely many degrees of freedom is succinctly captured by the symplectic
2-form\cite{Abraham}\cite{Arnold}\cite{Woodhouse}
$$
\omega\;=\;dp_i\wedge dq^i\;\,,\EQN 2-form-omega
$$
where ${q^i\,}$, ${\scriptstyle{i,\,j\,=\,1,\,\dots,\,N\,}}$, are the
generalized coordinates describing the configuration of the system, and
${p_j}$ are the corresponding conjugate momenta (here Einstein's summation
convention is understood). The collection of all possible values of coordinates
and momenta,
${(q_{{\!}_1},\,\dots,\,q_{{\!}_N}\,;\,p_{{\!}_1},\,\dots,\,p_{{\!}_N})\,}$, is
referred to as the {\it phase-space} of the system, on which the dynamical
evolution is determined by a Hamiltonian function ${\boh}$ via the Hamiltonian
equations of motion. If we combine the ${p_i}$ and ${q^j}$ in a single variable
${Q^{\scriptscriptstyle I}\,}$, ${\scriptstyle{I\,=\,1,\,\dots\,,\,2N\,}}$,
with ${Q^i:=p_i}$ for ${\scriptstyle{i\,\leq\,N}}$ and ${Q^i:=q^{i-N}}$ for
${\scriptstyle{i\,>\,N}\,}$, then we can think of ${\omega}$ as an
antisymmetric ${\scriptstyle{2N\times 2N}}$ matrix ${\omega_{{\!}_{IJ}}\,}$,
with inverse ${\omega^{\scriptscriptstyle{IJ}}\,}$, whose nonzero
matrix elements are
$$
\omega_{i,\,i+N}= -\,\omega_{i+N,\,i} = 1\;.\EQN
$$ 
With the help of this
invertible matrix ${\omega^{\scriptscriptstyle{IJ}}\,}$,
the Hamiltonian equations of motion can be succinctly expressed as
$$
{dQ^{\scriptscriptstyle{I}}\over{dt\;\,}}\;=\;
\omega^{\scriptscriptstyle{IJ}}\,{{\partial\boh}
\over{\partial Q^{\scriptscriptstyle{J}}}}\,,\EQN integral-curves
$$
and the Poisson bracket of any two functions
${U(Q^{\scriptscriptstyle I})}$ and ${V(Q^{\scriptscriptstyle I})}$ as
$$
\{U,\,V\}\;=\;\omega^{\scriptscriptstyle{IJ}}\,
{{\partial U}\over{\partial Q^{\scriptscriptstyle I}}}\,
{{\partial V}\over{\partial Q^{\scriptscriptstyle J}}}\;.\EQN
$$
Since the essential features of the symplectic form ${\omega}$ can be described
in a coordinate independent manner\cite{Abraham}\cite{Arnold}\cite{Woodhouse},
it can be viewed as the invariant geometric structure that underlies the
definitions of Hamiltonian equations and Poisson brackets of
classical mechanics. If we now view our phase-space as the cotangent-bundle
${T^*Z}$ on a configuration space ${Z}$ with coordinates ${q^i\,}$, and let
${p_j}$ be the corresponding components of a covector ${p}$ at ${q\in Z\,}$,
then the 2-form \Ep{2-form-omega} is a {\it closed} 2-form on ${T^*Z\,}$,
$$
d\omega\;=\;0\;,\EQN
$$
because its components in this particular coordinate system are constant.
Conversely, if ${\omega}$ is any closed 2-form on a ${\scriptstyle{2N}}$
dimensional manifold ${\cal Z}$ such that it is non-degenerate (i.e., the
matrix ${\omega_{{\!}_{IJ}}(z)}$ is invertible at each point ${z\in{\cal Z}}$),
then, according to Darboux's theorem\cite{Woodhouse}, locally one can always
introduce coordinates ${(p_i,\,q^j)}$ on ${\cal Z\,}$, called the
{\it canonical coordinates}, which put ${\omega}$ in the standard form
\Ep{2-form-omega}. This immediately suggests a generalization:
the cotangent-bundle form of phase-space is
only a special form of phase-space (but, of
course, of a fundamentally important kind) since, in general, the `isomorphism'
${{\cal Z}\cong T^*Z}$ holds only in a local neighborhood of a point
${z\in{\cal Z}\,}$. In other words, globally the phase-space may not be
anything like a cotangent-bundle. In general phase-space
is defined as a pair ${({\cal Z},\,\omega)}$ in which ${\cal Z}$ is a smooth
manifold, called a {\it symplectic manifold}, and ${\omega\,}$, defined
everywhere on ${\cal Z\,}$, is a closed, non-degenerate and real-valued
2-form called the
{\it symplectic structure} on ${\cal Z\,}$. The `observables' are then
functions of the form ${U:{\cal Z}\rightarrow\real}$ generating a one-parameter
family of canonical transformations on ${\cal Z\,}$. In
particular, the Hamiltonian of a classical system generating the time evolution
is a function ${\boh:{\cal Z}\rightarrow\real}$ on the phase-space
${{\cal Z}}$ of the system, 
and the dynamical trajectories governed by the Hamiltonian equations of motion
correspond to integral curves on ${\cal Z}$ of the Hamiltonian vector-field,
${X_{\scriptscriptstyle{\rm{I\!H}}}:{\cal Z}\rightarrow{\cal Z}\,}$, defined by
$$
{X_{\scriptscriptstyle{\rm{I\!H}}}}\cont\omega\,+\,d\,{\boh}
\;:=\;0\;.\EQN h-flow
$$
It is easy to verify that the flow of ${X_{\scriptscriptstyle{\rm{I\!H}}}}$
preserves ${\omega}$ in the sense that
${\hbox{\it\char36}{\!}_{X_{\scriptscriptstyle{\rm{I\!H}}}}\,\omega = 0\,}$.
Moreover, for a curve ${Q(t)}$ whose tangent vector at every point
coincides with ${X_{\scriptscriptstyle{\rm{I\!H}}}\,}$, the component form
of equation \Ep{h-flow} in local coordinates ${(p_i,\,q^j)}$ is precisely the
equation \Ep{integral-curves}.

The above abstract definition of phase-space quite elegantly encapsulates
the important fact that it is the assignment of a symplectic structure
${\omega}$ on the classical phase-space ${\cal Z\,}$, rather than a particular
choice of coordinates ${(p_i,\,q^j)}$ on ${\cal Z\,}$, that is more intrinsic
to the canonical formulation of a classical theory. Yet, {\it prima facie}, the
very concept of phase-space appears to be non-covariant because phase-spaces
are usually constructed by first decomposing spacetime into spacelike
hypersurfaces at instants of time, and then specifying the initial data
${(p_i,\,q^j)}$ on one of these hypersurfaces. However, thanks to the
one-to-one correspondence between each dynamically allowed trajectory of a
given classical system and its initial data, such a manifestly non-covariant
route to phase-space is not indispensable. It is clear from equation
\Ep{h-flow} that each point ${z\in{\cal Z}}$ is an appropriate initial data for
uniquely determining a complete Hamiltonian trajectory of the system. This
establishes an isomorphism between the space of solutions to the dynamical
equations and the canonical phase-space of the system, and, thereby,
allows one to pull-back the symplectic structure
from the phase-space to the space of solutions. As a result, we can identify
our phase-space ${\cal Z}$ with the manifold of solutions to the equations of
motion, which now we may also denote by ${\cal Z\,}$. This space of solutions
${\cal Z}$ equipped with the symplectic structure ${\omega}$ is then our
desired manifestly covariant phase-space ${({\cal Z},\,\omega)\,}$.
Of course, as allowed by Darboux's theorem, one can always choose a coordinate
system on ${\cal Z\,}$, at least locally, and identify the solutions to the
dynamical equations with their Cauchy data in that coordinate system; but there
is no fundamental necessity to violate covariance in this manner. Besides,
globally such an identification may not be possible in general, because, as
noted above, globally the manifold ${({\cal Z},\,\omega)}$ may not
be isomorphic to a cotangent-bundle. A cotangent-bundle can, of course, be
`polarized'\cite{Woodhouse}${\;}$--- i.e., foliated by the individual cotangent
spaces of constant ${q}$ --- and, thereby, the configuration and momentum
variables can be distinguished. But a general symplectic manifold may not be
polarizable in this manner, and, consequently, the global identification of
solutions to the equations of motion with integral curves of the Hamiltonian
vector-field becomes untenable in general. Instead, the `time-development' of
a classical system is understood within this framework by interpreting the
Hamiltonian flow as a mapping between {\it entire histories} of the system.
Given a notion of time, the mapping from one dynamical trajectory to another
infinitesimally distant one, whose initial data at time ${t+\epsilon}$ are the
same as the initial data of the first trajectory at time ${t\,}$, is
interpreted as the `time-evolution' from ${t}$ to ${t+\epsilon}$ generated by
the Hamiltonian function ${\boh:{\cal Z}\rightarrow\real\,}$.

The (pre)symplectic structure in the covariant description of an
{\it infinite}-dimensional phase-space ${({\cal Z},\,\omega)}$ of a field
theory is defined by a real valued function of the field variables,
$$
\omega\;=\;\int_{\Sigma_t}\omega^{\mu}\,n_{\mu}\;
d{\scriptstyle{\Sigma_t}}\;,\EQN omega-defi
$$
where ${\Sigma_t}$ is some spacelike hypersurface, ${n_{\mu}}$ is the inverse
of an `outward-pointing' unit normal to this surface, and ${\omega^{\mu}\,}$
--- called the {\it presymplectic current density} ---
is a {\it conserved} and {\it closed}
2-form (which may be degenerate, and hence the prefix `pre' meaning `before the
removal of degeneracy'; see below). As discussed above, if the manifold
${\cal Z}$ is polarizable so that we can distinguish between configuration and
momentum variables, then we can express it as a cotangent-bundle of the
configuration space. The above general expression for ${\omega}$
can then be reduced (at least in the case of a first-order action)
to the standard canonical presymplectic structure,
$$
\omega\;=\;\int_{\Sigma_t}dp_i\wedge dq^i
\;d{\scriptstyle{\Sigma_t}}\;,\EQN stand-omega
$$
once a choice of global coordinates ${(p_i,\,q^j)}$ is made.
Despite its appearance in \Ep{omega-defi} and \Ep{stand-omega}, ${\omega}$
is independent of the choice of hypersurface ${\Sigma_t}$ (provided that either
${\Sigma_t}$ is chosen to be compact or the field variables on it are subjected
to satisfy suitable boundary conditions at spatial infinity).

This manifest covariance of ${\omega}$ may require some convincing. Let us
look more closely at how it can be
shown by considering a dynamical theory for a collection of smooth fields
${{\cal Q}^r(x)}$ on a spacetime ${\M}$ equipped with a derivative operator
${\nabla_{\!\mu}\,}$; here $r$ is a collective label for fields representing
spacetime indices as well as internal and discrete
indices. We assume that our spacetime
${\M}$ is globally hyperbolic, i.e., topologically ${\real\times\Sigma\,}$,
where each image ${\Sigma_t}$
of ${\Sigma}$ for any ${t\in\real}$ is diffeomorphic to ${\real^3\,}$. Let
${Z}$ denote the infinite-dimensional manifold constituted by the fields
${{\cal Q}^r}$ on ${\M\,}$. Since functions on ${Z}$
are functionals of the form ${f[{\cal Q}(x)]\,}$, an action functional
${{\cal I}:Z\rightarrow\real}$ may be constructed over some measurable region
of $\M$ such that equations of motion for a field ${{\cal Q}^r}$ can be
obtained by extremizing the action under any variation
${\delta{\cal Q}^r\equiv{\tt Y}^r}$ of the field which vanishes on the
boundary of the region. The variations ${\delta{\cal Q}}$ here are tangent
vectors ${{\tt Y}\in T_{\cal Q}Z}$ at points in space ${Z}$ corresponding to
the fields ${{\cal Q}\,}$, the tangent spaces ${T_{\cal Q}Z}$ at ${\cal Q}$ are
vector spaces of the variations ${{\tt Y}\,}$, and the first variation
${\delta\,{\cal I}}$ of the action can be expressed as an exterior derivative
of ${\cal I}$ in ${Z}$ applied to ${\tt Y}$:
$$
\delta\,{\cal I}\;=\;d_z\,{\cal I}({\tt Y})\;;\EQN
$$
i.e., the variational derivative `${\delta}$' is equivalent to the exterior
derivative `${d_z}$' on the space ${Z\,}$. Then, for a given measurable region
${{\scriptstyle{\cal O}}\subset\M}$ with a non-null boundary
${\scriptstyle{\partial{\cal O}}}$ and a collection of fields ${{\cal Q}\in
Z\,}$, the variation of action
$$
{\cal I}[{\cal Q}]\;=\;\int_{\cal O}
d{\scriptstyle V}\;{\cal L}({\cal Q}^r,\,\nabla_{\!\mu}{\cal Q}^r,\,\dots,\,
\nabla_{\!\mu}\nabla_{\!\mu_{2}}\,\dots\nabla_{\!\mu_{k}}{\cal Q}^r)\EQN
$$
is equal to 
$$
d_z\,{\cal I}({\tt Y})\;=\;\int_{\cal O}
d{\scriptstyle V}\;\Leftcases{37pt}
{{\delta{\cal L}}\over{\delta{\cal Q}^r}}\,\delta{\cal Q}^r\;+\;
\nabla_{\!\mu}{\cal J}^{\mu}_{\!{\scriptscriptstyle{\cal Q}}}
\Rightcases{37pt}\;,\EQN pull-of-this
$$
where ${d{\scriptstyle V}}$ is the volume element on ${\M}$ compatible with
the derivative operator ${\nabla_{\!\mu}\,}$,
$$
{\cal J}^{\mu}_{\!{\scriptscriptstyle{\cal Q}}}\;:=\;\Leftcases{37pt}
{{\delta{\cal L}}\over{\delta(\nabla_{\!\mu}{\cal Q}^r)}}
\Rightcases{37pt}\delta{\cal Q}^r\,+\,\dots\,+\,
\Leftcases{37pt}{{\delta{\cal L}}\over
{\delta(\nabla_{\!\mu\mu_{2}\dots\mu_{k}}{\cal Q}^r)}}\Rightcases{37pt}
\nabla_{\!\mu_{2}\dots\mu_{k}}\delta{\cal Q}^r\EQN PPCD
$$
is what is called the {\it presymplectic potential current density}, and
the variational derivative ${{\delta\,}\over{\delta l}}$ of a local function
is defined as
$$
{{\delta\,}\over{\delta l}}\;:=\;{{\partial\,}\over{\partial l}}\;-\;
\nabla_{\!\mu}\Leftcases{37pt}
{{\partial\;\;\;}\over{\partial(\nabla_{\!\mu}l)}}
\Rightcases{37pt}\,+\;\nabla_{\!\mu}\nabla_{\!\nu}\Leftcases{37pt}
{{\partial\;\;\;\;\;}\over{\partial(\nabla_{\!\mu}\nabla_{\!\nu}l)}}
\Rightcases{37pt}\,-\;....\EQN
$$
Under the requirement that the action ${\cal I}$ remains stationary for any
variation ${\tt Y}$ of ${\cal Q}$ which vanishes on the boundary, the subspace
${{\cal Z}\subset Z}$ of solutions to the dynamical equations is defined by the
condition ${{{\delta{\cal L}}\over{\delta{\cal Q}}}=0\,}$. Conversely, on the
space ${\cal Z}$ consisting of fields ${\cal Q}$ which extremize
${{\cal I}\,}$, the pull-back of equation \Ep{pull-of-this} reduces to the
boundary term
$$
i^*\,d_z\,{\cal I}({\tt Y})\;=\;\oint_{\scriptstyle{\partial{\cal O}}}
{\cal J}^{\mu}_{\!{\scriptscriptstyle{\cal Q}}}\!({\tt Y})\,n_{\mu}\;
d{\scriptstyle S}\;,\EQN sigm
$$
where ${i:{\cal Z}\hookrightarrow Z}$ is the natural embedding of the
submanifold ${\cal Z}$ into ${Z\,}$, ${n_{\mu}}$ is the inverse of
an `outward-pointing'
unit normal to the boundary ${\scriptstyle{\partial{\cal O}}}$ as before,
${d{\scriptstyle S}}$ is the surface element on
${\scriptstyle{\partial{\cal O}}\,}$, and now the tangent vector ${{\tt Y}\in
T_{\cal Q}{\cal Z}}$ is a solution to the {\it linearized} equations of motion
at ${\cal Q}$ (i.e., both ${\cal Q}$ and ${{\cal Q}+\epsilon {\tt Y}}$ satisfy
the equations of motion to lowest order in ${\epsilon\,}$). Clearly, the
surface term \Ep{sigm} does not contribute to the variation of the action
because
${{\cal J}^{\mu}_{\!{\scriptscriptstyle{\cal Q}}}({\tt Y})}$ vanishes when
${{\tt Y}\equiv\delta{\cal Q}=0}$ is assumed on the boundary.
Nevertheless, it is precisely this term that captures the manifestly covariant
essence of the canonical structure of phase-space. In order to see this,
consider a volume segment ${{\scriptstyle{\cal O}(\Sigma_{t'}-\Sigma_t)}
\subset\M}$ which
is bounded by two Cauchy surfaces ${\Sigma_{t'}}$ and ${\Sigma_t}$ connected by
a timelike world tube ${{\cal T}_{\infty}}$ at spatial infinity, and define a
1-form ${\theta_{\Sigma_t}({\tt Y})}$ on ${\cal Z}$ for all tangent vectors
${{\tt Y}\in T_{\cal Q}{\cal Z}}$ by
$$
\theta_{\Sigma_t}({\tt Y})\;:=\;\int_{\Sigma_t}
{\cal J}^{\mu}_{\!{\scriptscriptstyle{\cal Q}}}\!({\tt Y})\,n_{\mu}\;
d{\scriptstyle{\Sigma_t}}\;,\EQN pre-potential
$$
so that in terms of this 1-form equation \Ep{sigm} becomes
$$
i^*\,d_z\,{\cal I}({\tt Y})\;=\;\theta_{\Sigma_{t'}}({\tt Y})\;-\;
\theta_{\Sigma_t}({\tt Y})\;+\;\theta_{{\cal T}_{\infty}}({\tt Y})\;.
\EQN onup
$$
Note that, in general, this {\it presymplectic potential}
${\theta_{\Sigma_t}({\tt Y})}$ (as it is sometimes called in suggestive analogy
with the electromagnetic vector-potential) is not unique and
{\it depends} on the choice of
hypersurface ${\Sigma_t\,}$. However, a presymplectic 2-form on ${\cal Z}$ can
now be defined as the exterior derivative of ${\theta_{\Sigma_t}({\tt Y})\,}$,
$$
\omega_{\Sigma_t}({{\tt Y}\suba},\,{{\tt Y}\subb})\;:=\;
d_z\,\theta_{\Sigma_t}({{\tt Y}\suba},\,{{\tt Y}\subb})\;,\EQN cova-omeg
$$
which in the light of equation \Ep{onup} is immediately seen to
behave as
$$
0\;\equiv\;i^*\,d_z^2\,{\cal I}\;=\;\omega_{\Sigma_{t'}}\,-\,
\omega_{\Sigma_t}\,+\,\omega_{{\cal T}_{\infty}}\EQN 4.60
$$
under changes of the Cauchy surface ${\Sigma_t\,}$.
Thus, in case ${\Sigma_t}$ is non-compact,
simply by a suitable choice of boundary condition (e.g., 
${{\cal J}^{\mu}_{\!{\scriptscriptstyle{\cal Q}}}\rightarrow 0}$ at spatial
infinity ensuring the vanishing of the term ${\omega_{{\cal T}_{\infty}}}$)
we obtain the desired covariant behavior of the presymplectic structure as
claimed above.

In the view of equation \Ep{cova-omeg}, the closed-ness of the 2-form
${\omega}$ is immediate: ${d_z\,\omega=0\,}$. Whereas comparisons of equations
\Ep{omega-defi}, \Ep{pre-potential}, \Ep{cova-omeg},
and \Ep{4.60} show that the corresponding {\it presymplectic current density}
$$
\omega^{\mu}({{\tt Y}\suba},\,{{\tt Y}\subb})\;:=\;d_z\,
{\cal J}^{\mu}_{\!{\scriptscriptstyle{\cal Q}}}
({{\tt Y}\suba},\,{{\tt Y}\subb})\;,\EQN j-to-w
$$
whose integral over ${\Sigma_t}$ gives the 2-form ${\omega_{\Sigma_t}}$ on
${{\cal Z}\,}$, is conserved,
$$
\nabla_{\!\mu}\,\omega^{\mu}\;=\;0\;,\EQN
$$
thanks to the equations of motion ${{{\delta{\cal L}}\over{\delta{\cal Q}}}=0}$
and their linearized versions satisfied, respectively, by the fields
${{\cal Q}}$ and their variations ${{\tt Y}\,}$. Thus, at each spacetime point
${x\,}$, ${\omega^{\mu}(x)}$ is a vector-valued
2-form on the space ${{\cal Z}}$ of classical solutions, but in its
dependence on ${x\,}$, it is a conserved current density. Furthermore,
in case ${\Sigma_t}$ is non-compact, if a 4-divergence term is added to the
Lagrangian, ${{\cal L}\rightarrow{\cal L}+\nabla_{\!\mu}\lambda^{\mu}\,}$,
then the potential current density
${{\cal J}^{\mu}_{\!{\scriptscriptstyle{\cal Q}}}}$ changes only by an exact
form ${d_z\lambda^{\mu}}$ plus an identically conserved vector density,
changing the 2-form ${\omega}$ only by a `surface term at infinity'\cite{Lee}.

It is clear from our notation in the equation \Ep{cova-omeg} that
${\omega}$ can also be viewed as a skew-symmetric bilinear function
${\omega:T_z{\cal Z}\times T_z{\cal Z}\rightarrow\real}$ on the tangent
vector-space ${T_z{\cal Z}}$ at some point $z$ on the phase-space
${{\cal Z}\,}$. The phase-space of a linear (i.e., non-self-interacting)
dynamical system is the prime example of such a symplectic vector-space with
a bilinear form\cite{Wald-94}.
The field equations of linear dynamical systems are
`already linearized'; i.e., the equations of motion for such systems are
linear in the linear canonical coordinates because their
Hamiltonians are at most quadratic functions on ${{\cal Z}\,}$. 
Consequently, their phase-spaces, viewed as manifolds ${\cal Z}$
of solutions, have a natural vector-space structure, and one may identify the
tangent space ${T_z{\cal Z}}$ at any point ${z\in{\cal Z}}$ with the space
${\cal Z}$ itself. The symplectic form ${\omega({\cal Q}\suba,\,
{\cal Q}\subb)\,}$, with ${{\cal Q}\suba,{\cal Q}\subb\in{\cal Z}\,}$, then
becomes a bilinear map on the symplectic vector-space ${{\cal Z}\,}$,
${\,\omega:{\cal Z}\times {\cal Z}\rightarrow\real\,}$,
which is independent of the choice of point $z$ used to make
this identification.

The presymplectic 2-form ${\omega}$ on ${\cal Z}$ we have defined so far is
necessarily degenerate if the action functional ${\cal I}$ admits any
gauge-arbitrariness. Conversely, a degenerate symplectic structure would
necessitate gauge-type symmetries in a theory (the issue, of course, is closely
related to the question of whether or not the Cauchy problem for
${{{\delta{\cal L}}\over{\delta{\cal Q}}}=0}$ is well posed). However,
at least in simple cases, it is possible to obtain
a genuine (i.e, non-degenerate) symplectic 2-form ${\widetilde\omega}$ on a
reduced phase-space ${\widetilde{\cal Z}}$ by the so-called {\it reduction
procedure}\cite{Lee}\cite{Woodhouse}. A 2-form ${\widetilde\omega}$ of the pair
${({\widetilde{\cal Z}},\,{\widetilde\omega})}$ is said to be
{\it non-degenerate} (or {\it weakly non-degenerate}, to be more precise) if
${{\widetilde\omega}({{\tt Y}\suba},\,{{\tt Y}\subb}) = 0}$ for all
${{{\tt Y}\subb}\in{\widetilde{\cal Z}}}$ implies ${{{\tt Y}\suba} = 0\,}$,
and it is said to be {\it strongly non-degenerate} if the map
${T_z{\widetilde{\cal Z}}\rightarrow T^*_z{\widetilde{\cal Z}}\,:\,
{\tt Y}\mapsto {\tt Y}\cont\omega}$ is a linear isomorphism at each point
${z\in{\widetilde{\cal Z}}}$ (the criteria of weak and strong non-degeneracy
are clearly equivalent when ${\widetilde{\cal Z}}$ is finite dimensional).
The reduction procedure amounts to projecting down the
presymplectic 2-form ${\omega}$ from the space ${\cal Z}$ of solutions of the
equations of motion to the {\it physical} phase-space ${\widetilde{\cal Z}}$
--- the space of solutions to the equations of motion {\it modulo
gauge-transformations}. The projection map ${\proj:{\cal Z}\rightarrow
{\widetilde{\cal Z}}}$ which accomplishes this, assigns each element of
${\cal Z}$ to its gauge-equivalence class. It can be shown that every
infinitesimal gauge transformation corresponds to a degenerate direction of the
2-form ${\omega}$ on ${{\cal Z}}$ by showing that ${\omega}$ is actually a
unique pull-back ${\proj^*({\widetilde{\omega}})}$ from ${\widetilde{\cal Z}}$
to ${\cal Z}$ of a {\it non-degenerate} 2-form ${\widetilde{\omega}}$ on the
reduced phase-space ${{\widetilde{\cal Z}}\,}$. This reduced phase-space, in
turn, is just the space of orbits in ${\cal Z}$ of the group ${\cal K}$ of
gauge transformations; i.e., the reduced phase-space is the
quotient space ${{\widetilde{\cal Z}}:={\cal Z}/{\cal K}\,}$, where ${\cal K}$
is the {\it characteristic distribution} of ${\omega}$ with fibers
${{\cal K}_z}$ at ${z\in{\cal Z}}$ defined as\cite{Woodhouse}
$$
{\cal K}_z\;\;:=\;\;\{{\tt Y}\in T_z{\cal Z}\,|\;
{\tt Y}\cont\omega\,=\,0\,\}\;\;\subset\;\;T_z{\cal Z}\;.\EQN distribution
$$
In other words, any nonvanishing vector field ${{\tt Y}\!\subk}$ tangent to
the ${\cal K}$-orbits on ${\cal Z}$ is a degenerate direction of ${\omega\,}$:
${{{\tt Y}\!\subk}\in {V\!\!\subk}({\cal Z})
\Rightarrow\omega({{\tt Y}\!\subk}\,,
{\tt Y})=0}$ ${\,\forall\,\;{\tt Y}\in T_z{\cal Z}\,}$, where
${{V\!\!\subk}({\cal Z})}$ denotes the set of vector fields tangent to
${{\cal K}\,}$. Then, for one thing, the Lie derivative
of ${\omega}$ along the degenerate directions is always zero:
$$
{\tt Y}\;\in\;{V\!\!\subk}({\cal Z})\;\;\;\Rightarrow\;\;\;
\hbox{\it\char36}_{{}_{\rm Y}}\,\omega\;=\;
d_z(\,{\tt Y}\cont\omega)\;+\;{\tt Y}\cont d_z\,\omega\;=\;0\;.\EQN
$$
Agreeably, the quotient space ${{\cal Z}/{\cal K}={\widetilde{\cal Z}}}$
is a Hausdorff manifold since the characteristic
distribution ${\cal K}$ on ${\cal Z}$ is an integrable submanifold: if
${{\tt Y}\suba}$ and ${{\tt Y}\subb}$ are degeneracy vector fields with values
in ${{\cal K}\,}$, then so is ${[{{\tt Y}\suba},\,{{\tt Y}\subb}]\,}$.
Clearly, the gauge-directions are eliminated in passing from ${\cal Z}$ to
${\widetilde{\cal Z}}$, rendering ${\widetilde{\omega}}$ on
${\widetilde{\cal Z}}$ manifestly gauge-invariant.
In practice, all one has to do to ensure gauge-invariance is to make sure
that a given presymplectic (i.e., degenerate) structure ${\omega}$ on
${\cal Z}$ is a pull-back of the genuine (i.e., non-degenerate)
symplectic structure ${\widetilde{\omega}}$ from the reduced phase-space
${{\widetilde{\cal Z}}\,}$: ${\omega=\proj^*({\widetilde{\omega}})\,}$.
This is guaranteed if and only if ${\widetilde{\omega}}$ on
${{\widetilde{\cal Z}}}$ has vanishing components in the gauge-directions:
$$
{{\tt Y}\!\subk}\cont{\widetilde{\omega}}\;=\;0\,.\EQN gauge-directions
$$

Finally, note that, although ${\omega}$ is exact by definition
(${\omega=d\theta}$; cf. equation \Ep{cova-omeg}), ${\widetilde{\omega}}$ need
not satisfy this property, because, in general, the quotient space 
${\widetilde{\cal Z}}$ could be topologically much more subtle compared to the
space ${{\cal Z}\,}$. It is always possible, however, to find a
local neighborhood on ${\widetilde{\cal Z}}$ such that
${{\widetilde{\omega}}=d{\widetilde{\theta}}}$ within it. If
${\widetilde{\cal Z}}$ happens to be symplectically diffeomorphic to
a cotangent-bundle, then, of course,
${\widetilde{\omega}}$ is its natural exact symplectic structure with
${\widetilde{\theta}}$ being the standard canonical 1-form\cite{Woodhouse}.
For convenience, let us illustrate the relations between various spaces
we have encountered by the following diagram:

$$
\matrix{Z&\;\;\mapright{{}i^*}\;\;&{\cal Z}\cr
&&\cr
\mapdown{\rm{I\!P}}&&\mapdown{\rm{I\!P}}\cr
&&\cr
&&\cr
{\widetilde{Z}}&\;\;\mapright{{}i^*}\;\;&{\widetilde{\cal Z}}\cr}\;\;\;\;
\EQN Monique
$$
Here ${\widetilde{Z}}$ denotes collection of all kinematically possible field
configurations modulo gauge-transformations (cf. equation
\Ep{configuration-space}), whereas 
${{\widetilde{\cal Z}}\subseteq{\widetilde{Z}}}$ denotes collection of purely
dynamically possible field configurations modulo gauge-transformations. If
constraints are present, however, then
${{\widetilde{\cal Z}}\subset{\widetilde{Z}}\,}$, and not all
kinematically possible states are dynamically possible.

\bigskip
\vskip 0.4in
\nosechead{\bf References}
\bigskip\ListReferences
\end